\documentclass[ aps,
                physrev,
                preprint,
                onecolumn,
                floatfix,
                superscriptaddress]{revtex4-2}
\usepackage[left=1in, right=1in, top=1in, bottom=1in]{geometry}
\usepackage{mlmodern}
\usepackage{graphicx}
\usepackage{bm}
\usepackage{physics}
\usepackage[unicode=true,
    psdextra,
    colorlinks=true,
    pdfstartview=FitV,
    linkcolor=blue,
    citecolor=blue,
    urlcolor=blue,
    anchorcolor = blue]{hyperref}
\usepackage{cleveref}
\begin{document}
%%%%%%%%%%%%%%%%%%%%%%%%%%%%%%%% TITLE %%%%%%%%%%%%%%%%%%%%%%%%%%%%%%%%%%%%%%%%%%%%
\title{Deformation dynamics of Oldroyd B drop in alternating electric field}

%%%%%%%%%%%%%%%%%%%%%%%%%%%%%%%%% AUTHORS %%%%%%%%%%%%%%%%%%%%%%%%%%%%%%%%%%%%%%%%%
\author{Sarika Shivaji Bangar}
\email[Corresponding author: ]{sarikabangar@iisc.ac.in}
\affiliation{Department of Mechanical Engineering, Indian Institute of Science, Bengaluru-560012, India}

\author{Gaurav Tomar}%
\email[Corresponding author: ]{gtom@iisc.ac.in}
\affiliation{Department of Mechanical Engineering, Indian Institute of Science, Bengaluru-560012, India}%

\date{\today}

%%%%%%%%%%%%%%%%%%%%%%%%%%%%%%%% ABSTRACT %%%%%%%%%%%%%%%%%%%%%%%%%%%%%%%%%%%%%%%%%
\begin{abstract}
    The deformation of viscoelastic drops in alternating electric fields underlies electrohydrodynamic applications such as microfluidics, inkjet printing, and drop manipulation. In this work, we investigate the dynamics of a neutrally buoyant Oldroyd-B drop subjected to a uniform alternating electric field using asymptotic analysis and direct numerical simulations with the open-source solver Basilisk. An analytical solution is derived in the limits of small deformation and weak elasticity for an Oldroyd-B drop suspended in an Oldroyd-B medium, with both fluids modeled as leaky dielectrics under axisymmetric Stokes flow. Depending on the conductivity and permittivity ratios, six electrohydrodynamic regions are identified, distinguished by deformation mode, flow direction, and nonlinear response to the applied field; while some regions exhibit stable spheroidal deformation, others transition to pointed-tip, multi-lobed, or oblate breakup states beyond a critical electric capillary number. Across high-, intermediate-, and low-frequency regimes, the drop deformation oscillates at twice the applied field frequency, with its mean value and amplitude governed jointly by the field frequency and viscoelasticity. For parameter combinations that deform the drop into a prolate shape under a steady field, the drop remains prolate at high frequency, is predominantly prolate with brief oblate excursions at intermediate frequency, and undergoes large-amplitude oscillations between near-spherical and highly deformed -- lobed or pointed -- states at low frequency; the mean deformation decreases monotonically or varies non-monotonically with $De$ depending on the field frequency and electrical property ratios. In contrast, for parameter combinations that deform the drop into an oblate shape under a steady field, the drop oscillates between oblate shapes at high frequency, develops dimples at maximum deformation at intermediate frequency, and breaks up at low frequency once $De$ and $Ca_E$ are sufficiently large; here the mean deformation increases monotonically with $De$.
\end{abstract}

%%%%%%%%%%%%%%%%%%%%%%%%%%%%%%%% KEYWORDS %%%%%%%%%%%%%%%%%%%%%%%%%%%%%%%%%%%%%%%

\maketitle
\newpage

\section{Introduction}\label{sec:introduction}
The study of drop deformation under electric fields underpins numerous technological and natural phenomena. Applications span inkjet printing (\citet{basaran2013nonstandard, lau2017ink}), electrostatic coating (\citet{hines1966electrostatic}), electrospraying and atomization (\citet{kelly1984electrostatic, law2018electrostatic}), and microfluidic drop manipulation (\citet{laser2004review, stone2004engineering, phan2025demand}). Electric-field-driven deformation is also critical in oil de-emulsification processes (\citet{eow2002electrostatic, alvarado2010enhanced, zhang2011application}) and electric propulsion technologies (\citet{moreau2015electrohydrodynamic, huh2019numerical}). Beyond engineered systems, these mechanisms provide insight into natural processes such as rain electrification and drop fragmentation in thunderstorms (\citet{simpson1909electricity, wilson1921iii, blanchard1963electrification}).

When a dielectric drop is exposed to an external electric field, it generally elongates into a prolate spheroid at weak fields and disintegrates once the field strength becomes sufficiently high (\citet{o1953distortion, taylor1964disintegration}). The underlying mechanism was first attributed to the competition between interfacial electric stresses and surface tension by \citet{allan1962particle}, while \citet{o1953distortion} independently reached similar conclusions through an energy-based formulation. To account for the appearance of oblate shapes observed in experiments, \citet{o1957electric} emphasized the crucial role of electrical conductivity in determining whether a drop assumes a prolate or oblate configuration.
Building on these foundational studies, \citet{taylor1966studies} formulated the classical leaky dielectric model (LDM), which demonstrated that tangential electric stresses generate internal circulatory flows, with the direction and magnitude of deformation governed by the relative conductivity and permittivity of the fluids. The model was subsequently refined by \citet{melcher1969electrohydrodynamics}, extended to alternating electric fields by \citet{torza1971electrohydrodynamic}, and corrected for higher-order contributions by \citet{ajayi1978note}. Despite these advancements, experimental results by \citet{torza1971electrohydrodynamic} consistently showed greater deformation than predicted, a discrepancy not resolved even after incorporating higher-order effects. A comprehensive review of the LDM can be found in \citet{saville1997electrohydrodynamics}.

The electrohydrodynamics of Newtonian drops subjected to alternating current (AC) electric fields has been the focus of extensive investigation. The pioneering work of \citet{torza1971electrohydrodynamic} extended the leaky dielectric model to AC conditions by solving the axisymmetric Stokes and Laplace equations. Their analysis revealed that the electric stress, velocity field, and deformation amplitude of the drop oscillate at twice the applied frequency, with deformation oscillations becoming more pronounced as the frequency decreases. Subsequently, \citet{sozou1972electrohydrodynamics} enhanced this framework by incorporating the unsteady inertial term ($\rho\pdv{\bm{u}}{t}$) into the momentum equation, demonstrating that its omission leads to significant errors at high frequencies or in systems involving low-viscosity fluids.
\citet{vlahovska2009electrohydrodynamic} formulated an electrohydrodynamic theory for vesicles in AC fields, showing that vesicles with a more conductive interior deform into a prolate shape, whereas the inverse conductivity ratio induces a transition from prolate to oblate morphology at a critical frequency. \citet{thaokar2012dielectrophoresis} conducted an asymptotic study on dielectrophoretic motion and deformation of drops in non-uniform AC fields, while \citet{mohanty2024analytical} analytically examined drop dynamics in leaky dielectric media under combined alternating and steady electric fields. Their findings indicated that the time-averaged deformation follows the same scaling as that produced by a DC field of equivalent root-mean-square (RMS) magnitude, decreasing with increasing frequency toward a high-frequency plateau. They also observed that while mean deformation is only weakly sensitive to the viscosity ratio, oscillation amplitude depends strongly on it.
To overcome the limitations of asymptotic approximations, \citet{esmaeeli2018electrohydrodynamics} performed direct numerical simulations of leaky dielectric drops in AC fields and confirmed that the mean deformation closely matches that caused by an RMS-equivalent DC field. However, \citet{sahu2020simulations} later showed that this AC–DC equivalence holds strictly only when the conductivity and permittivity ratios are identical.
Recent studies have expanded this understanding to more complex systems. \citet{soni2018electrohydrodynamics} combined analytical and experimental methods to investigate compound drops in AC fields, while \citet{song2024electrokinetic} explored electro-capillary-driven Marangoni flows at the interface of liquid metal drops. \citet{santra2024modulating} provided numerical and theoretical insights into drops in extensional flow subjected to AC fields, showing that asymptotic predictions underestimate deformation at high electric capillary numbers. They demonstrated that extensional flow induces prolate deformation through viscous stresses, whereas the AC field imposes oscillatory electric and hydrodynamic stresses that can alternately promote prolate or oblate shapes depending on the phase of the field. Their study revealed that the deformation dynamics, breakup length, and breakup time are governed by the interplay between viscous stresses and time-varying electric forces.
Most recently, \citet{kireev2025influence} examined the effects of both steady and alternating electric fields on drop deformation and breakup. Their results showed that the steady-state deformation primarily depends on the electric capillary number and is only weakly influenced by viscosity, whereas transient maximum deformation increases as the viscosity ratio decreases due to reduced damping. Simulations of a water drop in air revealed resonance behavior near 120 Hz, delayed deformation at 50 Hz, and breakup above a critical drop size. At very high frequencies (500 Hz), electric field effects diminished, as the oscillation period became shorter than the characteristic mechanical relaxation time of the drop.

Despite the ubiquity of viscoelastic drops in natural and industrial systems, studies on their electrohydrodynamics remain limited. \citet{ha1999deformation} first analyzed viscoelastic effects on drop deformation and stability in electric fields using a second-order fluid model. \citet{ha2000deformation} compared Newtonian and non-Newtonian conducting drops in electric field. Their results showed that elasticity, whether shear-independent or shear-rate-dependent, reduces deformation and enhances stability; however, elasticity in the suspending medium stabilizes drops only at low viscosity ratios and may promote instability at higher ones. \citet{lima2014numerical} numerically examined Giesekus drops, demonstrating that longer polymer relaxation times suppress deformation through elastic resistance but also lower effective viscosity via shear thinning. More recent studies have extended this understanding. \citet{zhao2025electrohydrodynamic} reported that elasticity decreases both deformation and rotation in combined electric and shear flows, while \citet{das2026effect} developed analytical and numerical results for Oldroyd-B drops, showing deformation trends across small to large Deborah number regimes. In a recent study, \citet{bangar2026large} examined the large deformation of an Oldroyd-B drop under a steady external electric field. In a subsequent study, \citet{BANGAR2026105645} investigated the influence of a uniform steady electric field on the deformation of a shear-thinning drop using the linear PTT model, revealing intriguing non-monotonic variations of deformation with Deborah number that are not observed in Oldroyd-B drops.

Despite these advances, a comprehensive understanding of viscoelastic drop electrohydrodynamics under alternating electric fields remains limited. In this work, the deformation and breakup dynamics of an Oldroyd-B drop subjected to a uniform AC electric field are investigated through analytical and numerical approaches. The Oldroyd-B model captures the key viscoelastic effects associated with polymer elasticity while excluding shear-thinning and finite-extensibility, making it a fundamental framework for assessing elastic influences on drop behavior in oscillatory fields. In the first part of the study, an asymptotic analysis is carried out to obtain the deformation of a weakly viscoelastic drop under small deformations and low Deborah numbers. To explore regimes of strong elasticity and large deformations, numerical simulations are subsequently performed, enabling detailed characterization of nonlinear electrohydrodynamic responses and breakup behavior.

%%%%%%%%%%%%%%%%%%%%%%%%%%%%%%%%%%%% PROBLEM FORMULATION %%%%%%%%%%%%%%%%%%%%%%%%%%%%%%%%%%%%%%%
\section{Problem formulation}\label{sec:problemFormulation}
We consider a fluid drop of radius $\tilde{R}_0$, suspended in another immiscible fluid and subjected to an oscillatory electric field of the form $\tilde{\bm{E}}_\infty \cos(\omega \tilde{t})$. The drop and surrounding medium are denoted by the subscripts $i$ and $e$, respectively. The electrical properties of each phase are characterized by conductivity $\sigma$ and permittivity $\epsilon$, while $\omega$ represents the frequency of the imposed alternating field. The viscoelasticity of the fluid is modeled using the Oldroyd-B constitutive relation, with polymer relaxation time $\lambda$, solvent viscosity $\mu_s$, and polymeric viscosity $\mu_p$. The ratio of solvent to total viscosity of a fluid is defined as $\beta = \mu_s / (\mu_s + \mu_p)$, and the interfacial tension between the two fluids is denoted by $\gamma$.

\subsection{Governing equations}
The mass and momentum conservation for incompressible flow is given by
\begin{equation}
	\bm{\nabla}.\bm{u} = 0
    \label{Eqn_Continuity}
\end{equation}
and 
\begin{equation}
\rho\frac{D \bm{u}}{D t} = -\bm{\nabla}p + \bm{\nabla}.\bm{\tau} + \bm{F}
    \label{Eqn_NS1}
\end{equation}
respectively, where $\bm{u}$ is the fluid velocity vector, $\rho$ is the density, $p$ is pressure and $\bm{\tau}$ is stress arising from deformation of the fluid and $\bm{F}$ comprises all external body forces.
The stress tensor $\bm{\tau}$ is related to the fluid velocity gradient tensor through a constitutive relation.
For a Newtonian fluid, the stress is directly proportional to the strain rate, thus,
\begin{equation}
	\bm{\tau} = 2\mu \bm{D}
\end{equation}
where $\mu$ is the dynamic viscosity of the fluid and $\bm{D}$ is given by,
\begin{equation}
	\bm{D} = \frac{\bm{\nabla u} + (\bm{\nabla u})^T}{2}
\end{equation}
is the deformation tensor. For the viscoelastic model used in this study (Oldroyd-B), the stress tensor consists of solvent and polymeric stress, given by,
\begin{equation}
	\bm{\tau} = \bm{\tau}_s + \bm{\tau}_p
    \label{Eqn_Stress_split}
\end{equation}
where the solvent stress tensor is $\bm{\tau}_s = 2\mu_s \bm{D}$ where $\mu_s$ is the dynamic viscosity of the solvent.
The polymeric stress tensor is governed by the evolution equation,
\begin{equation}
    \bm{\tau}_p + \lambda\left(\frac{\partial \bm{\tau}_p}{\partial t} + (\bm{u}.\bm{\nabla})\bm{\tau}_p - \bm{\tau}_p(\bm{\nabla u}) - (\bm{\nabla u})^T\bm{\tau}_p\right) = \mu_p(\bm{\nabla u} + (\bm{\nabla u})^T)
    \label{Eqn_PolymericStress}
\end{equation}
where $\lambda$ is the relaxation time of the polymeric fluid and $\mu_p$ is the polymeric viscosity given by $\mu_p = \mu - \mu_s$. 

For a fluid in presence of an electric field, $\bm{E}$,  the external body force $\bm{F}$, is given by,
\begin{equation}
    \bm{F} = \bm{\nabla} . \bm{\tau}^{E}
\end{equation}
where $\bm{\tau}^{E}$ is the Maxwell's stress tensor (neglecting the magnetic field)
given by,
\begin{equation}
    \bm{\tau}^{E} = \epsilon \Big( \bm{E}\bm{E} - \frac{1}{2} (\bm{E}.\bm{E}) \bm{I}  \Big).
\end{equation}
Here, $\epsilon$ is the permittivity of the fluid medium. 
Since characteristic time scale for magnetic phenomenon ($t_M = \mu_M \sigma L^2$, where $\mu_M$ is magnetic permeability, $\sigma$ is electric conductivity and $L$ is the characteristic length scale) is much smaller than electric relaxation time ($t_E = \frac{\epsilon}{\sigma}$, where $\epsilon$ is the electric permittivity), magnetic effects can be ignored. Thus the electric field $\bm{E}$ is governed by Gauss's law and Faraday's law as follows.
From Gauss's law in a dielectric material,
\begin{equation}
	\bm{\nabla}.(\epsilon \bm{E}) = q_f
    \label{Eqn_GaussLaw}
\end{equation}
where $\epsilon$ is the permittivity of the medium and $q_f$ is the free charge density. 
From Faraday's law, 
\begin{equation}
	\bm{\nabla} \times \bm{E} = 0.
         \label{Eqn_FaradayLaw}
\end{equation}
Thus electric field can be represented in terms of scalar electric potential $V$, leading to equation,
\begin{equation}
\label{Eqn_Poisson_potential}    
	\bm{\nabla}.(\epsilon \bm{\nabla}V) = -q_f
\end{equation}
Conservation equation for bulk free charge is given by,
\begin{equation}
	\pdv{q_f}{t} + \bm{\nabla}.\bm{J}_f = 0
\end{equation}
where the current density $\bm{J}_f$ is given by
\begin{equation}
	\bm{J}_f = \sigma \bm{E} + q_f\bm{u}
\end{equation}
where first term is the current density due to Ohmic conduction and second term is because of convection of charges.
Thus we have,
\begin{equation}
\label{Eqn_Governing_charge}
    \pdv{q_f}{t} + \bm{u} \cdot \grad{q_f}  =  \bm{\nabla}.\Big( \sigma \grad{V}  \Big).
\end{equation}
Solving equations \autoref{Eqn_Poisson_potential} and \ref{Eqn_Governing_charge} with appropriate boundary conditions, we get the charge density and electric potential, using which we find the electric field $\bm{E}$. Thus one can find the Maxwell's stress tensor and get the external body force due to electric field.

%%%%%%%%%%%%%%%%%%%%%%%%%%%%%%%%%%%%%%%%%%%%%%%%%%%%%%%%%%%%%%%%%%%%%%%%%%%%%%%%%%%%%%%%
%%%%%%%%%%%%%%%%%%%%%%%%%%%%%%%%%%%% ASYMPTOTIC %%%%%%%%%%%%%%%%%%%%%%%%%%%%%%%%%%%%%%%
\section{Asymptotic analysis}\label{sec:asymptotic}
\begin{figure}[ht!]
    \centering
    \includegraphics{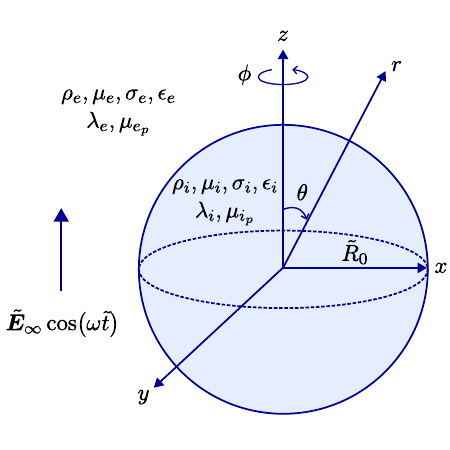}
    \caption{Schematic of the problem showing a drop of radius $\tilde{R}_0$ is subjected to the alternating electric field $\tilde{\bm{E}}\cos(\omega\tilde{t})$. Problem is formulated in the spherical axisymmetric coordinate system with origin at the center of the drop. Physical properties of drop and ambient phases are denoted by subscripts $i$ and $e$ respectively.}
    \label{Fig_schematic_analytical}
\end{figure}
For asymptotic analysis, the problem is formulated in spherical axisymmetric coordinates, with the $z$-axis as the axis of symmetry (\autoref{Fig_schematic_analytical}), and $\theta$ and $\phi$ denoting the polar and azimuthal angles, respectively. The electric potentials in the drop and ambient phases are expressed in complex form as  
\begin{equation}
    \tilde{V}_i' = \tilde{V}_i^* e^{i\omega \tilde{t}}, \qquad 
    \tilde{V}_e' = \tilde{V}_e^* e^{i\omega \tilde{t}},
\end{equation}
with their real parts $\tilde{V}_i$ and $\tilde{V}_e$ representing the physical potentials. 
Similarly, the interfacial charge density is written as 
\begin{equation}
    \tilde{q}' = \tilde{q}^* e^{i\omega \tilde{t}}
\end{equation}
\citep{torza1971electrohydrodynamic}. 
The drop interface deforms under the applied electric field, and its shape is not known a priori. In the small-deformation limit, the interface is approximated as  
\begin{equation}
    \tilde{r}_s = \tilde{R}_0 \big(1 + f(\theta,\tilde{t})\big).
\end{equation}
The governing equations and boundary conditions are non-dimensionalized to reduce the number of parameters. Dimensional variables are denoted with a tilde, while non-dimensional variables are written without tilde. In the Stokes-flow limit, viscous effects dominate; accordingly, the viscous scale is used to non-dimensionalize pressure and stress. Characteristic scales are based on the ambient fluid properties, and the velocity scale is determined by balancing electric and viscous stresses:
\begin{equation}
\tilde{U} = \frac{\epsilon_e \tilde{E}_\infty^2 \tilde{R}_0}{\mu_e}.
\end{equation}
Non-dimensional variables are defined as:
\begin{align*}
&\bm{E} = \frac{\tilde{\bm{E}}}{\tilde{E}_\infty}, \quad V = \frac{\tilde{V}}{\tilde{E}_\infty \tilde{R}_0}, \quad 
\bm{\nabla} = \tilde{R}_0 \tilde{\bm{\nabla}}, \quad \kappa = \tilde{R}_0 \tilde{\kappa}, \quad t = \frac{\tilde{t} \tilde{U}}{\tilde{R}_0},\\
&\bm{u}_{i,e} = \frac{\tilde{\bm{u}}_{i,e}}{\tilde{U}}, \quad 
p_{i,e} = \frac{\tilde{p}_{i,e} \tilde{R}_0}{\mu_e \tilde{U}}, \quad 
\bm{\tau}_{i,e} = \frac{\tilde{\bm{\tau}}_{i,e} \tilde{R}_0}{\mu_e \tilde{U}}.
\end{align*}

Under the leaky-dielectric assumption, charge resides only at the interface and the bulk is essentially charge-free. The non-dimensional electric potential in the drop and ambient phases, satisfies
\begin{equation}
    \nabla^2 V^* = 0 \quad \text{where} \quad \bm{E}^* = -\nabla V^*.
\end{equation}
Electric field satisfies the following boundary conditions:
\begin{subequations}
\begin{equation}
    V_e^* = - r \cos\theta \quad \text{as} \quad r \to \infty
\end{equation}
\begin{equation}
    \frac{\partial V_i^*}{\partial r} = 0 \quad \text{at} \quad  r = 0
\end{equation}
\begin{equation}
    V_i^* = V_e^* \quad \text{at} \quad  r = 1 + f(\theta,t)
\end{equation}
\begin{equation}
    (1 + i \omega a) \bm{E}_e^* \cdot \bm{n} = (\sigma_r + i a \omega \epsilon_r) \bm{E}_i^* \cdot \bm{n} \quad \text{at} \quad  r = 1 + f(\theta,t)
\end{equation}
\end{subequations}
where $t_R = \omega \tilde{R}_0 / \tilde{U}$, $\sigma_r = \sigma_i / \sigma_e$, $\epsilon_r = \epsilon_i / \epsilon_e$, and $a = \epsilon_e / \sigma_e$.  
The non-dimensional Maxwell stress is
\begin{equation}
    \bm{\tau}^E = \mathcal{E} \Big(\bm{E} \bm{E} - \frac{1}{2} (\bm{E} \cdot \bm{E}) \bm{I} \Big),
\end{equation}
with non-dimensional electrical permittivity $\mathcal{E} = 1$ for the ambient fluid and $\mathcal{E} = \epsilon_r$ for the drop.

The non-dimensional flow equations are
\begin{align}
    \nabla \cdot \bm{u} &= 0, \\
    Re \left( \frac{\partial \bm{u}}{\partial t} + \nabla \cdot (\bm{u}\bm{u}) \right) &= - \nabla p + \nabla \cdot \bm{\tau},
\end{align}
which reduce to Stokes flow for $Re \to 0$:
\begin{equation}
    \nabla p = \nabla \cdot \bm{\tau}, \quad \bm{\tau} = \bm{\tau}_s + \bm{\tau}_p.
\end{equation}
The solvent and polymeric stress tensors are
\begin{equation}
    \bm{\tau}_s = M \beta (\nabla \bm{u} + (\nabla \bm{u})^T)
\end{equation}
\begin{equation}
    \bm{\tau}_p + \Lambda De \Big( \frac{\partial \bm{\tau}_p}{\partial t} + \bm{u} \cdot \nabla \bm{\tau}_p - \bm{\tau}_p \nabla \bm{u} - (\nabla \bm{u})^T \bm{\tau}_p \Big) = M (1-\beta) (\nabla \bm{u} + (\nabla \bm{u})^T)
\end{equation}
where for the drop: $M = \mu_i / \mu_e$, $\beta = \mu_{i_s}/\mu_i$, $\Lambda = \lambda_i / (\lambda_i + \lambda_e)$;  
for the ambient fluid: $M=1$, $\beta = \mu_{e_s}/\mu_e$, $\Lambda = \lambda_e / (\lambda_i + \lambda_e)$.  
The Deborah number is $De = (\lambda_i + \lambda_e) \tilde{U} / \tilde{R}_0$.
Boundary conditions satisfied by the flow field are:
\begin{subequations}
\begin{equation}
    \bm{u}_e = 0 \quad \text{as} \quad r \to \infty
\end{equation}
\begin{equation}
    \bm{u}_i \cdot \bm{n} = \bm{u}_e \cdot \bm{n} = \frac{\partial r}{\partial t} \quad \text{at} \quad r = 1 + f(\theta,t)
\end{equation}
\begin{equation}
    \bm{u}_i \cdot \bm{t} = \bm{u}_e \cdot \bm{t} \quad \text{at} \quad r = 1 + f(\theta,t)
\end{equation}
\begin{equation}
    \tau_{i,nt} + \tau^E_{i,nt} = \tau_{e,nt} + \tau^E_{e,nt} \quad \textbf{at} \quad r = 1 + f(\theta,t)
\end{equation}
\begin{equation}
    (-p_i + \tau_{i,nn} + \tau^E_{i,nn}) - (-p_e + \tau_{e,nn} + \tau^E_{e,nn}) = \frac{\kappa}{Ca_E} \quad \text{at} \quad r = 1 + f(\theta,t)
\end{equation}
\end{subequations}
where $Ca_E = \epsilon_e \tilde{E}_\infty^2 \tilde{R}_0 / \gamma$.

\subsection{Asymptotic expansion}
The governing equations are inherently nonlinear and generally require numerical solution. 
However, in the limit of small deformation and weak viscoelastic effects, the problem can be treated using asymptotic expansions. 
When a drop is subjected to an electric field, the electric stresses act as the deforming force, while surface tension provides the restoring force. 
Consequently, the deformation remains small when the electric capillary number, defined as the ratio of electric to surface tension forces,  
\begin{equation}
    Ca_E = \frac{\epsilon_e \tilde{E}_\infty^2 \tilde{R}_0}{\gamma},
\end{equation}
is small. The influence of viscoelasticity is quantified by the Deborah number,  
\begin{equation}
    De = \frac{\lambda \tilde{U}}{\tilde{R}_0},
\end{equation}
which represents the ratio of the polymer relaxation time to the characteristic flow time scale. 
In the asymptotic regime, both $Ca_E$ and $De$ are treated as small parameters, and the relevant physical quantities are expanded in a two-parameter asymptotic series:
\begin{subequations}
\begin{equation}
	V = V_{00} + Ca_E V_{10} + De V_{01} + Ca_E De V_{11} + ...
\end{equation}
\begin{equation}
	\bm{E} = \bm{E}_{00} + Ca_E\bm{E}_{10} + De\bm{E}_{01} + Ca_E De\bm{E}_{11} + ...
\end{equation}
\begin{equation}
	\bm{u} = \bm{u}_{00} + Ca_E\bm{u}_{10} + De\bm{u}_{01} + Ca_E De\bm{u}_{11} + ...
\end{equation}
\begin{equation}
	p = p_{00} + Ca_E p_{10} + De p_{01} + Ca_E De p_{11} + ...
\end{equation}
\begin{equation}
	\bm{\tau} = \bm{\tau}_{00} + Ca_E\bm{\tau}_{10} + De\bm{\tau}_{01} + Ca_E De\bm{\tau}_{11} + ...
\end{equation}
\end{subequations}
Since the electric stress acts as the driving force, the drop remains undeformed for $Ca_E = 0$. 
Accordingly, the interface is expressed as a perturbation series in both $Ca_E$ and $De$:
\begin{multline}
\label{Eqn_InterfaceForm}
    r_s = 1 + Ca_E \left\{ f_{10}(\theta, t) + De\, f_{11}(\theta, t) + \order{De^2} \right\} \\
    + Ca_E^2 \left\{ f_{20}(\theta, t) + De\, f_{21}(\theta, t) + \order{De^2} \right\} + \order{Ca_E^3}.
\end{multline}
The expanded variables are then substituted into the governing equations and boundary conditions, and terms of different orders are collected. 
This procedure yields the system of equations and corresponding boundary conditions at orders $\order{1}$, $\order{Ca_E}$, and $\order{De}$.  
To determine the boundary conditions at each order, the normal and tangential components of relevant vectors and tensors are computed using the method of domain perturbations.

\subsection{Method of domain perturbations}
When an electric field is applied, the initially spherical interface of the drop becomes deformed. 
Since the exact shape of the deformed surface is unknown \emph{a priori}, 
the interfacial boundary conditions must be imposed at the perturbed interface, 
which is simultaneously determined as part of the solution. 
If the interface is represented by
\begin{equation}
    F = r_s - [1 + f(\theta)] = 0 \; ,
\end{equation}
the outward unit normal to the surface is given by 
$\boldsymbol{n} = \nabla F / \lvert \nabla F \rvert$. 
The curvature of the deformed interface, obtained from the divergence of the normal vector, 
can be expressed as
\begin{multline}
    \kappa = \nabla \cdot \boldsymbol{n} = 2 - Ca_E \Big[ 2 f^{(Ca_E)} + \cot\theta~f^{\prime (Ca_E)} + f^{\prime\prime (Ca_E)} \\
    + De \big( 2f^{(Ca_E De)} + \cot\theta~f^{\prime (Ca_E De)} + f^{\prime\prime (Ca_E De)} \big) \quad + O(De^2) \Big] \\ 
    + Ca_E^2 \Big[ 2(f^{(Ca_E)})^2 - 2 f^{(Ca_E^2)} + 2\cot\theta~f^{(Ca_E)}f^{\prime (Ca_E)} - 2\cot\theta~f^{\prime (Ca_E^2)} \\
    \quad + 2 f^{(Ca_E)} f^{\prime\prime (Ca_E)} - f^{\prime\prime (Ca_E^2)} + O(De) \Big] + O(Ca_E^3) \; .
\end{multline}

The boundary conditions at the interface involve both normal and tangential components of velocity and stress, expressed in spherical axisymmetric coordinates. To evaluate these components on the deformed interface, vector and tensor quantities are rotated through a small angle $\alpha$, representing the deviation between the normals to the deformed and spherical surfaces. Since $\alpha$ is small, the approximations $\sin\alpha \approx \alpha$ and $\cos\alpha \approx 1$ are applied. For any vector $\boldsymbol{a}$ with components $(a_r, a_\theta)$, the components normal and tangential to the deformed surface are expressed as
\begin{align}
    a_n &= a_r\cos\alpha - a_\theta \sin\alpha \approx a_r - a_\theta f'(\theta) \\
    a_t &= a_r\sin\alpha + a_\theta \cos\alpha \approx a_\theta + a_r f'(\theta).
\end{align}
Using the same transformation for the stress tensor, the corresponding components are given by
\begin{align}
    \tau_{nn} &= \tau_{rr} - 2\tau_{r\theta} f'(\theta), \\
    \tau_{nt} &= \tau_{r\theta} + (\tau_{rr} - \tau_{\theta\theta}) f'(\theta), \\
    \tau_{tt} &= \tau_{\theta\theta} + 2\tau_{r\theta} f'(\theta).
\end{align}
To apply the boundary conditions at the perturbed interface, all quantities are expanded about the spherical surface using Taylor series. 

\subsection{Asymptotic solution}
Substitution of asymptotic expansions into the governing equations and interfacial conditions yields systems of equations and corresponding boundary conditions at orders $\order{1}$, $\order{Ca_E}$, and $\order{De}$.
The leading-order electric field is first determined for a spherical interface, yielding the interfacial electric stress at $\order{1}$. The corresponding flow field, obtained by solving the hydrodynamic equations at this order, is decomposed using linearity as  $\bm{u}_{00} = \bm{u}_{00}^{I} + \bm{u}_{00}^{II}$,
where $\bm{u}_{00}^{I}$ represents the flow induced by the tangential electric stress at the interface, and $\bm{u}_{00}^{II}$ accounts for the effect of the rate of change of the drop radius on the normal velocity. The component $\bm{u}_{00}^{I}$ contributes to the deformation at $\order{Ca_E}$, while $\bm{u}_{00}^{II}$ governs the higher-order deformation at $\order{Ca_E^2}$.
Flow field $\bm{u}_{00}^{I}$ satisfies velocity and tangential stress continuity across the interface but not the normal stress balance. To satisfy the latter, a deformation of $\order{Ca_E}$ is introduced, yielding the shape correction $f_{10}(\theta)$. The deformed interface modifies the electric field at $\order{Ca_E}$, from which the corresponding electric stress and flow field are determined. The subsequent deformation at $\order{Ca_E^2}$, denoted by $f_{20}(\theta)$, follows from the normal stress balance at $\order{Ca_E}$.
At $\order{De}$, electric field equations and corresponding boundary conditions are homogeneous. Thus, no additional electric stress contribution occurs. At this order, the flow equations contain inhomogeneous terms arising from the $\order{1}$ flow field. Solving these provides the $\order{De}$ flow field and the viscoelastic deformation at $\order{Ca_E De}$, represented by $f_{11}(\theta)$. Detailed derivation of asymptotic solution is given in Appendix \ref{appA}.

\subsection{Drop deformation}
The resulting deformed interface obtained from the asymptotic solution is expressed as  
\begin{equation}
    r_s(\theta) = 1 + Ca_E f_{10}(\theta)+ Ca_E De f_{11}(\theta) + Ca_E^2 f_{20}(\theta).
\end{equation}
where
\begin{align}
f_{10}(\theta) &= B^S_{10} P_2(\cos\theta), \\
f_{11}(\theta) &= B^S_{11} P_2(\cos\theta) + C^S_{11} P_4(\cos\theta), \\
f_{20}(\theta) &= A^S_{20} + B^S_{20} P_2(\cos\theta) + C^S_{20} P_4(\cos\theta)
\end{align}
where $P_2(\cos\theta)$ and $P_4(\cos\theta)$ are second and fourth order Legendre polynomials in $\cos\theta$, respectively.
The expressions for the constants in the above equations, together with those for the electric and flow fields and the detailed solution of the system of equations, are presented in Appendix \ref{appA}.
The deformation is quantified using Taylor’s deformation parameter,
\begin{equation}
    D = \frac{L - B}{L + B},
\end{equation}
where \( L = r_s|_{\theta=0} \) represents the drop’s axis length along the direction of the electric field, and \( B = r_s|_{\theta=\pi/2} \) denotes the axis length perpendicular to the electric field.
For a steady electric field, the $(\sigma_r, \epsilon_r)$ phase plot (refer to \citet{bangar2026large}) characterizes distinct regions based on the deformation type, flow direction, and the sign of the second-order deformation coefficient. As demonstrated in \citet{das2026effect}, the viscoelasticity of the drop exerts a more significant effect on deformation than that of the surrounding medium. Consequently, the present analysis considers an Oldroyd-B drop suspended in a Newtonian ambient and subjected to an alternating electric field.

\begin{figure}[ht!]
    \centering
    \includegraphics[width=0.95\textwidth]{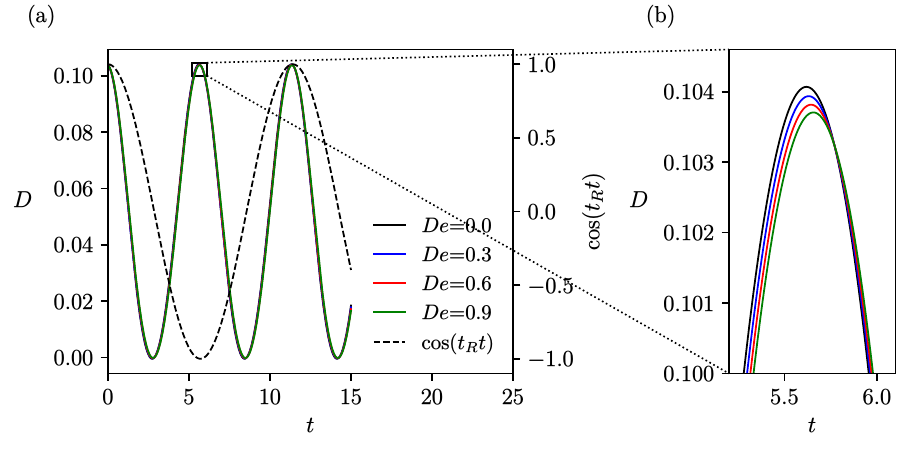}
    \caption{(a) Deformation vs time for various $De$ for $\mu_r=1$, $\sigma_r=10$, $\epsilon_r=1.37$, $Ca_E=0.2$, $\beta_i = 1/9$, $a = 4.425 \times 10^{-6}$, $t_R=0.550792$, $\omega=62.8$. Dotted black line shows the imposed external electric field. (b) Magnified view of the marked region in (a).}
    \label{Fig_DvsTime_prolate}
\end{figure}
\begin{figure}[ht!]
    \centering
    \includegraphics[width=0.5\textwidth]{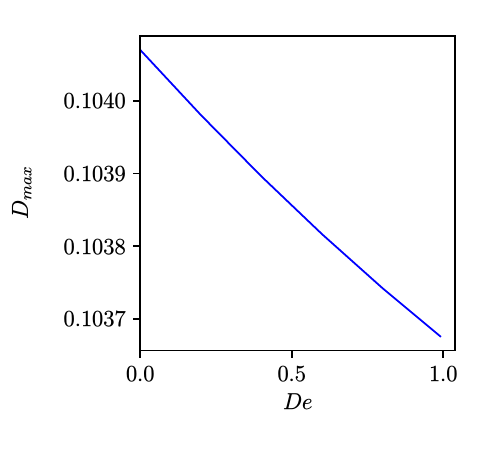}
    \caption{Maximum deformation vs $De$ for $\mu_r=1$, $\sigma_r=10$, $\epsilon_r=1.37$, $Ca_E=0.2$, $\beta_i = 1/9$, $a = 4.425 \times 10^{-6}$, $t_R=0.550792$, $\omega=62.8$.}
    \label{Fig_Dmax_prolate}
\end{figure}
\autoref{Fig_DvsTime_prolate}(a) shows the variation in drop deformation over time for various Deborah numbers $De$ for a case corresponding to the prolate deformation (parameters: $\mu_r=1$, $\sigma_r=10$, $\epsilon_r=1.37$, $Ca_E=0.2$, $\beta_i = 1/9$, $a = 4.425 \times 10^{-6}$, $t_R=0.550792$ and $\omega=62.8$). The black dotted line shows the external electric field in the figure. As expected, we observe that the deformation of the drop exhibits a frequency of oscillations that is twice that of the imposed external electric field. 
\autoref{Fig_DvsTime_prolate}(b) shows the magnified view of the region marked in \autoref{Fig_DvsTime_prolate}(a). We observe that the maximum drop deformation decreases with increasing $De$ and the position of the maximum shifts toward the right.
The drop deformation can be represented by $D = D_{max}\cos({2t_Rt + \alpha})$, where $D_{max}$ is the maximum deformation and $\alpha$ is the phase angle. \autoref{Fig_Dmax_prolate} represents the variation of the maximum drop deformation $D_{max}$ with $De$. We note that the maximum drop deformation decreases with $De$.
Thus, viscoelasticity acts to oppose the drop deformation in the case of prolate deformation.

\begin{figure}[ht!]
    \centering
    \includegraphics[width=0.95\textwidth]{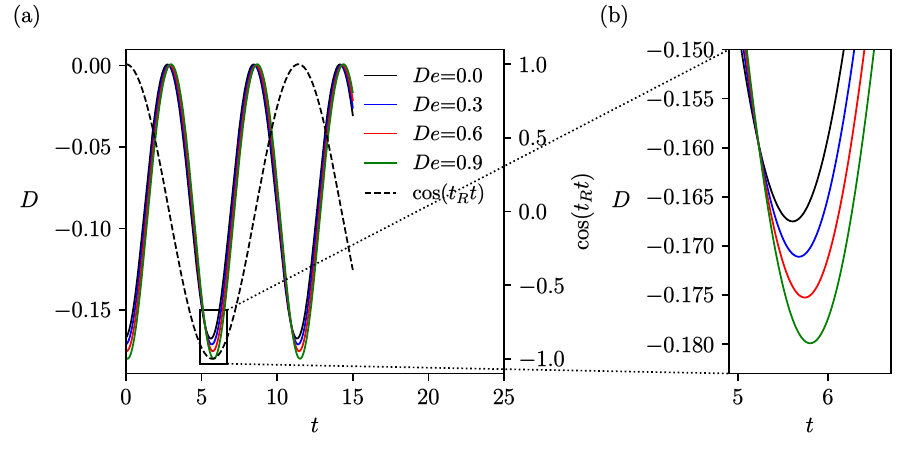}
    \caption{(a) Deformation vs time for various $De$ for $\mu_r=1$, $\sigma_r=0.1$, $\epsilon_r=2$, $Ca_E=0.2$, $\beta_i = 1/9$, $a = 4.425 \times 10^{-6}$, $t_R=0.550792$, $\omega=62.8$. Dotted black line shows the imposed external electric field. (b) Magnified view of the marked region in (a). }
    \label{Fig_DvsTime_oblate}
\end{figure}
\begin{figure}[ht!]
    \centering
    \includegraphics[width=0.5\textwidth]{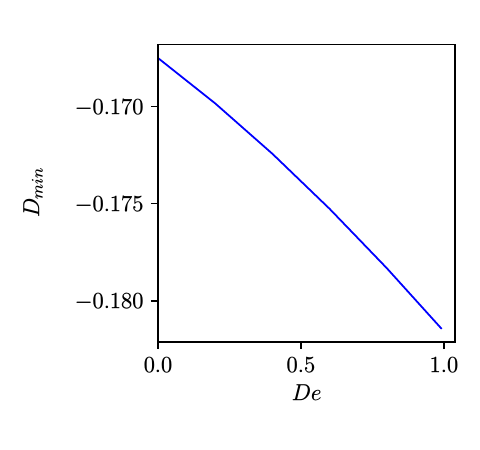}
    \caption{Maximum deformation vs $De$ for $\mu_r=1$, $\sigma_r=0.1$, $\epsilon_r=2$, $Ca_E=0.2$, $\beta_i = 1/9$, $a = 4.425 \times 10^{-6}$, $t_R=0.550792$, $\omega=62.8$.}
    \label{Fig_Dmax_oblate}
\end{figure}
\autoref{Fig_DvsTime_oblate}(a) shows the variation of a drop deformation over time for various Deborah numbers for a case of oblate deformation ($\mu_r=1$, $\sigma_r=0.1$, $\epsilon_r=2$, $Ca_E=0.2$, $\beta_i = 1/9$, $a = 4.425 \times 10^{-6}$, $t_R=0.550792$ and $\omega=62.8$). The drop deformation oscillates with a frequency twice that of the imposed electric field, like the prolate case. \autoref{Fig_DvsTime_oblate}(b) represents the magnified view of the marked region in \autoref{Fig_DvsTime_oblate}(a). The magnified view shows that the minimum of a drop deformation decreases as $De$ increases and the position of the minimum shifts towards the right. 
The minimum deformation variation with $De$, for the oblate case, is shown in \autoref{Fig_Dmax_oblate}. We observe that the minimum drop deformation decreases with $De$. Since the drop deformation is negative for the oblate case, we note that the magnitude of the drop deformation is enhanced by the increase in $De$. 
From \autoref{Fig_DvsTime_prolate} and \autoref{Fig_DvsTime_oblate}, we can observe that the effect of viscoelasticity is more prominent for an oblate case than a prolate case.

From physical reasoning, one expects the amplitude of oscillations of the deformation parameter to decrease with increasing frequency of the applied electric field, since at high frequencies the drop does not have sufficient time to respond to the rapidly varying field. However, this trend is not captured by the deformation obtained from the asymptotic solution. This discrepancy arises because, in the asymptotic analysis, bulk charge relaxation is neglected, assuming instantaneous charge accumulation at the interface, and the electric potential in both phases is obtained from Laplace equation. As a result, the predicted oscillation amplitude fails to reflect the physically expected frequency dependence of amplitude.
Moreover, the asymptotic analysis is valid only in the limit of small electric capillary number and small Deborah number. To address these limitations, numerical simulations are performed to investigate the deformation dynamics of a viscoelastic drop subjected to an alternating electric field.

%%%%%%%%%%%%%%%%%%%%%%%%%%%%%%%%%%%%%%%%%%%%%%%%%%%%%%%%%%%%%%%%%%%%%%%%%%%%%%%%%%%%%%%%
%%%%%%%%%%%%%%%%%%%%%%%%%%%%%%%%%%%% NUMERICAL %%%%%%%%%%%%%%%%%%%%%%%%%%%%%%%%%%%%%%%
\section{Numerical simulations}\label{sec:numerical}
\begin{figure}[ht!]
    \centering
    \includegraphics[width=0.6\textwidth]{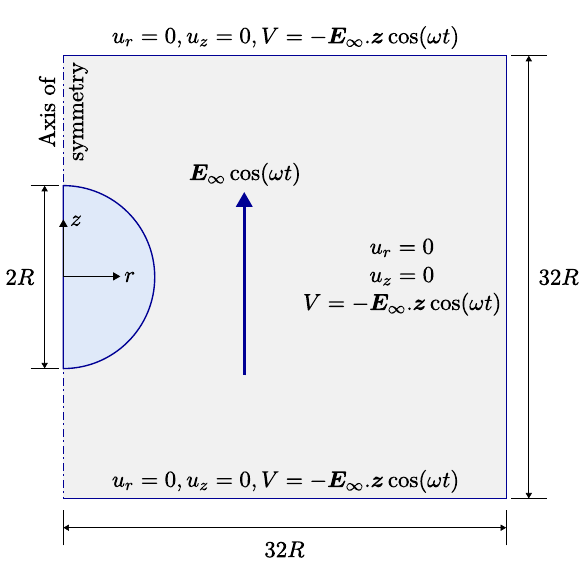}
    \caption{Schematic (not to the scale) of the simulation domain for drop in an alternating electric field. Drop of radius $R$ is imposed with external electric field $\bm{E}_\infty\cos(\omega t)$. Axisymmetric domain of $32R$ is considered for the simulations. Origin of coordinate system is at the center of the drop. Boundary conditions on the velocity components and electric potential are shown.}
    \label{Fig_Schematic}
\end{figure}

Axisymmetric simulations are performed using the open-source solver \href{https://basilisk.fr}{Basilisk}, which solves two-phase flows via a geometric Volume of Fluid (VOF) method. The solver employs a second-order time-splitting projection scheme with cell-centered variables; diffusion is treated implicitly, while advection is discretized using the second-order Bell-Colella-Glaz (BCG) scheme. The governing equations for electric potential and charge density are solved by the numerical approach as presented by \citet{lopez2011charge}.
Viscoelastic constitutive equation is solved using the log-conformation tensor approach which was proposed by \citet{fattal2004constitutive}.  Polymeric stress is written in terms of conformation tensor as,
\begin{equation}
    \bm{\tau_p} = \frac{\mu_p}{\lambda}(\bm{A}-\bm{I}),
\end{equation}
where $\bm{A}$ is a symmetric positive definite tensor. The evolution of $\bm{\Psi} = \log \bm{A}$ satisfies
\begin{equation}
    \pdv{\bm{\Psi}}{t} + \bm{u}\cdot\nabla\bm{\Psi} = \bm{\Omega}\bm{\Psi} - \bm{\Psi}\bm{\Omega} + 2\bm{B} + \frac{1}{\lambda}(e^{-\bm{\Psi}} - \bm{I}),
\end{equation}
where $\bm{\Omega}$ is an anti-symmetric tensor and $\bm{B}$ is a symmetric tensor that commutes with $\bm{A}$.
Equation for $\bm{\Psi}$ is solved numerically by split scheme as explained by \citet{lopez2019adaptive}:
\begin{align}
	&\pdv{\bm{\Psi}}{t} = \bm{\Omega}\bm{\Psi} - \bm{\Psi}\bm{\Omega} + 2\bm{B} \label{Eqn_Psi_upperConv}\\
	&\pdv{\bm{\Psi}}{t} + \grad.(\bm{u\Psi}) = 0 \label{Eqn_Psi_Advection}\\
    &\pdv{\bm{A}}{t} = \frac{1}{\lambda}(\bm{I} - \bm{A}) \label{Eqn_Psi_Model}
\end{align}
where the upper-convected derivative (\autoref{Eqn_Psi_upperConv}) is treated explicitly, advection (\autoref{Eqn_Psi_Advection}) is solved using the BCG scheme, and the model term (\autoref{Eqn_Psi_Model}) is integrated analytically.

The computational domain is defined in $(r,z)$ coordinates, centered on the drop as shown in \autoref{Fig_Schematic}. The extent is $-L/2 \leq z \leq L/2$ and $0 \leq r \leq L$. The outer boundaries of the computational domain are subjected to free-slip boundary conditions.
\begin{subequations}
\begin{align}
    u_z(r,-L/2) &= 0, \quad
    \frac{\partial u_r}{\partial z}(r,-L/2) = 0, \\
    u_z(r,L/2) &= 0, \quad
    \frac{\partial u_r}{\partial z}(r,L/2) = 0, \\
    u_r(L,z) &= 0, \quad
    \frac{\partial u_z}{\partial r}(L,z) = 0.
\end{align}
\end{subequations}
The applied electric field corresponds to the potential distribution $V = -E_\infty\cos(\omega t) z$. Thus, the boundary conditions for $V$ are,
\begin{equation}
    V(r, -L/2) = -E_\infty z \cos(\omega t)
\end{equation}
\begin{equation}
    V(r, L/2) = -E_\infty z \cos(\omega t)
\end{equation}
\begin{equation}
    V(L, z) = -E_\infty z \cos(\omega t)
\end{equation}
Polymeric stress components satisfy homogeneous Neumann boundary conditions at the outer boundaries, which introduce only localized errors and do not affect the global solution \citep{alves2021numerical}. Symmetry conditions are imposed along $r=0$. Symmetry conditions for velocity field are given by,
\begin{subequations}
\begin{equation}
u_r(0,z) = 0, \quad \pdv{u_z}{r}\Big|_{(0,z)} = 0, \quad \pdv{V}{r}\Big|_{(0,z)} = 0,
\end{equation}
\end{subequations}
and symmetry conditions for the polymeric stress are,
\begin{equation}
     \pdv{\tau_{p_{rr}}}{r}\Big|_{(0, z)} = \pdv{\tau_{p_{zz}}}{r}\Big|_{(0, z)} = \tau_{p_{rz}}(0, z) = \pdv{\tau_{p_{\theta\theta}}}{r}\Big|_{(0, z)} = 0.
\end{equation}
\begin{figure}[ht!]
    \centering
    \includegraphics[width=0.6\textwidth]{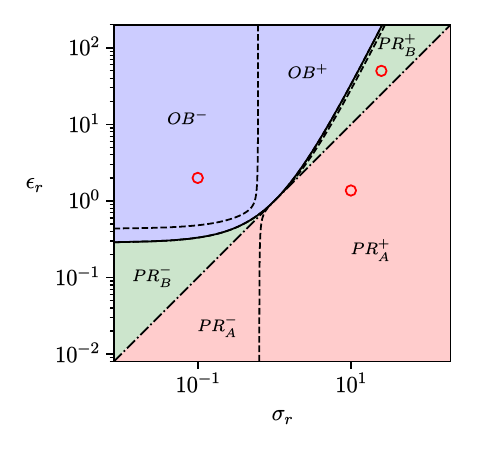}
    \caption{Phase diagram in the $(\sigma_r,\epsilon_r)$ plane (log--log scale) for $\mu_r=1$, illustrating the deformation regimes of a drop subjected to an external electric field. The deformation parameter is determined using an asymptotic expansion in the electric capillary number $Ca_E$ and the Deborah number $De$, yielding \(D = Ca_E\, D_{Ca_E} + Ca_E^2\, D_{Ca_E^2} + Ca_E De\, D_{Ca_EDe}.\) The quantities $D_{Ca_E}$, $D_{Ca_E^2}$, and $D_{Ca_EDe}$ represent the contributions at $\order{Ca_E}$, $\order{Ca_E^2}$, and $\order{Ca_EDe}$, respectively. Regions denoted by $PR$ and $OB$ correspond to prolate and oblate deformation. For the prolate regime, subscripts $A$ and $B$ identify equator-to-pole and pole-to-equator circulations, respectively, while superscripts $+$ and $-$ indicate positive and negative contributions from the $\order{Ca_E^2}$ term. The representative $(\sigma_r,\epsilon_r)$ combinations investigated in this work are highlighted by red circle markers. The asymptotic formulation used to obtain the phase diagram is a special case of the general formulation developed in the present work. The phase diagram is recovered by setting the angular frequency of the applied electric field to zero ($\omega=0$). The derivation of the general formulation is presented in Appendix \ref{appA}, while the $\omega=0$ limit corresponds to the results of \citet{das2026effect}.}
    \label{Fig_QR_selection}
\end{figure}
The electrohydrodynamic deformation of an Oldroyd-B drop is analyzed for $\rho_r = 1$, $\mu_r = 1$, $Re = 1$, and $\beta_i = 1/9$. Among the six regions of the $(\sigma_r,\epsilon_r)$ phase plot, deviations from Newtonian behavior are negligible in the $PR_A^-$, $PR_B^-$, and $OB^+$ regions, whereas significant viscoelastic effects are observed in the $PR_A^+$, $PR_B^+$, and $OB^-$ regions. In the latter three regions, the steady deformation decreases with $De$ in $PR_A^+$ and $PR_B^+$, while it increases with $De$ in $OB^-$, with corresponding changes in the critical $Ca_E$ (\citet{bangar2026large}). Therefore, representative points from the $PR_A^+$, $PR_B^+$, and $OB^-$ regions are selected for detailed investigation. The selected $(\sigma_r,\epsilon_r)$ pairs, also indicated on the $(\sigma_r,\epsilon_r)$ phase plot in \autoref{Fig_QR_selection}, are:
\begin{enumerate}
\item $PR_A^+$: $(10, 1.37)$,
\item $PR_B^+$: $(25, 50)$,
\item $OB^-$: $(0.1, 2)$.
\end{enumerate}
Numerical simulations are performed for each pair across a range of $Ca_E$ and $De$. Three distinct, non-zero angular frequencies ($\omega$) were selected for the imposed alternating electric field: $628 \text{ rad/s}$ (corresponding to $100 \text{ Hz}$), $62.8 \text{ rad/s}$ ($10 \text{ Hz}$), and $6.28 \text{ rad/s}$ ($1 \text{ Hz}$).

The dimensional parameters are fixed as: drop radius $R = 0.001$, ambient density $\rho_e = 1000$, permittivity $\epsilon_e = 4.425\times10^{-11}$, conductivity $\sigma_e = 10^{-5}$, and interfacial tension $\gamma = 0.065$, with all remaining quantities derived from the specified non-dimensional values. The characteristic velocity, based on the balance of electric and viscous stresses, is 
$U = 0.2549 \sqrt{Ca_E}$, yielding an electric Reynolds number $Re_E = \frac{\epsilon_e U}{R \sigma_e} = 1.128\times10^{-3}\sqrt{Ca_E}$.

%%%%%%%%%%%%%%%%%%%%%%%%%%%%%%%%%%%%%%%%%%%%%%%%%%%%%%%%%%%%%%%%%%%%%%%%%%%%%%%%%%%%%%%%
    
    %%%%%%%%%%%%%%%%%%%%%%%%%%%%%%%%%%%%%% RESULTS %%%%%%%%%%%%%%%%%%%%%%%%%%%%%%%%%%%%%%%%%
\section{Results \& discussion}\label{sec:results}
\subsection{\texorpdfstring{$(\sigma_r, \epsilon_r)$ from $PR_A^+$ region}{Lg}}
\label{Sec_chap6_PRA_plus}
The point $(\sigma_r, \epsilon_r) = (10, 1.37)$ is chosen from the $PR_A^+$ region for a comprehensive analysis of the electrohydrodynamic deformation of a viscoelastic drop under an alternating electric field.
\subsubsection{High angular frequency regime \texorpdfstring{($\omega = 628 \text{ rad/s}$)}{Lg}}
\begin{figure}[ht!]
    \centering
    \includegraphics[width=0.95\textwidth]{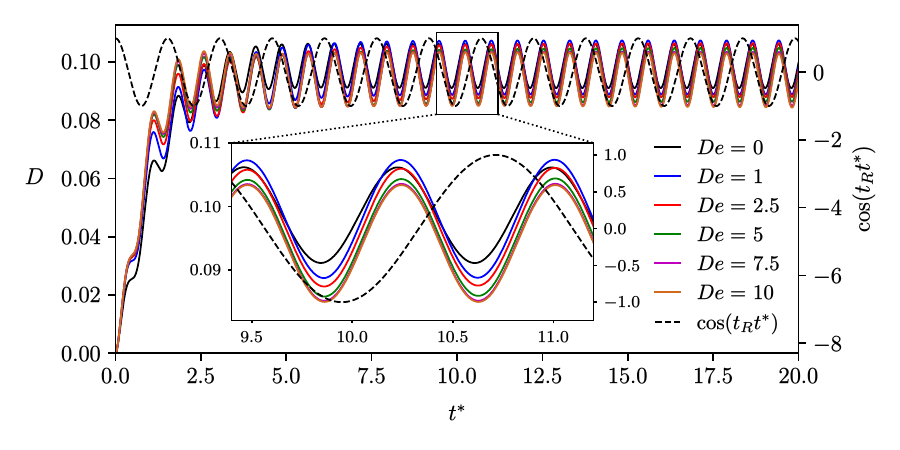}
    \caption{Variation of the deformation parameter with non-dimensional time for $\sigma_r=10$, $\epsilon_r=1.37$ ($PR_A^+$), $\omega=628$, and $Ca_E=0.36$ at different Deborah numbers ($De$).}
    \label{Fig_DvsTime_PRA_plus_omega628}
\end{figure}
\begin{figure}[ht!]
    \centering
    \includegraphics[width=0.85\textwidth]{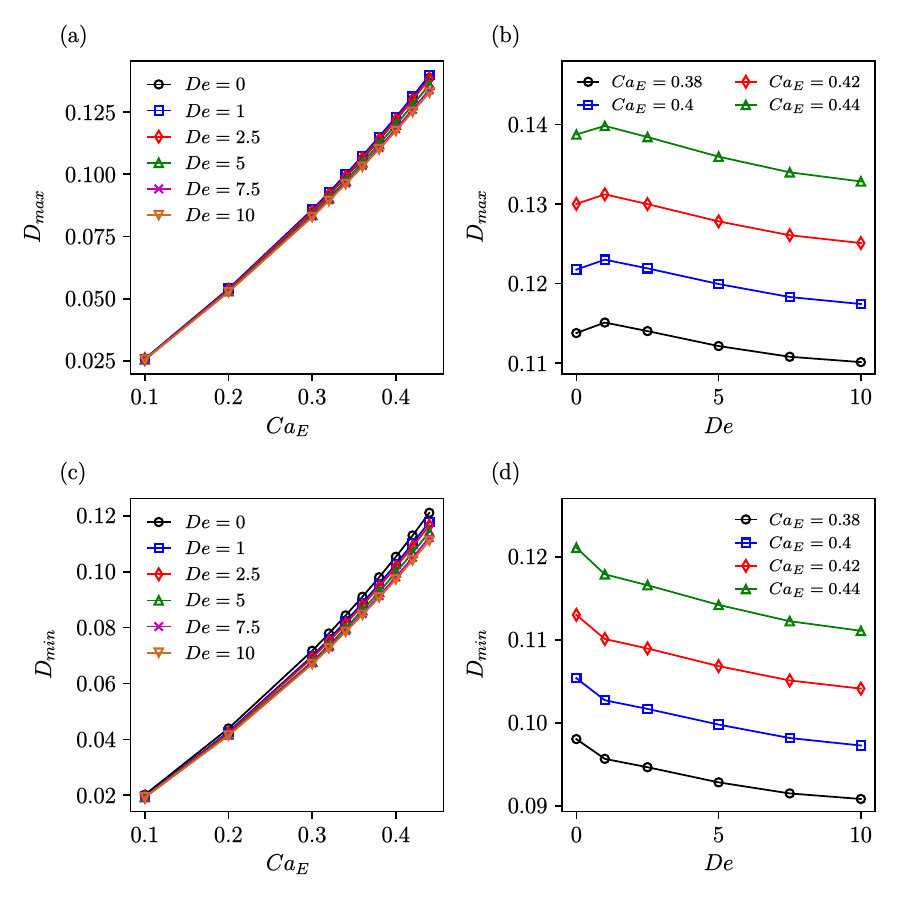}
    \caption{(a) Maximum deformation versus $Ca_E$, (b) maximum deformation versus $De$, (c) minimum deformation versus $Ca_E$, and (d) minimum deformation versus $De$ for $\sigma_r=10$, $\epsilon_r=1.37$ ($PR_A^+$) at $\omega=628$.}
    \label{Fig_Dmax_Dmin_PRA_plus_omega628}
\end{figure}
\begin{figure}
    \centering
    \includegraphics[width=0.82\textwidth]{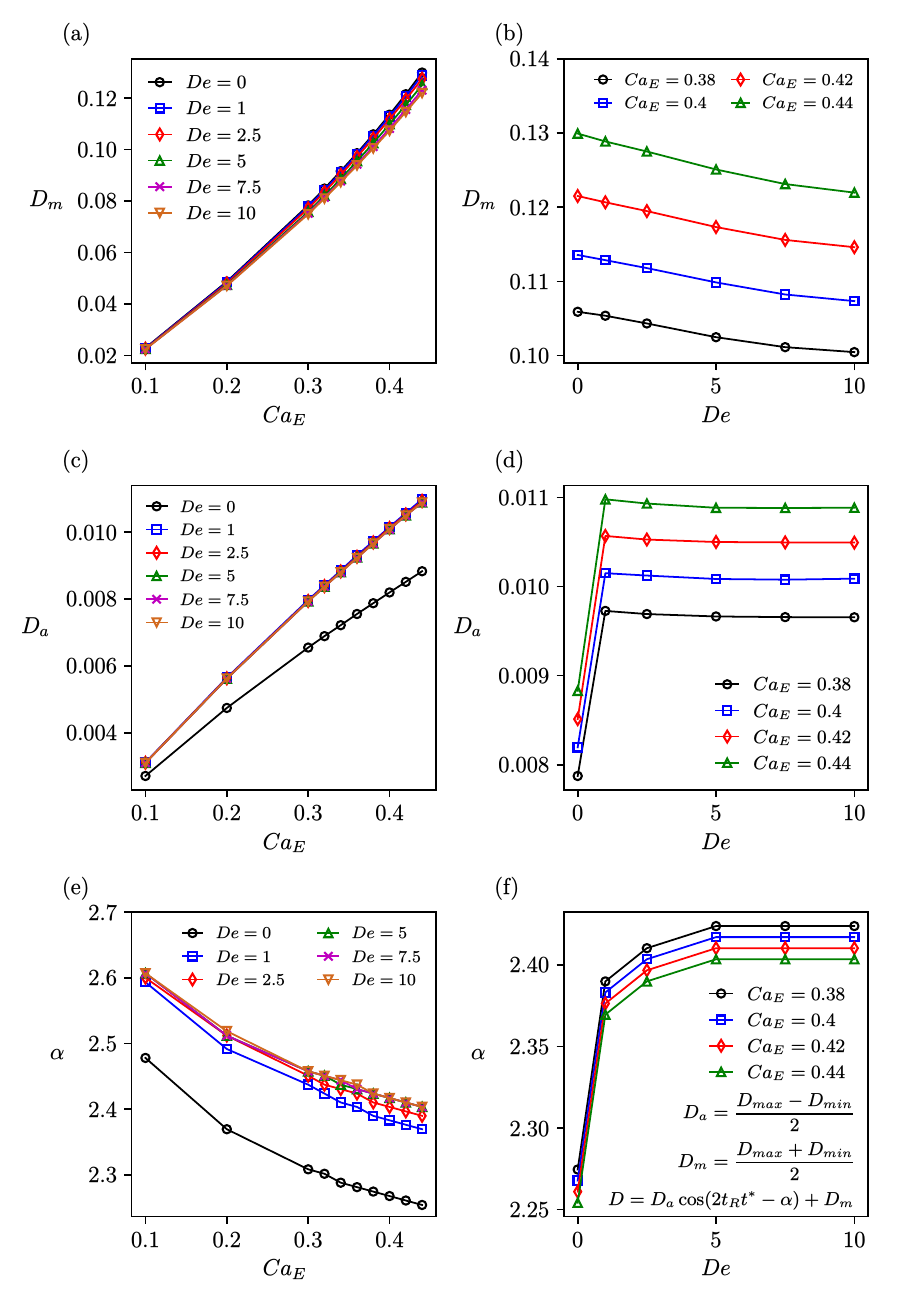}
    \caption{(a) Variation of the mean deformation with $Ca_E$, (b) variation of the mean deformation with $De$, (c) variation of the oscillation amplitude with $Ca_E$, (d) variation of the oscillation amplitude with $De$, (e) variation of the phase difference with $Ca_E$, and (f) variation of the phase difference with $De$ for $\sigma_r=10$, $\epsilon_r=1.37$ ($PR_A^+$) at $\omega=628$.}
    \label{Fig_alpha_amp_PRA_plus_omega628}
\end{figure}
\autoref{Fig_DvsTime_PRA_plus_omega628} shows the temporal evolution of the drop deformation as a function of non-dimensional time (scaled by $R/U$) for different Deborah numbers ($De$) at a fixed electric capillary number ($Ca_E=0.36$) under a high-frequency alternating electric field ($\omega=628~\mathrm{rad/s}$). As expected, the drop deformation oscillates at twice the frequency of the applied electric field owing to the quadratic dependence of the electric stress on the electric field strength. The deformation oscillates between well-defined maximum and minimum values with a noticeable phase lag relative to the applied electric field, reflecting the finite response time of the viscoelastic drop to the rapidly oscillating electric forcing. Furthermore, the drop retains a spheroidal shape throughout the oscillation cycle for all the investigated values of $Ca_E$ and $De$, with no evidence of multi-lobed deformations. This indicates that the rapid oscillation of the electric field suppresses the development of interfacial shape instabilities, even at the instant of maximum deformation.

\autoref{Fig_Dmax_Dmin_PRA_plus_omega628} illustrates the influence of the electric capillary number ($Ca_E$) and Deborah number ($De$) on the deformation extrema. 
\autoref{Fig_Dmax_Dmin_PRA_plus_omega628}(a) shows that the maximum deformation ($D_{max}$) increases monotonically with increasing $Ca_E$ for all $De$, indicating that stronger electric fields generate larger electric stresses, resulting in greater stretching of the drop during each oscillation cycle. In contrast, \autoref{Fig_Dmax_Dmin_PRA_plus_omega628}(b) reveals a non-monotonic dependence of $D_{max}$ on $De$. Specifically, $D_{max}$ initially increases as $De$ increases from 0 to 1, followed by a gradual decrease with further increase in $De$. This behavior reflects the increasing inability of the polymeric stresses to relax over the oscillation period. As $De$ increases beyond unity, the accumulated elastic stresses increasingly oppose the electrically induced deformation, thereby reducing the maximum deformation attained during each cycle.
\autoref{Fig_Dmax_Dmin_PRA_plus_omega628}(c) demonstrates that the minimum deformation ($D_{min}$) increases monotonically with increasing $Ca_E$, indicating that a stronger electric field maintains the drop in a more deformed state throughout the oscillation cycle. Conversely, \autoref{Fig_Dmax_Dmin_PRA_plus_omega628}(d) shows that $D_{min}$ decreases monotonically with increasing $De$, suggesting that the accumulated elastic stresses increasingly suppress the deformation during the entire oscillation cycle.

\autoref{Fig_alpha_amp_PRA_plus_omega628} presents the variation of the mean deformation, oscillation amplitude, and phase difference with $Ca_E$ and $De$. As shown in \autoref{Fig_alpha_amp_PRA_plus_omega628}(a), the mean deformation increases monotonically with increasing $Ca_E$, indicating that larger electric stresses produce greater time-averaged deformation. In contrast, \autoref{Fig_alpha_amp_PRA_plus_omega628}(b) shows that the mean deformation decreases with increasing $De$, reflecting the increasing opposition offered by the accumulated elastic stresses to the electrically driven deformation.
Regarding the oscillatory response, \autoref{Fig_alpha_amp_PRA_plus_omega628}(c) shows that the oscillation amplitude ($D_a$) increases with $Ca_E$ for all $De$. This is consistent with the corresponding increase in $D_{max}$ and the larger deformation sustained throughout the oscillation cycle at higher electric field strengths. \autoref{Fig_alpha_amp_PRA_plus_omega628}(d) exhibits a more complex dependence of $D_a$ on $De$. The oscillation amplitude initially increases as $De$ increases from 0 to 1 owing to the simultaneous increase in $D_{max}$ and decrease in $D_{min}$. However, for $De \geq 1$, $D_a$ becomes nearly independent of $De$, indicating saturation of the oscillation amplitude. \autoref{Fig_alpha_amp_PRA_plus_omega628}(e) shows that the phase difference decreases with increasing $Ca_E$ for all $De$, indicating that stronger electric forcing enables the drop to respond more rapidly to the applied field. Conversely, \autoref{Fig_alpha_amp_PRA_plus_omega628}(f) demonstrates that the phase difference increases with increasing $De$, reflecting the slower evolution of the polymeric stresses in increasingly viscoelastic fluids. Similar to the oscillation amplitude, the phase difference also approaches a limiting value for $De \geq 5$. 

For the high electric field frequency of $\omega=628~\mathrm{rad/s}$, the deformation oscillates at twice the applied frequency, giving an oscillation period of
\[t_{\omega}=\frac{2\pi}{2\omega}=0.005.\]
The characteristic deformation time scale is estimated as
\[t_{def}=\frac{\mu_eR}{\gamma}=\frac{\sqrt{\gamma Ca_E}\,R}{\gamma},\]
which varies from $0.00124$ to $0.00260$ over the range of $Ca_E$ considered. The polymer relaxation time is given by
\[\lambda=\frac{De\,R}{U}=\frac{0.003923\,De}{\sqrt{Ca_E}},\]
which ranges from $0.00591$ to $0.01240$ even for $De=1$ and increases further with increasing $De$. Thus, for the high-frequency forcing considered here,
\[t_{def}<t_{\omega}\lesssim\lambda,\]
with the separation between $\lambda$ and $t_{\omega}$ increasing as $De$ increases. The interface can therefore respond on a time scale shorter than the forcing period, but the polymeric stresses cannot fully relax within an oscillation cycle. As $De$ increases, the polymeric stresses retain memory over a progressively larger portion of the forcing cycle. Once $\lambda$ becomes substantially larger than $t_{\omega}$, further increases in $De$ have a limited effect on the evolution of the polymeric stresses within a single cycle, providing a physical basis for the observed saturation of the oscillation amplitude at higher $De$. The finite deformation time of the interface also contributes to the phase difference at this high frequency. Since the forcing period is relatively short, the interface cannot adjust its shape instantaneously to the rapidly varying electric stress, resulting in a finite phase lag even in the Newtonian limit. For the viscoelastic drop, the additional polymeric memory further modifies this response because the polymeric stresses cannot adjust instantaneously to the imposed forcing. Thus, the observed phase difference results from the combined effects of the finite interfacial response time and the delayed evolution of the polymeric stresses. At the high frequency considered here, these effects produce a substantial phase difference between the applied electric stress and the drop deformation. Thus, both the oscillation amplitude and the phase lag exhibit saturation at high Deborah numbers because the polymer relaxation time is already much larger than the oscillation period.

\subsubsection{Intermediate angular frequency regime \texorpdfstring{($\omega = 62.8 \text{ rad/s}$)}{Lg}}
\begin{figure}[ht!]
    \centering
    \includegraphics[width=0.95\textwidth]{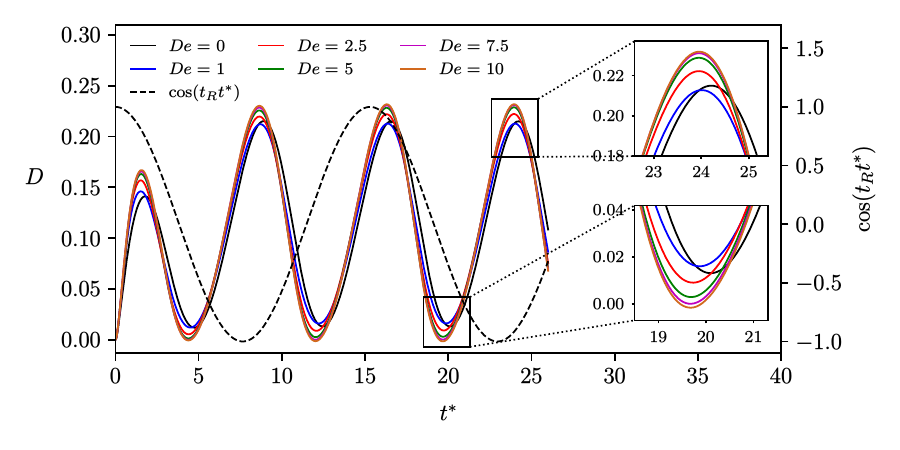}
    \caption{Deformation parameter as a function of non-dimensional time for various Deborah numbers ($De$) at $\sigma_r=10$, $\epsilon_r=1.37$ ($PR_A^+$), $\omega=62.8$, and $Ca_E=0.36$.}
    \label{Fig_DvsTime_PRA_plus_omega62p8}
\end{figure}
For an intermediate angular frequency of the imposed electric field, $\omega=62.8~\mathrm{rad/s}$, the temporal evolution of the drop deformation is presented in \autoref{Fig_DvsTime_PRA_plus_omega62p8} for various Deborah numbers ($De$) at a fixed electric capillary number of $Ca_E=0.36$. Consistent with the high-frequency case, the drop deformation oscillates at twice the frequency of the applied electric field owing to the quadratic dependence of the electric stress on the electric field. However, the larger oscillation period at this intermediate frequency allows the drop to respond more effectively to the time-varying electric stresses within each forcing cycle. Consequently, the deformation dynamics differ significantly from those observed at high frequency. For each $De$, the deformation oscillates between a distinct maximum and a distinct minimum with a considerably smaller phase lag than that observed at $\omega=628~\mathrm{rad/s}$.

\begin{figure}[ht!]
    \centering
    \includegraphics[width=0.85\textwidth]{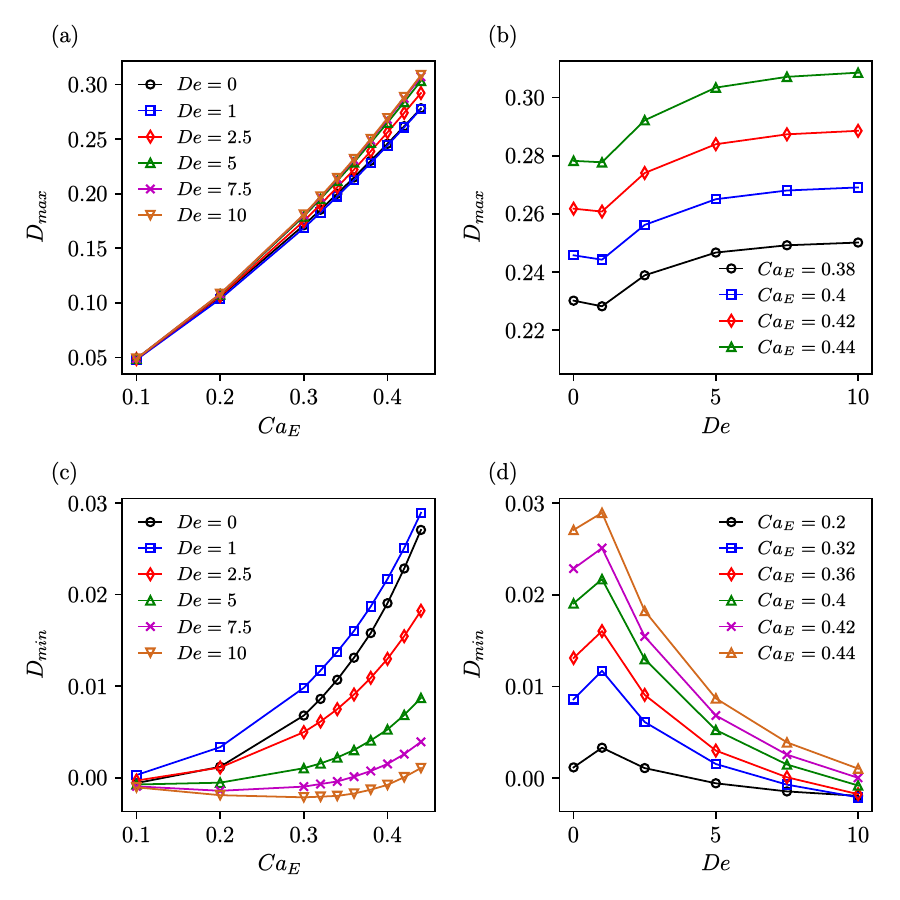}
    \caption{(a) Variation of the maximum deformation with $Ca_E$, (b) variation of the maximum deformation with $De$, (c) variation of the minimum deformation with $Ca_E$, and (d) variation of the minimum deformation with $De$ for $\sigma_r=10$, $\epsilon_r=1.37$ ($PR_A^+$) at $\omega=62.8$.}
    \label{Fig_Dmax_Dmin_PRA_plus_omega62p8}
\end{figure}
\begin{figure}
    \centering
    \includegraphics[width=0.82\textwidth]{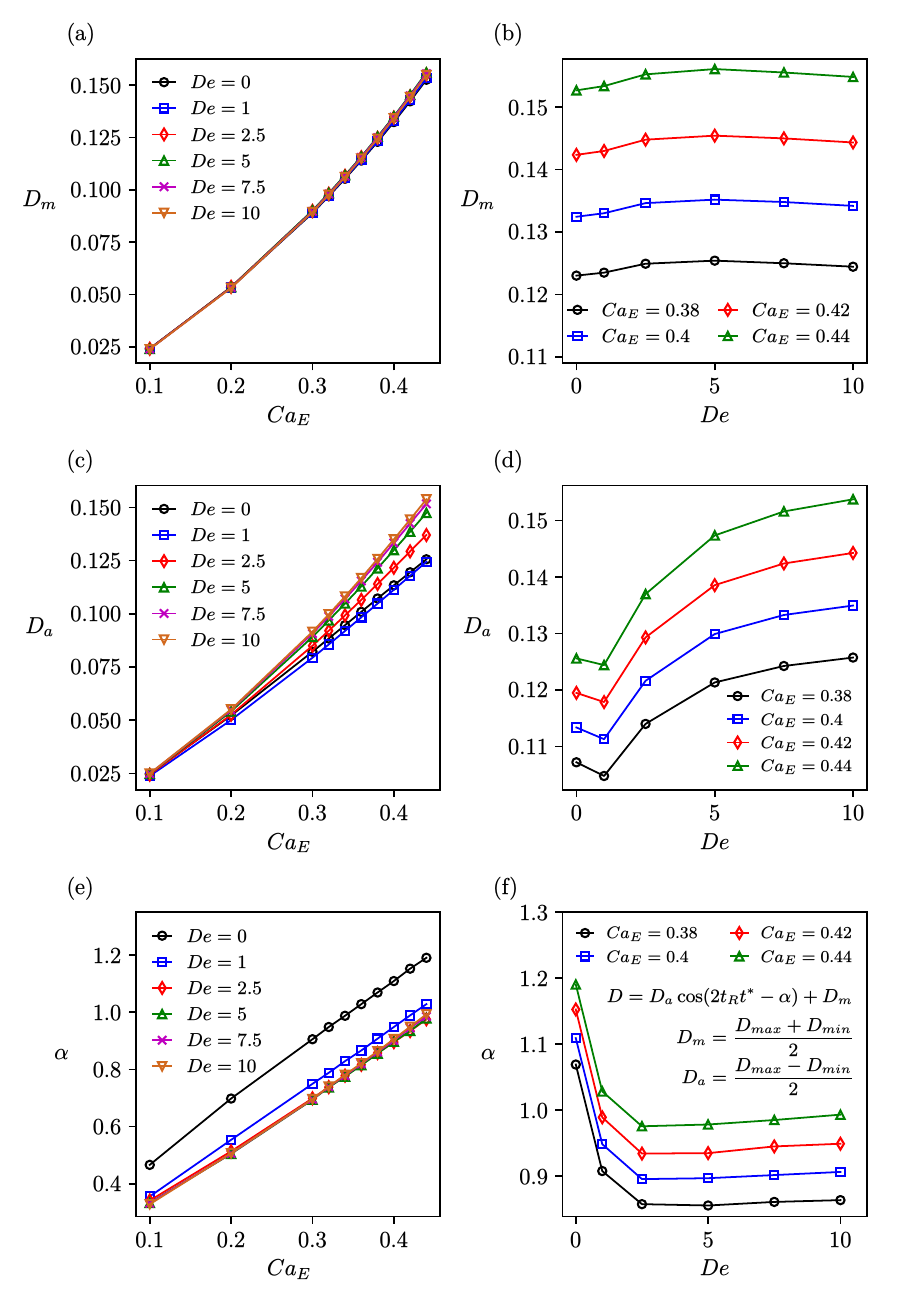}
    \caption{(a) Variation of the mean deformation with $Ca_E$, (b) variation of the mean deformation with $De$, (c) variation of the oscillation amplitude with $Ca_E$, (d) variation of the oscillation amplitude with $De$, (e) variation of the phase difference with $Ca_E$, and (f) variation of the phase difference with $De$ for $\sigma_r=10$, $\epsilon_r=1.37$ ($PR_A^+$) at $\omega=62.8$.}
    \label{Fig_alpha_amp_PRA_plus_omega62p8}
\end{figure}

\autoref{Fig_Dmax_Dmin_PRA_plus_omega62p8} illustrates the influence of $Ca_E$ and $De$ on the deformation extrema ($D_{max}$ and $D_{min}$). \autoref{Fig_Dmax_Dmin_PRA_plus_omega62p8}(a) shows that $D_{max}$ increases monotonically with $Ca_E$ for all investigated $De$, as the stronger electric stresses associated with increasing electric field strength drive greater peak deformation. In contrast, \autoref{Fig_Dmax_Dmin_PRA_plus_omega62p8}(b) reveals a non-monotonic dependence of $D_{max}$ on $De$. Specifically, $D_{max}$ decreases slightly as $De$ increases from $0$ to $1$, followed by a gradual increase with further increase in $De$. Thus, the drop exhibits the greatest resistance to peak deformation around $De=1$. At larger $De$, $D_{max}$ gradually approaches a limiting value for each $Ca_E$. 
\autoref{Fig_Dmax_Dmin_PRA_plus_omega62p8}(c) presents the variation of the minimum deformation ($D_{min}$) with $Ca_E$ for different $De$. For smaller $De$, $D_{min}$ increases with $Ca_E$, whereas at larger $De$ its dependence becomes non-monotonic. At the lowest $Ca_E$, $D_{min}$ becomes slightly negative for all $De$ except $De=1$, indicating that the drop transiently adopts a slightly oblate configuration during part of the oscillation cycle. The range of $Ca_E$ over which this occurs increases with $De$. For example, at $De=10$, $D_{min}$ remains negative for $Ca_E\leq0.4$, whereas at $De=7.5$ it remains negative for $Ca_E\leq0.34$. For $De=5$, negative values occur at $Ca_E=0.1$ and $0.2$, while for $De=0$ and $2.5$, negative $D_{min}$ is observed only at $Ca_E=0.1$. Thus, at intermediate frequency, sufficiently strong viscoelastic effects combined with weak electric forcing can cause the deformation to cross from prolate to slightly oblate during an oscillation cycle. The dependence of $D_{min}$ on $De$ is shown in \autoref{Fig_Dmax_Dmin_PRA_plus_omega62p8}(d). $D_{min}$ initially increases as $De$ increases from $0$ to $1$ and subsequently decreases with further increase in $De$. The decrease for $De>1$ becomes more pronounced as $Ca_E$ increases, indicating that the interaction between the viscoelastic stress response and the electric forcing becomes increasingly important at stronger electric fields.

\autoref{Fig_alpha_amp_PRA_plus_omega62p8} presents the dependence of the mean deformation, oscillation amplitude, and phase difference on $Ca_E$ and $De$. As shown in \autoref{Fig_alpha_amp_PRA_plus_omega62p8}(a), the mean deformation increases monotonically with $Ca_E$, reflecting the increase in electric stress with field strength. In contrast, \autoref{Fig_alpha_amp_PRA_plus_omega62p8}(b) shows only a weak non-monotonic dependence of the mean deformation on $De$, with a slight initial increase followed by a gradual decrease. Overall, the mean deformation remains nearly insensitive to $De$. This weak dependence is consistent with the relevant time scales: the forcing period, $t_\omega=0.05$, is substantially larger than the deformation time scale, $t_{def}=0.00124$--$0.00283$. The interface therefore has sufficient time to deform and recover during each cycle. At the same time, the polymer relaxation time ranges from approximately $0.0059$--$0.0124$ at $De=1$ to $0.0591$--$0.1240$ at $De=10$. Hence, polymeric stresses can undergo appreciable relaxation at lower $De$, while at higher $De$ they persist over a significant portion of the forcing cycle. Because the interface can respond substantially within each forcing cycle, variations in the elastic stress primarily affect the instantaneous deformation response, while their effect on the cycle-averaged deformation remains weak.

Regarding the oscillatory response, \autoref{Fig_alpha_amp_PRA_plus_omega62p8}(c) shows that the oscillation amplitude ($D_a$) increases monotonically with $Ca_E$, consistent with the stronger electric forcing producing a larger variation in deformation during a cycle. \autoref{Fig_alpha_amp_PRA_plus_omega62p8}(d) shows a non-monotonic dependence of $D_a$ on $De$, with an initial decrease from $De=0$ to $De=1$, followed by an increase at larger $De$. This behavior follows directly from the corresponding changes in $D_{max}$ and $D_{min}$. For $0\leq De\leq1$, $D_{max}$ decreases while $D_{min}$ increases, reducing the difference between the two extrema and hence the oscillation amplitude. For $De>1$, $D_{max}$ increases while $D_{min}$ decreases, increasing the difference between the extrema and consequently increasing $D_a$.
The phase response is also affected by the intermediate forcing time scale. Since $t_\omega=0.05$ is much larger than $t_{def}$, the interface has sufficient time to respond to the evolving electric stress, resulting in a substantially smaller phase difference than in the high-frequency regime. Nevertheless, the response is not instantaneous because the deformation is governed by the coupled evolution of electric, viscous, capillary, and polymeric stresses. As shown in \autoref{Fig_alpha_amp_PRA_plus_omega62p8}(e), the phase difference increases with $Ca_E$. This indicates that increasing the electric forcing does not simply increase the deformation magnitude; it also modifies the rate at which the interface evolves during each cycle, producing a larger temporal offset between the forcing and the deformation response.
Finally, \autoref{Fig_alpha_amp_PRA_plus_omega62p8}(f) shows a weak non-monotonic dependence of the phase difference on $De$. The phase difference initially decreases up to approximately $De=2.5$ and then increases gradually with further increase in $De$. This behavior reflects the changing relative importance of polymer relaxation and stress persistence within the forcing cycle. At lower $De$, the polymeric stresses relax more readily and modify the instantaneous deformation response, while at larger $De$, the persistent elastic stresses increasingly influence the temporal evolution of the interface. The resulting competition produces the observed minimum in phase difference rather than a monotonic variation with $De$.

\subsubsection{Low angular frequency regime \texorpdfstring{($\omega = 6.28 \text{ rad/s}$)}{Lg}}
\begin{figure}[ht!]
    \centering
    \includegraphics[width=0.85\textwidth]{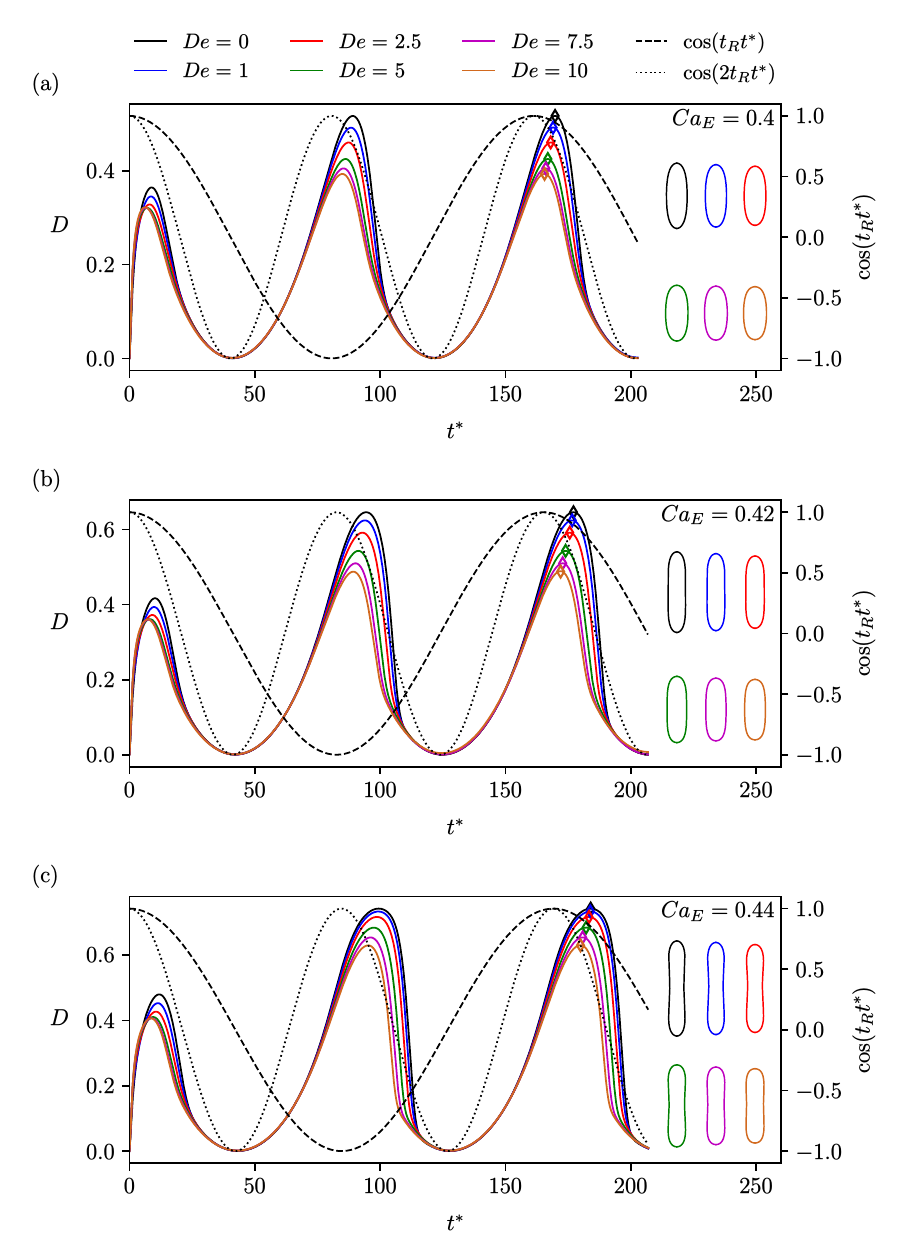}
    \caption{Deformation parameter as a function of non-dimensional time for various Deborah numbers ($De$) at $\sigma_r=10$, $\epsilon_r=1.37$ ($PR_A^+$), $\omega=6.28$, and (a) $Ca_E=0.40$, (b) $Ca_E=0.42$, and (c) $Ca_E=0.44$.}
    \label{Fig_DvsTime_PRA_plus_omega6p28}
\end{figure}
\begin{figure}[b!]
    \centering
    \includegraphics[width=0.85\textwidth]{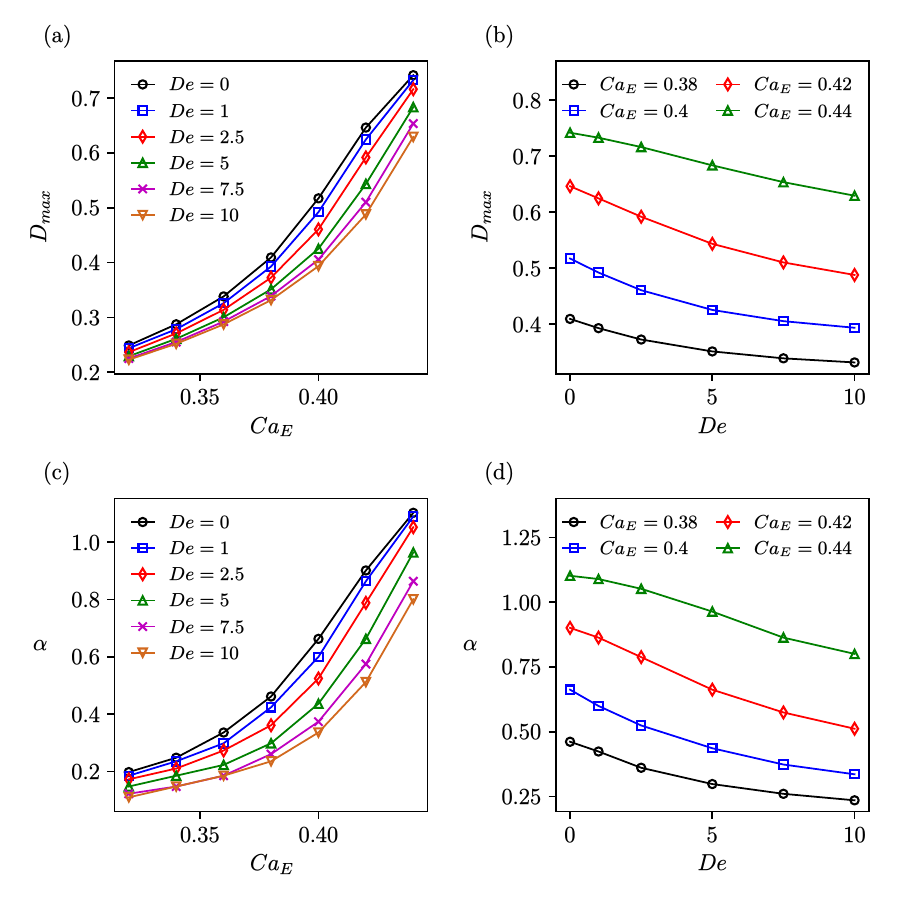}
    \caption{(a) Variation of the maximum deformation with $Ca_E$, (b) variation of the maximum deformation with $De$, (c) variation of phase difference with $Ca_E$, and (d) variation of phase difference with $De$ for $\sigma_r=10$, $\epsilon_r=1.37$ ($PR_A^+$) at $\omega=6.28$.}
    \label{Fig_Dmax_Dmin_PRA_plus_omega6p28}
\end{figure}
For the lowest imposed electric field frequency, $\omega=6.28~\mathrm{rad/s}$, \autoref{Fig_DvsTime_PRA_plus_omega6p28}(a)--(c) show the temporal evolution of the drop deformation for various Deborah numbers ($De$) at $Ca_E=0.4$, 0.42, and 0.44, respectively. At this low frequency, the forcing period $t_\omega = 0.5$  is much larger than both the characteristic deformation time and the polymer relaxation time. Consequently, the drop is able to respond almost completely to the slowly varying electric stresses, and the deformation closely follows a quasi-steady evolution during each oscillation cycle. As expected, the deformation oscillates at twice the frequency of the applied electric field. Unlike the high- and intermediate-frequency cases, the deformation oscillates between a nearly spherical configuration and a prolate shape, with the minimum deformation remaining close to zero throughout the cycle.
Furthermore, the deformation profile deviates from the nearly sinusoidal behavior observed at high and intermediate frequencies. The deformation exhibits a relatively slow increase followed by a comparatively rapid decrease within each oscillation cycle, reflecting the nonlinear response of the interface to the slowly varying electric forcing. Another noteworthy observation, consistent across all investigated $Ca_E$, is that the peak deformation occurs progressively earlier in the cycle with increasing $De$, indicating a systematic reduction in the phase difference between the applied electric field and the drop deformation.

The interfacial shapes corresponding to the maximum deformation, marked by diamonds in \autoref{Fig_DvsTime_PRA_plus_omega6p28}, further illustrate the influence of the electric field strength. For $Ca_E=0.4$, the drop remains spheroidal at the instant of maximum deformation for all investigated Deborah numbers, as shown in \autoref{Fig_DvsTime_PRA_plus_omega6p28}(a). Increasing the electric capillary number to $Ca_E=0.42$ results in a more elongated interface, with the onset of a weak two-lobed structure at lower Deborah numbers (see \autoref{Fig_DvsTime_PRA_plus_omega6p28}(b)). A further increase to $Ca_E=0.44$ leads to the formation of a pronounced two-lobed shape at the deformation peak for all investigated Deborah numbers, as shown in \autoref{Fig_DvsTime_PRA_plus_omega6p28}(c). Since the electric field varies slowly, the drop remains under strong electric forcing for a sufficiently long duration during each cycle, allowing the interface to undergo substantial stretching before recovery begins. Consequently, large deformations and the associated multi-lobed structures are observed only in the low-frequency regime. Simultaneously, the maximum deformation decreases with increasing Deborah number, indicating the increasing resistance of the polymeric stresses to electrically driven deformation.

\autoref{Fig_Dmax_Dmin_PRA_plus_omega6p28} summarizes the variation of the maximum deformation ($D_{max}$) and the phase difference ($\alpha$) with the electric capillary number ($Ca_E$) and the Deborah number ($De$). As shown in \autoref{Fig_Dmax_Dmin_PRA_plus_omega6p28}(a), $D_{max}$ increases monotonically with increasing $Ca_E$ for all investigated Deborah numbers, confirming that stronger electric fields generate larger electric stresses and therefore greater peak deformation. In contrast, \autoref{Fig_Dmax_Dmin_PRA_plus_omega6p28}(b) shows that $D_{max}$ decreases monotonically with increasing $De$. Since the forcing period is considerably larger than the polymer relaxation time, the polymeric stresses have sufficient time to relax during each oscillation cycle, and the deformation approaches a sequence of quasi-steady states. Under these conditions, increasing $De$ primarily enhances the elastic resistance to deformation, thereby reducing the maximum deformation attained by the drop.

The variation of the phase difference is presented in \autoref{Fig_Dmax_Dmin_PRA_plus_omega6p28}(c, d). \autoref{Fig_Dmax_Dmin_PRA_plus_omega6p28}(c) shows that the phase difference increases monotonically with increasing $Ca_E$. At this low frequency, the drop is able to closely follow the slowly varying electric field. However, larger values of $Ca_E$ produce greater deformation, and the interface consequently requires a longer time to attain its maximum deformation, resulting in a larger phase difference. Conversely, \autoref{Fig_Dmax_Dmin_PRA_plus_omega6p28}(d) shows that the phase difference decreases monotonically with increasing $De$. As the Deborah number increases, the maximum deformation decreases, allowing the drop to attain its peak deformation earlier within the forcing cycle. Consequently, the phase lag between the applied electric field and the deformation decreases with increasing viscoelasticity.

Overall, the low-frequency response is fundamentally different from the high- and intermediate-frequency regimes. Since the forcing period is much larger than both the deformation and polymer relaxation time scales, the drop responds in a nearly quasi-steady manner to the oscillating electric field. Consequently, the primary influence of increasing viscoelasticity is a monotonic reduction in both the maximum deformation and the phase difference, whereas increasing the electric field strength provides sufficient time for the interface to develop highly elongated and eventually multi-lobed shapes before recovering during the latter part of the oscillation cycle.

\subsection{\texorpdfstring{$(\sigma_r, \epsilon_r)$ from $PR_B^+$ region}{Lg}}
\label{Sec_chap6_PRB_plus}
\begin{figure}[ht!]
    \centering
    \includegraphics[width=0.95\textwidth]{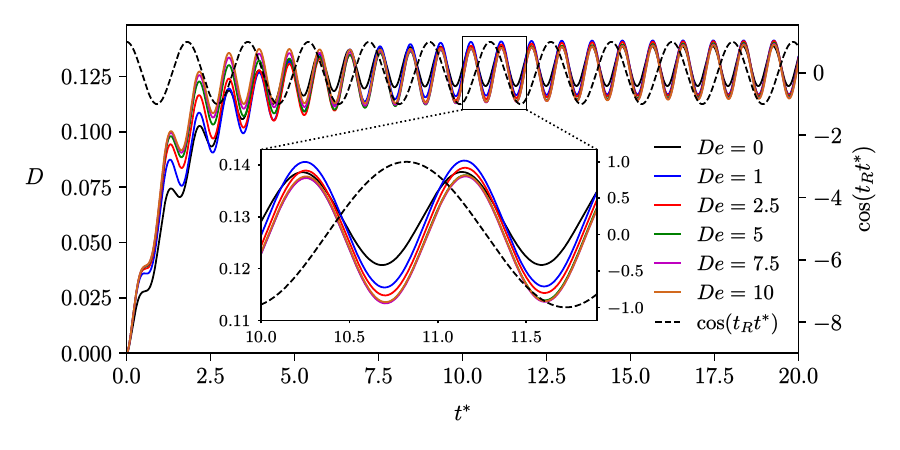}
    \caption{Deformation parameter as a function of non-dimensional time for various Deborah numbers ($De$) at $\sigma_r=25$, $\epsilon_r=50$ ($PR_B^+$), $\omega=628$, and $Ca_E=0.50$.}
    \label{Fig_DvsTime_PRB_plus_omega628}
\end{figure}
The point $(\sigma_r, \epsilon_r) = (25, 50)$, from the $PR_B^+$ region, is considered for a detailed analysis of a viscoelastic drop behavior under an alternating electric field.
\subsubsection{High angular frequency regime \texorpdfstring{($\omega = 628 \text{ rad/s}$)}{Lg}}
\autoref{Fig_DvsTime_PRB_plus_omega628} presents the temporal evolution of the deformation parameter for a viscoelastic drop subjected to an alternating electric field of angular frequency $\omega=628~\mathrm{rad/s}$ at a fixed electric capillary number of $Ca_E=0.5$ for various Deborah numbers. Similar to the $PR_A^+$ region, the deformation oscillates at twice the frequency of the applied electric field and, following the initial transients, varies periodically between well-defined non-zero maximum and minimum values. Furthermore, throughout the investigated range of $Ca_E$ and $De$, the drop retains a spheroidal shape at the instant of maximum deformation, with no evidence of pointed interfaces.

The dependence of the deformation characteristics on $Ca_E$ and $De$ is also qualitatively similar to that observed for the $PR_A^+$ region. The maximum and minimum deformations exhibit the same trends with both parameters, while the mean deformation increases with $Ca_E$ and decreases slightly with $De$. Likewise, the oscillation amplitude increases with increasing $Ca_E$ and becomes nearly independent of $De$ for $De\ge1$, whereas the phase difference decreases with increasing $Ca_E$ and increases with increasing $De$, approaching a limiting value at larger Deborah numbers. Therefore, the high-frequency response in the $PR_B^+$ region is governed by the same mechanism as that discussed for the $PR_A^+$ region, where the forcing period is much shorter than the polymer relaxation time, preventing significant stress relaxation within an oscillation cycle.

\subsubsection{Intermediate angular frequency regime \texorpdfstring{($\omega = 62.8 \text{ rad/s}$)}{Lg}}
\begin{figure}[ht!]
    \centering
    \includegraphics[width=0.95\textwidth]{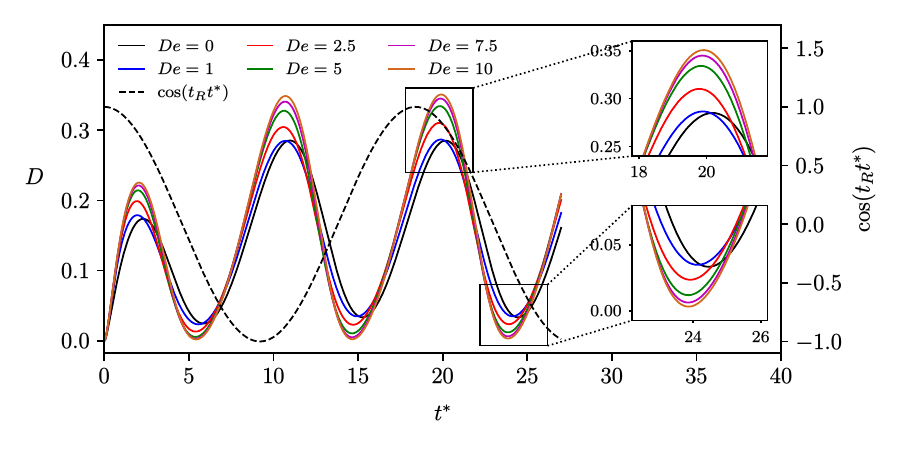}
    \caption{Deformation parameter as a function of non-dimensional time for various Deborah numbers ($De$) at $\sigma_r=25$, $\epsilon_r=50$ ($PR_B^+$), $\omega=62.8$, and $Ca_E=0.52$.}
    \label{Fig_DvsTime_PRB_plus_omega62p8}
\end{figure}
\autoref{Fig_DvsTime_PRB_plus_omega62p8} presents the temporal evolution of the drop deformation for various Deborah numbers ($De$) at the intermediate angular frequency, $\omega=62.8~\mathrm{rad/s}$. As in the $PR_A^+$ region, the deformation oscillates at twice the frequency of the applied electric field and varies periodically between well-defined maximum and minimum values. The dependence of the deformation extrema on $Ca_E$ and $De$ is also qualitatively similar to that reported for the $PR_A^+$ region, indicating that the same interplay between the forcing period and the polymer relaxation time governs the drop dynamics.

\begin{figure}
    \centering
    \includegraphics[width=0.82\textwidth]{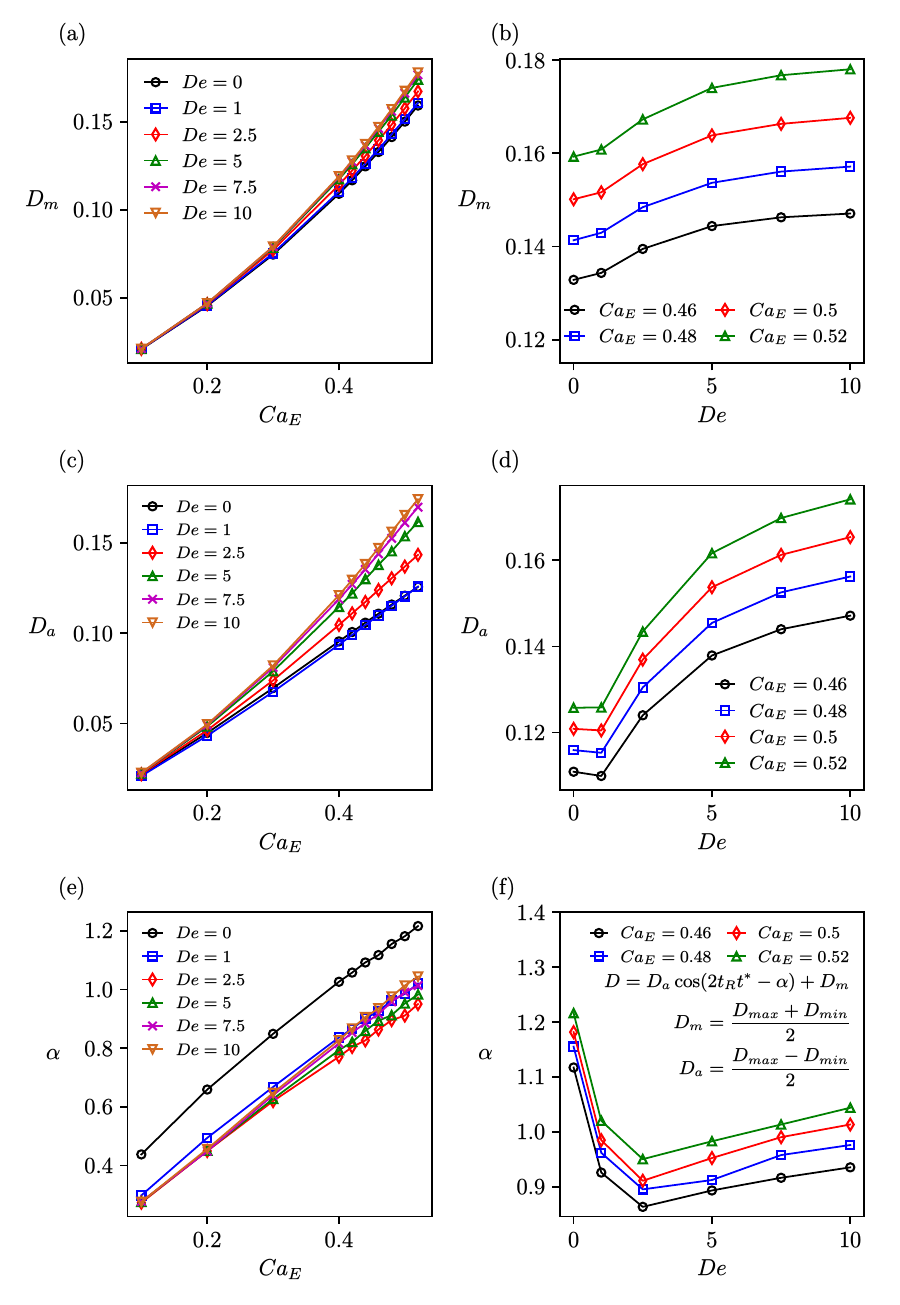}
    \caption{(a) Variation of the mean deformation with $Ca_E$, (b) variation of the mean deformation with $De$, (c) variation of the oscillation amplitude with $Ca_E$, (d) variation of the oscillation amplitude with $De$, (e) variation of the phase difference with $Ca_E$, and (f) variation of the phase difference with $De$ for $\sigma_r=25$, $\epsilon_r=50$ ($PR_B^+$) at $\omega=62.8$.}
    \label{Fig_alpha_amp_PRB_plus_omega62p8}
\end{figure}
\autoref{Fig_alpha_amp_PRB_plus_omega62p8}(a) shows that the mean deformation increases monotonically with increasing $Ca_E$, similar to the trend observed for the $PR_A^+$ region. However, unlike the $PR_A^+$ case, the mean deformation exhibits a slight increase with increasing $De$ (\autoref{Fig_alpha_amp_PRB_plus_omega62p8}(b)). This difference most likely arises from the distinct electrohydrodynamic stress distribution associated with the $(\sigma_r,\epsilon_r)$ pair corresponding to the $PR_B^+$ region. Nevertheless, the dependence of the mean deformation on $De$ remains weak.
The oscillation amplitude, shown in \autoref{Fig_alpha_amp_PRB_plus_omega62p8}(c, d), follows trends qualitatively identical to those observed for the $PR_A^+$ region. Specifically, $D_a$ increases monotonically with increasing $Ca_E$, while it decreases slightly as $De$ increases from 0 to 1 and subsequently increases for larger Deborah numbers. As discussed for the $PR_A^+$ region, this behavior results from the corresponding variations in the maximum and minimum deformation and reflects the comparable magnitudes of the forcing period and the polymer relaxation time at this intermediate frequency.
Similarly, the phase difference shown in \autoref{Fig_alpha_amp_PRB_plus_omega62p8}(e, f) exhibits the same qualitative behavior as in the $PR_A^+$ region. The phase difference increases with increasing $Ca_E$ and varies non-monotonically with $De$, decreasing initially before increasing gradually at larger Deborah numbers. These trends are governed by the same interplay between the forcing period and the polymer relaxation time discussed previously for the $PR_A^+$ region.

\subsubsection{Low angular frequency regime \texorpdfstring{($\omega = 6.28 \text{ rad/s}$)}{Lg}}

\begin{figure}[ht!]
    \centering
    \includegraphics[width=0.85\textwidth]{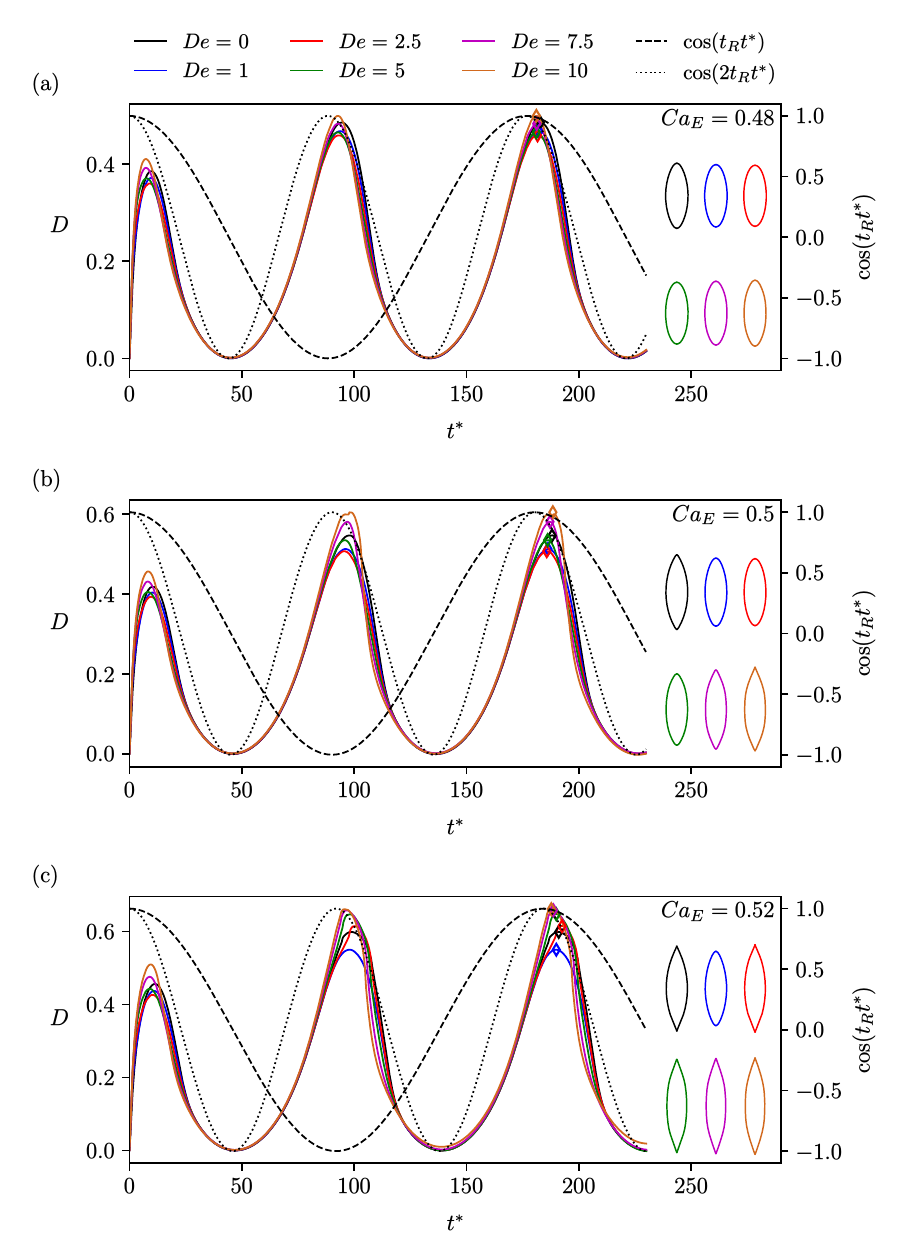}
    \caption{Deformation parameter as a function of non-dimensional time for various Deborah numbers ($De$) at $\sigma_r=25$, $\epsilon_r=50$ ($PR_B^+$), $\omega=6.28$, and (a) $Ca_E=0.48$, (b) $Ca_E=0.50$, and (c) $Ca_E=0.52$.}
    \label{Fig_DvsTime_PRB_plus_omega6p28}
\end{figure}
\autoref{Fig_DvsTime_PRB_plus_omega6p28}(a)--(c) present the temporal evolution of the drop deformation for various Deborah numbers ($De$) at $Ca_E=0.48$, 0.5, and 0.52, respectively, under a low-frequency alternating electric field ($\omega=6.28~\mathrm{rad/s}$). Similar to the $PR_A^+$ region, the deformation oscillates at twice the frequency of the applied electric field and varies between a nearly spherical configuration and a prolate shape. Consequently, unlike the high- and intermediate-frequency cases, the minimum deformation remains close to zero. Furthermore, the deformation profile departs from the nearly sinusoidal behavior observed at higher frequencies, reflecting the quasi-steady response of the drop to the slowly varying electric forcing.

The interfacial shapes corresponding to the maximum deformation, marked by diamonds in the figure, highlight the effect of increasing electric field strength. For $Ca_E=0.48$, the drop remains spheroidal at the instant of maximum deformation for all investigated Deborah numbers, as shown in \autoref{Fig_DvsTime_PRB_plus_omega6p28}(a). Increasing the electric capillary number to $Ca_E=0.5$ results in the development of pointed ends for $De=0$, 7.5, and 10 (see \autoref{Fig_DvsTime_PRB_plus_omega6p28}(b)). A further increase to $Ca_E=0.52$ leads to pointed interfaces for all investigated Deborah numbers except $De=1$, as shown in \autoref{Fig_DvsTime_PRB_plus_omega6p28}(c). Similar to the $PR_A^+$ region, the low forcing frequency allows the interface to remain under strong electric stresses for a sufficiently long duration within each oscillation cycle, enabling substantial stretching before recovery begins.

\begin{figure}[ht!]
    \centering
    \includegraphics[width=0.85\textwidth]{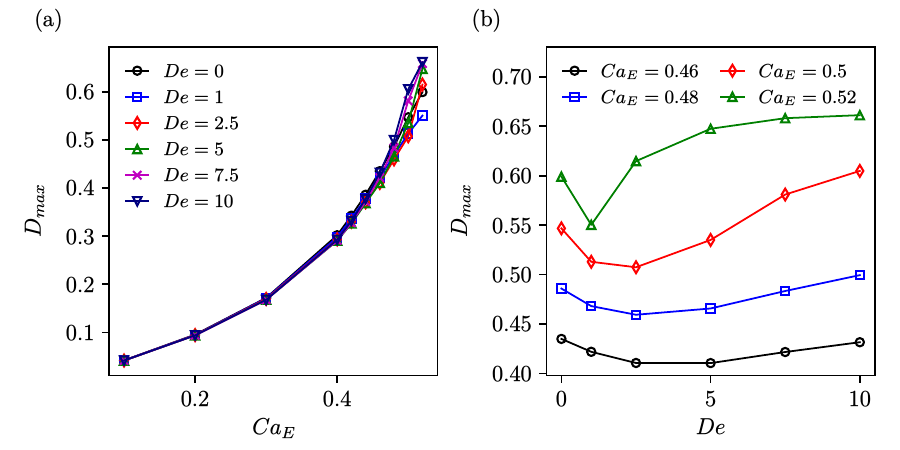}
    \caption{(a) Variation of the maximum deformation with $Ca_E$, (b) variation of the maximum deformation with $De$ for $\sigma_r=25$, $\epsilon_r=50$ ($PR_B^+$) at $\omega=6.28$.}
    \label{Fig_Dmax_Dmin_PRB_plus_omega6p28}
\end{figure}

\autoref{Fig_Dmax_Dmin_PRB_plus_omega6p28} illustrates the dependence of the maximum deformation ($D_{max}$) on the electric capillary number ($Ca_E$) and the Deborah number ($De$). As shown in \autoref{Fig_Dmax_Dmin_PRB_plus_omega6p28}(a), $D_{max}$ increases monotonically with increasing $Ca_E$ for all investigated Deborah numbers, confirming that stronger electric fields generate larger electric stresses and consequently greater peak deformation. In contrast, \autoref{Fig_Dmax_Dmin_PRB_plus_omega6p28}(b) reveals a non-monotonic dependence of $D_{max}$ on $De$. For $Ca_E\le0.5$, $D_{max}$ initially decreases as $De$ increases from 0 to 2.5 and subsequently increases with further increase in $De$, indicating that the deformation is minimum around $De=2.5$. For $Ca_E=0.52$, however, the minimum deformation occurs at $De=1$.
For $Ca_E=0.48$, the drop remains spheroidal at the instant of maximum deformation for all investigated Deborah numbers. Increasing $Ca_E$ to 0.5 results in the appearance of pointed ends at peak deformation for $De=0$, 7.5, and 10, whereas at $Ca_E=0.52$, pointed interfaces are observed for all Deborah numbers except $De=1$. Thus, the onset of pointed interfaces correlates well with larger values of $D_{max}$, while the drop remains spheroidal for the Deborah number corresponding to the smallest peak deformation. 
The non-monotonic variation of $D_{max}$ with $De$ may be associated with the competing effects of viscoelastic resistance and stress-history effects. The initial increase in $De$ increases the elastic resistance of the drop, leading to a slight reduction in $D_{max}$ from $De=0$ to $De=1$. At larger $De$, the stronger viscoelastic memory can influence the maximum deformation attained during a cycle. In the corresponding steady electric-field problem, pronounced transient maxima in deformation were observed at higher $De$ before the drop reached its steady state for $PR_B^+$ region (\citet{bangar2026large}). Although the present response is periodic, a similar history-dependent response may contribute to the recovery of $D_{max}$ at larger $De$ and hence to the observed non-monotonic behavior.

\subsection{\texorpdfstring{$(\sigma_r, \epsilon_r)$ from $OB^-$ region}{Lg}}
\label{Sec_chap6_OB_plus}
To investigate the influence of an alternating electric field on the drop of Oldroyd-B fluid for $(\sigma_r,\epsilon_r)$ from $OB^-$ region, we select $(\sigma_r,\epsilon_r)=(0.1,2)$.
\subsubsection{High angular frequency regime \texorpdfstring{($\omega = 628 \text{ rad/s}$)}{Lg}}
\begin{figure}[ht!]
    \centering
    \includegraphics[width=0.95\textwidth]{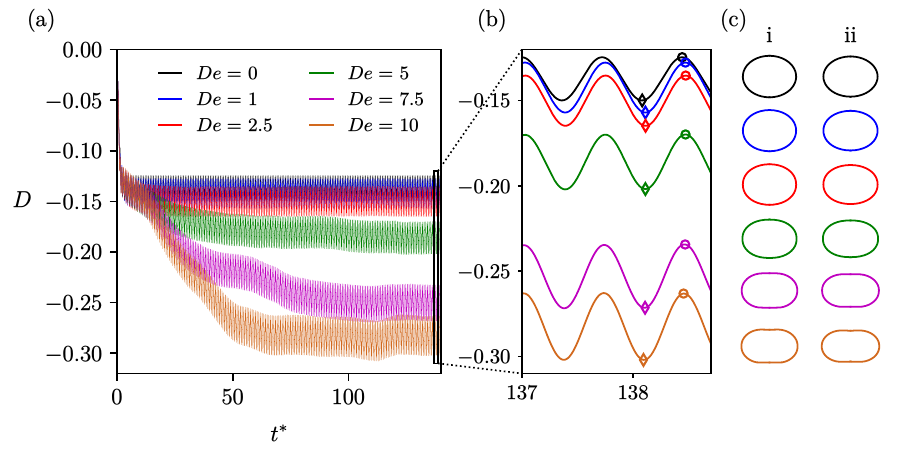}
    \caption{Deformation parameter as a function of non-dimensional time for various Deborah numbers ($De$) at $\sigma_r=0.1$, $\epsilon_r=2$ ($OB^-$), $\omega=628$, and $Ca_E=0.32$. (a) Deformation evolution for various $De$, (b) magnified view showing the shift of the deformation extrema toward more negative values with increasing $De$, and (c) corresponding drop interfaces at (i) $D_{max}$ and (ii) $D_{min}$, representing the least and most oblate states, respectively.}
    \label{Fig_DvsTime_OB_minus_omega628}
\end{figure}
Similar to the previously considered cases, the deformation parameter ($D$) oscillates at twice the frequency of the applied alternating electric field. \autoref{Fig_DvsTime_OB_minus_omega628}(a) presents the temporal evolution of the deformation of an Oldroyd-B drop corresponding to $(\sigma_r,\epsilon_r)=(0.1,2)$ in the $OB^-$ region, subjected to a high-frequency alternating electric field with $\omega=628~\mathrm{rad/s}$. Following the initial transient, $D$ settles into periodic oscillations entirely within the negative range, indicating that the drop remains oblate throughout the oscillation cycle. The magnified view in \autoref{Fig_DvsTime_OB_minus_omega628}(b) shows that, with increasing $De$, both extrema shift towards more negative values, corresponding to an increase in the magnitude of oblate deformation. The corresponding interfaces at $D_{max}$ and $D_{min}$ are shown in \autoref{Fig_DvsTime_OB_minus_omega628}(c)i and (c)ii, respectively, representing the least and most oblate states attained during a cycle. The interfaces become increasingly non-spherical with increasing $De$.

\begin{figure}
    \centering
    \includegraphics[width=0.82\textwidth]{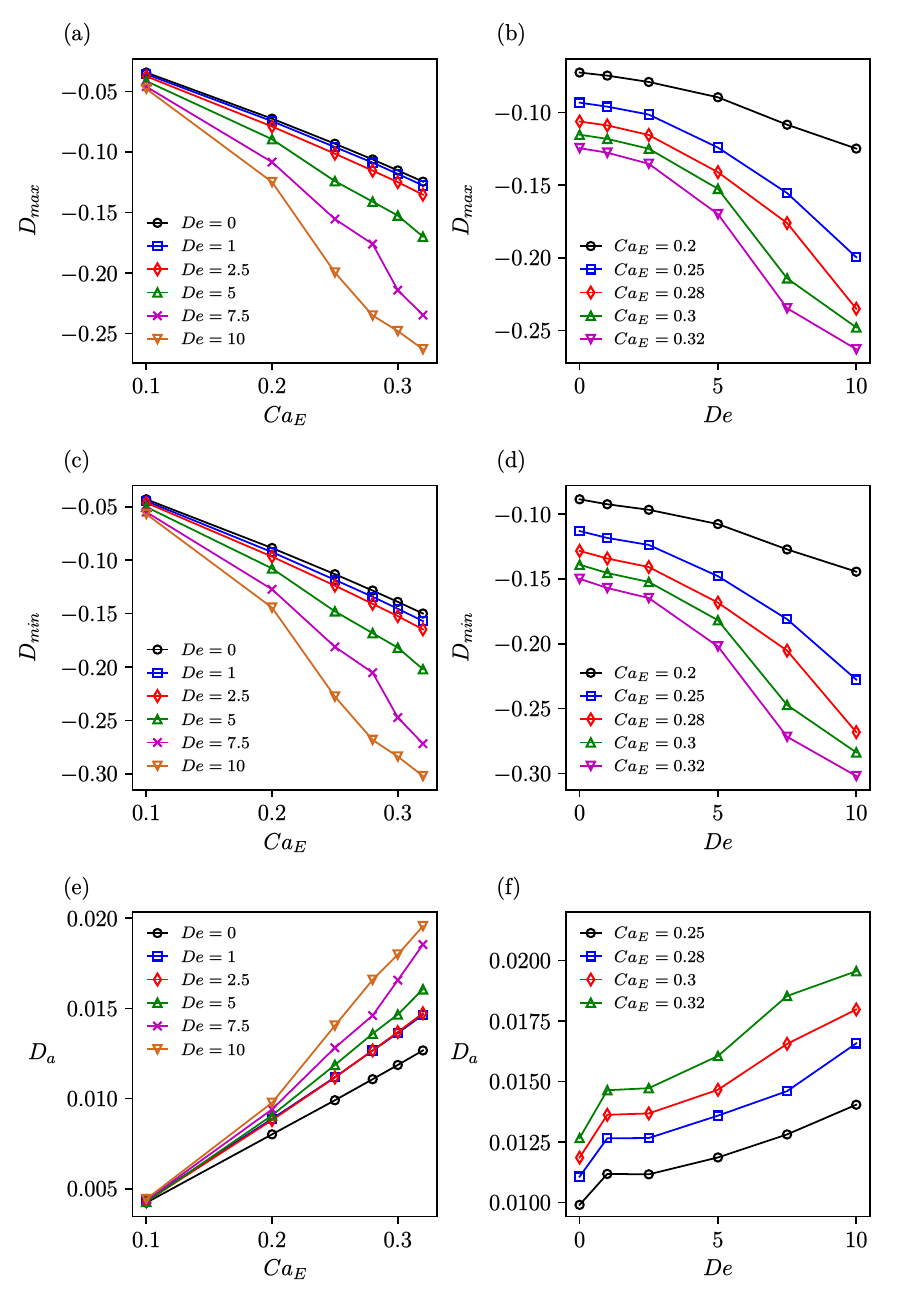}
    \caption{(a) Variation of the maximum deformation with $Ca_E$, (b) variation of the maximum deformation with $De$, (c) variation of the minimum deformation with $Ca_E$, (d) variation of the minimum deformation with $De$, (e) variation of the oscillation amplitude $D_a$ with $Ca_E$, and (f) variation of the oscillation amplitude $D_a$ with $De$ for $\sigma_r=0.1$, $\epsilon_r=2$ ($OB^-$) at $\omega=628$.}
    \label{Fig_Dmax_Dmin_OB_minus_omega628}
\end{figure}
The dependence of the deformation extrema on $Ca_E$ and $De$ is presented in \autoref{Fig_Dmax_Dmin_OB_minus_omega628}. As shown in \autoref{Fig_Dmax_Dmin_OB_minus_omega628}(a) and (b), $D_{max}$ decreases monotonically with both $Ca_E$ and $De$, indicating that the least oblate state becomes progressively more deformed. Similarly, \autoref{Fig_Dmax_Dmin_OB_minus_omega628}(c) and (d) show that $D_{min}$ decreases monotonically with both parameters, indicating an increase in the maximum magnitude of oblate deformation. Thus, increasing either the electric field strength or $De$ shifts the entire deformation cycle towards stronger oblate states. The increase with $Ca_E$ results from the larger electric stresses generated by stronger electric fields. The dependence on $De$ reflects the distinct viscoelastic response of the $OB^-$ regime: at the high frequency considered here, the forcing period is shorter than the polymer relaxation time, limiting polymer relaxation during each cycle. As $De$ increases, the viscoelastic contribution therefore has a stronger influence on the interfacial response, resulting in an enhanced magnitude of oblate deformation. This behavior is opposite to the reduction in prolate deformation observed in the $PR_A^+$ and $PR_B^+$ regimes.

The oscillation amplitude $D_a$, shown in \autoref{Fig_Dmax_Dmin_OB_minus_omega628}(e) and (f), increases monotonically with both $Ca_E$ and $De$. The increase with $Ca_E$ follows from the stronger variation of the electric stresses over a forcing cycle, while the increase with $De$ is consistent with the increasing separation of the two deformation extrema. Hence, increasing electric field strength or viscoelasticity produces both stronger oblate deformation and a larger variation in deformation over each cycle.

\begin{figure}[ht!]
    \centering
    \includegraphics[width=0.95\textwidth]{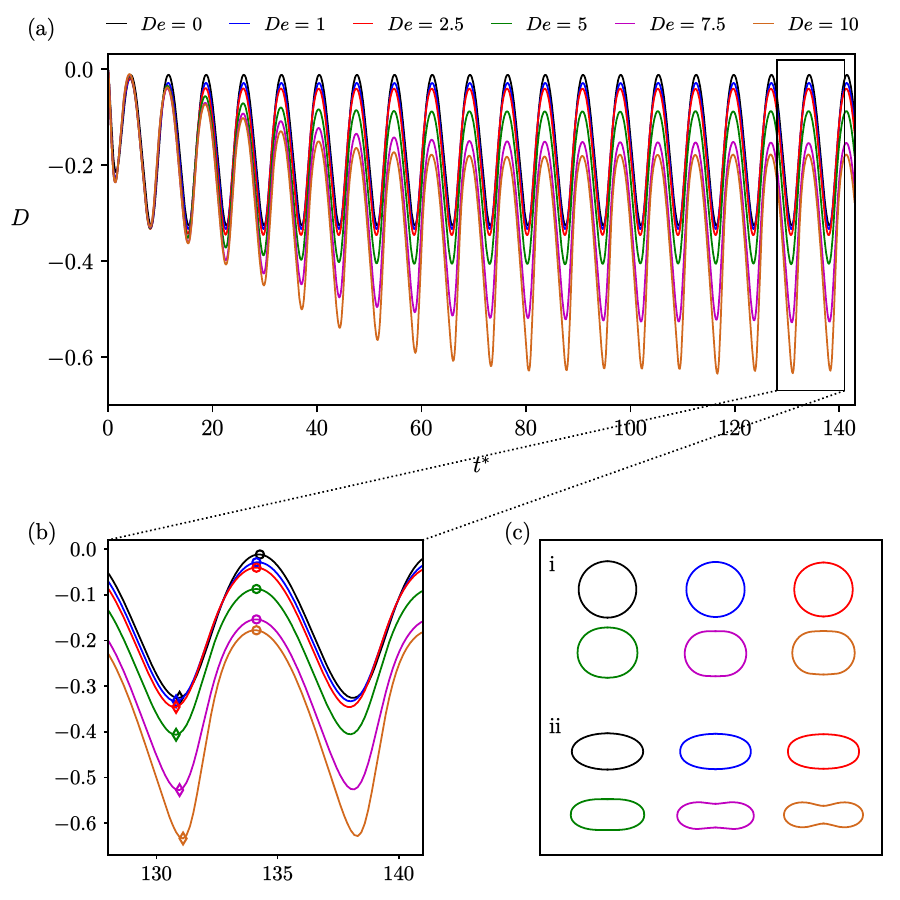}
    \caption{Temporal evolution of the deformation parameter $D$ for an Oldroyd-B drop in the $OB^-$ region, with $(\sigma_r,\epsilon_r)=(0.1,2)$, subjected to an intermediate-frequency alternating electric field with $\omega=62.8~\mathrm{rad/s}$ at $Ca_E=0.32$. (a) Deformation evolution for various $De$, (b) magnified view showing the shift of both deformation extrema toward more negative values with increasing $De$, and (c) corresponding drop interfaces at $D_{max}$ and $D_{min}$, representing the least and most oblate states, respectively.}
    \label{Fig_DvsTime_OB_minus_omega62p8}
\end{figure}
\subsubsection{Intermediate angular frequency regime \texorpdfstring{($\omega = 62.8 \text{ rad/s}$)}{Lg}}
\begin{figure}
    \centering
    \includegraphics[width=0.82\textwidth]{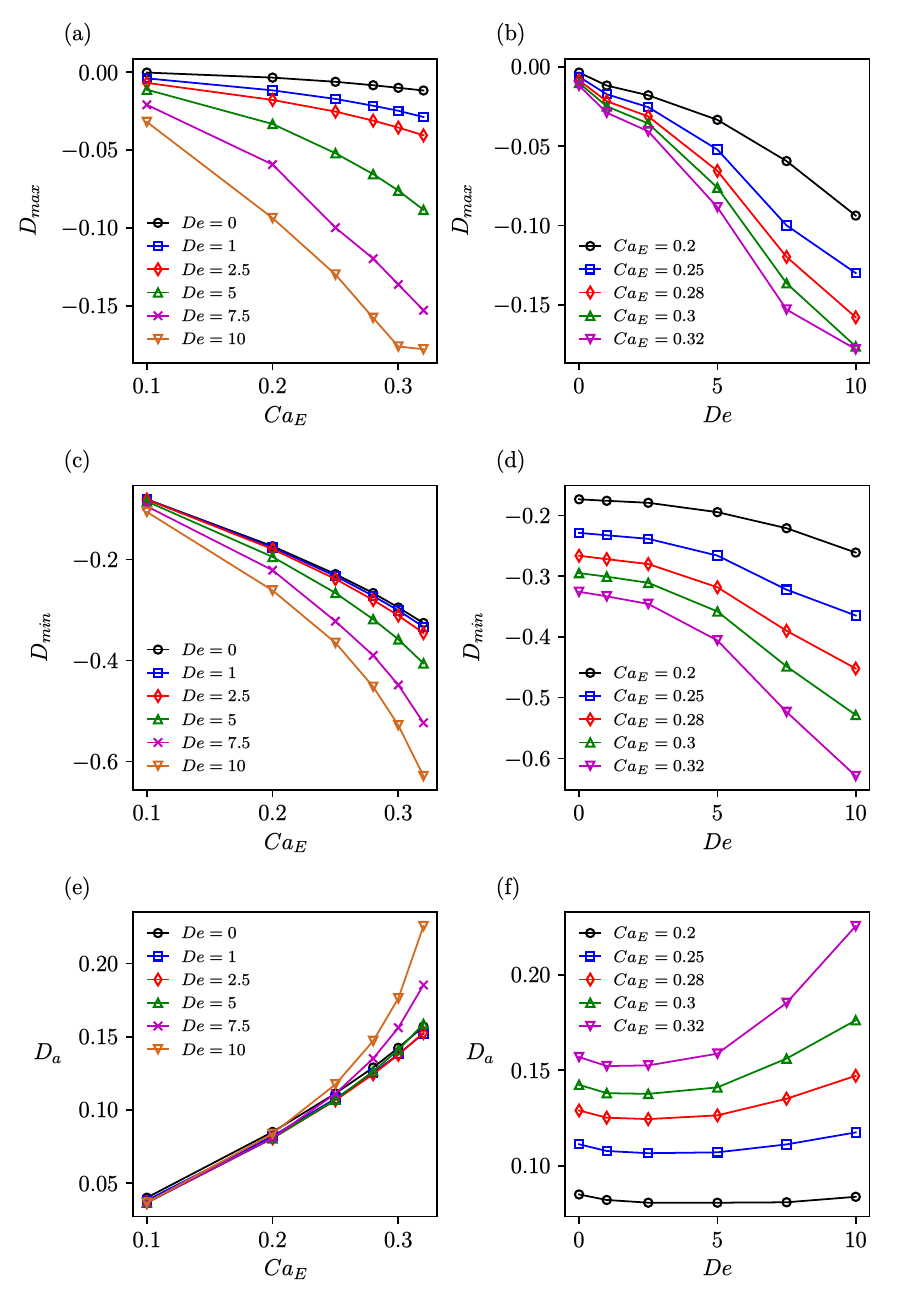}
    \caption{(a) Variation of the maximum deformation with $Ca_E$, (b) variation of the maximum deformation with $De$, (c) variation of the minimum deformation with $Ca_E$, (d) variation of the minimum deformation with $De$, (e) variation of the oscillation amplitude $D_a$ with $Ca_E$, and (f) variation of the oscillation amplitude $D_a$ with $De$ for $\sigma_r=0.1$, $\epsilon_r=2$ ($OB^-$) at $\omega=62.8$.}
    \label{Fig_Dmax_Dmin_OB_minus_omega62p8}
\end{figure}
This section examines the deformation of an Oldroyd-B drop subjected to an alternating electric field with an intermediate angular frequency of $\omega=62.8~\mathrm{rad/s}$ for $(\sigma_r,\epsilon_r)=(0.1,2)$ in the $OB^-$ region. \autoref{Fig_DvsTime_OB_minus_omega62p8}(a) presents the drop deformation as a function of non-dimensionalized time for various Deborah numbers at a constant $Ca_E=0.32$. Following the initial transient, the deformation settles into stable periodic oscillations between well-defined extrema, with the oscillation frequency equal to twice that of the applied electric field. The magnified view in \autoref{Fig_DvsTime_OB_minus_omega62p8}(b) shows that increasing $De$ shifts both extrema towards more negative values, corresponding to an increase in the magnitude of both the minimum and maximum oblate deformation. The corresponding interfaces are shown in \autoref{Fig_DvsTime_OB_minus_omega62p8}(c). At $D_{max}$, corresponding to the least oblate state during a cycle, the drop remains approximately spheroidal for all investigated $De$. At $D_{min}$, corresponding to the maximum magnitude of oblate deformation, increasingly pronounced dimples are observed at higher $De$.

The dependence of the deformation extrema and oscillation amplitude on $Ca_E$ and $De$ is presented in \autoref{Fig_Dmax_Dmin_OB_minus_omega62p8}. As shown in \autoref{Fig_Dmax_Dmin_OB_minus_omega62p8}(a) and (c), both $D_{max}$ and $D_{min}$ decrease monotonically with increasing $Ca_E$, indicating that the magnitude of both the least and most oblate states increases with electric field strength. The oscillation amplitude $D_a$ also increases monotonically with $Ca_E$, as shown in \autoref{Fig_Dmax_Dmin_OB_minus_omega62p8}(e). This behavior results from the increase in the electric stresses with increasing field strength, which drives the drop towards increasingly oblate configurations.

A similar monotonic decrease in $D_{max}$ and $D_{min}$ with increasing $De$ is observed in \autoref{Fig_Dmax_Dmin_OB_minus_omega62p8}(b) and (d), respectively. Thus, increasing $De$ increases the magnitude of oblate deformation throughout the oscillation cycle. However, unlike the monotonic behavior with $Ca_E$, \autoref{Fig_Dmax_Dmin_OB_minus_omega62p8}(f) shows that the oscillation amplitude initially decreases slightly with increasing $De$ and subsequently increases at larger $De$. This non-monotonic variation results from the different rates at which the two deformation extrema change with $De$: over the lower-$De$ range, the changes in $D_{max}$ and $D_{min}$ reduce the separation between the extrema, whereas at larger $De$ their separation increases, resulting in a larger oscillation amplitude.

The time scales provide a physical basis for this behavior. At $\omega=62.8~\mathrm{rad/s}$, the deformation oscillation period is $t_\omega=0.05$, while the characteristic deformation time is approximately $0.00124$--$0.00222$. Hence, $t_{def}\ll t_\omega$, allowing the interface to respond rapidly to the slowly varying electric forcing. The polymer relaxation time varies approximately from $0.00693$ to $0.00877$ over the investigated range of $Ca_E$ for $De = 1$ and $0.0693$ to $0.0877$ for $De=10$. Thus, $t_{def} < \lambda < t_\omega$ for lower $De$ and $t_{def} < t_\omega <\lambda$ for higher $De$.
Thus, at lower $De$, $\lambda$ is smaller than the forcing period and the polymer can undergo substantial relaxation within each cycle. As $De$ increases, $\lambda$ becomes comparable to and eventually larger than $t_\omega$, causing the polymeric phase to retain increasing memory of the deformation over a forcing cycle. The resulting change from a more rapidly relaxing to a more persistent viscoelastic response modifies the deformation dynamics and contributes to the non-monotonic variation of the oscillation amplitude with $De$. At sufficiently large $De$, the relaxation time exceeds the forcing period, and further increases in $De$ produce a progressively stronger persistent elastic response, leading to the observed increase in $D_a$.

\subsubsection{Low angular frequency regime \texorpdfstring{($\omega = 6.28 \text{ rad/s}$)}{Lg}}
\begin{figure}[ht!]
    \centering
    \includegraphics[width=0.85\textwidth]{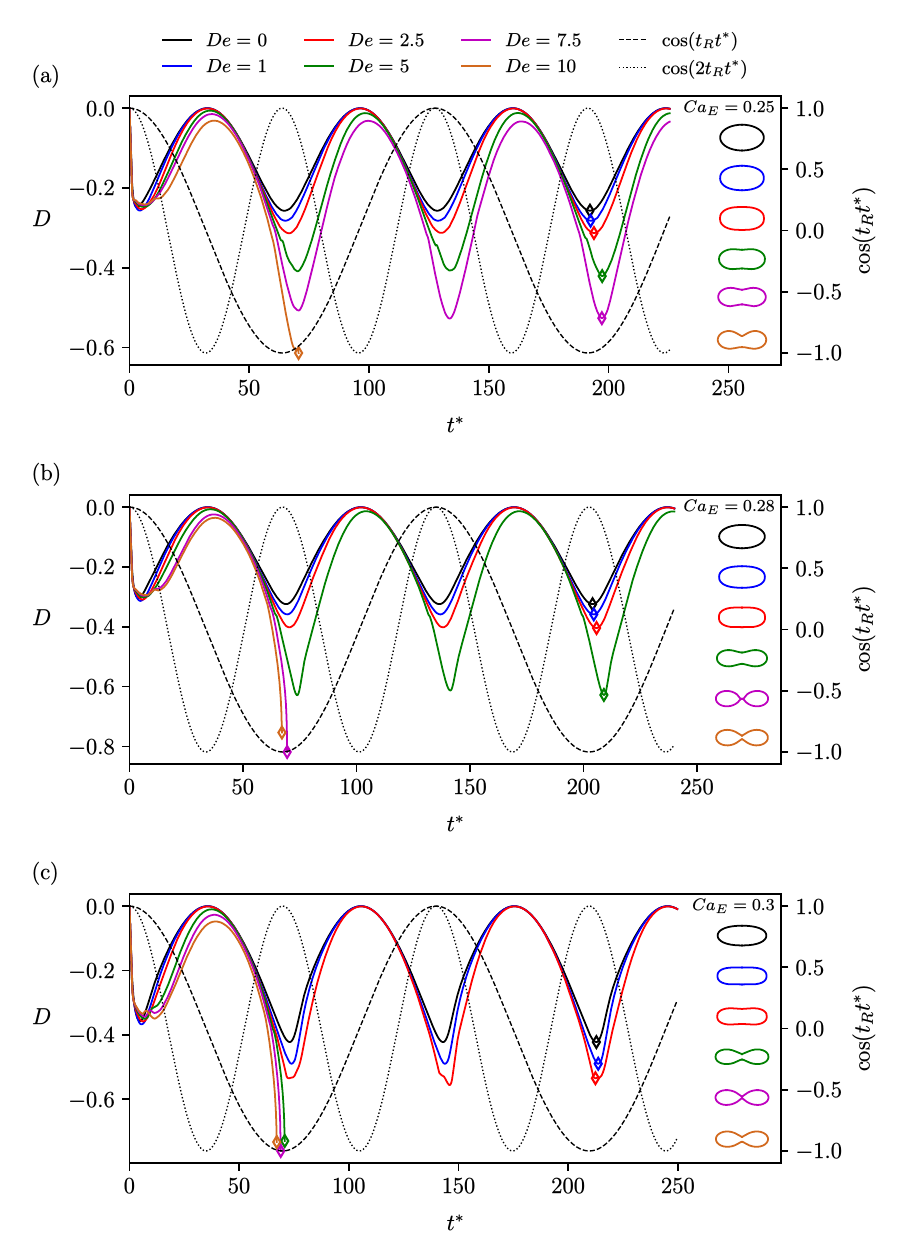}
    \caption{Deformation parameter as a function of non-dimensional time for various Deborah numbers ($De$) for $\sigma_r=0.1$, $\epsilon_r=2$ ($OB^-$), $\omega=6.28$, and (a) $Ca_E=0.25$, (b) $Ca_E=0.28$, and (c) $Ca_E=0.30$.}
    \label{Fig_DvsTime_OB_minus_omega6p28}
\end{figure}
\begin{figure}[ht!]
    \centering
    \includegraphics[width=0.85\textwidth]{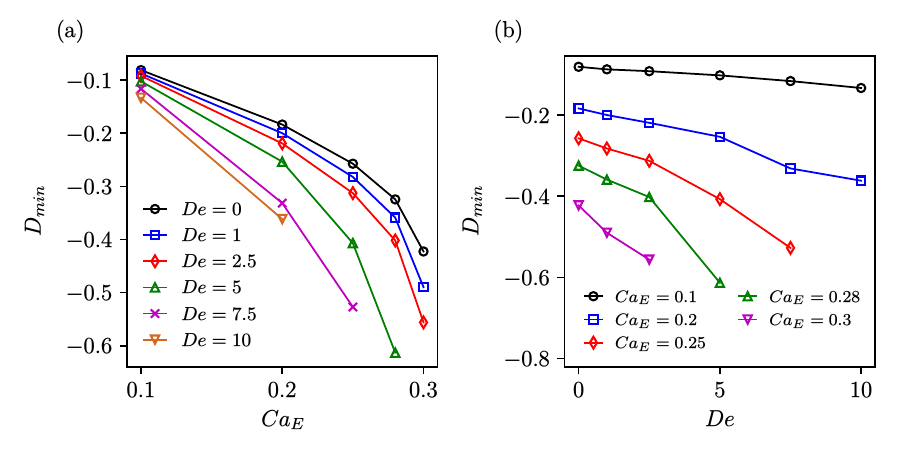}
    \caption{a) Variation of minimum deformation with $Ca_E$ and (b) Variation of minimum deformation with $De$ for $\sigma_r=0.1$, $\epsilon_r=2$ ($OB^-$), at $\omega=6.28$.}
    \label{Fig_Dmin_OB_minus_omega6p28}
\end{figure}
\autoref{Fig_DvsTime_OB_minus_omega6p28}(a), (b), and (c) present the temporal evolution of the drop deformation for $(\sigma_r,\epsilon_r)=(0.1,2)$ in the $OB^-$ region at three electric capillary numbers, $Ca_E=0.25$, $0.28$, and $0.3$, respectively. For a given $Ca_E$, increasing $De$ progressively promotes drop breakup, while increasing $Ca_E$ causes breakup to occur at lower values of $De$. At $Ca_E=0.25$, breakup occurs only at $De=10$, whereas at $Ca_E=0.28$, breakup is observed at $De=7.5$ and $10$. For $Ca_E=0.3$, breakup occurs at $De=5$, $7.5$, and $10$. At the still higher value of $Ca_E=0.32$, not shown in the figure, breakup occurs for all investigated $De$ values ($0$, $1$, $2.5$, $5$, $7.5$, and $10$). Thus, increasing either the electric field strength or the Deborah number promotes the transition from bounded deformation to breakup.

For the cases in which the drop remains intact, $D$ oscillates at twice the frequency of the applied electric field and remains predominantly negative, indicating oblate deformation throughout the cycle. The maximum deformation approaches zero at lower $De$, corresponding to a nearly spherical state, whereas at higher $De$ the maximum remains slightly negative, indicating that the drop does not completely recover its spherical shape. In contrast, $D_{min}$ becomes progressively more negative with increasing $De$, corresponding to an increase in the maximum magnitude of oblate deformation. The interface shapes shown at $D_{min}$ further demonstrate this enhancement, with increasingly pronounced dimples appearing at higher $De$. For cases that undergo breakup, the interface corresponding to the onset of breakup is shown.

The dependence of the maximum magnitude of oblate deformation on $Ca_E$ and $De$ is quantified in \autoref{Fig_Dmin_OB_minus_omega6p28}. As shown in \autoref{Fig_Dmin_OB_minus_omega6p28}(a), $D_{min}$ decreases monotonically with increasing $Ca_E$ for all investigated $De$, indicating that stronger electric stresses produce a larger magnitude of oblate deformation. Similarly, \autoref{Fig_Dmin_OB_minus_omega6p28}(b) shows that $D_{min}$ decreases monotonically with increasing $De$ for all $Ca_E$. Hence, increasing viscoelasticity enhances the maximum magnitude of oblate deformation and, at sufficiently large $Ca_E$ and $De$, drives the drop beyond the limit of stable deformation, resulting in breakup.

The observed trends can be interpreted in terms of the characteristic time scales at this low forcing frequency. For $\omega=6.28~\mathrm{rad/s}$, the deformation oscillation period is $t_\omega=0.5$, whereas the deformation time scale is approximately $0.00124$--$0.00283$ over the range of $Ca_E$ considered. The polymer relaxation time varies approximately from $0.00693$ to $0.00877$ over the investigated range of $Ca_E$ for $De = 1$ and $0.0693$ to $0.0877$ for $De=10$. Thus, $t_{def}<\lambda \ll t_\omega$, so both the interface and the polymeric stresses evolve on time scales much shorter than the forcing period. The drop therefore has sufficient time to respond to the slowly varying electric field during each cycle. Increasing $De$ increases the relaxation time and allows the viscoelastic stresses to persist for a larger portion of the deformation process, thereby modifying the electrically driven response and enhancing the magnitude of oblate deformation. When the combined effect of electric and viscoelastic stresses becomes sufficiently strong, the deformation exceeds the range over which the interface can remain intact, leading to breakup. This also explains why breakup occurs at progressively lower $De$ as $Ca_E$ is increased: stronger electric stresses require a smaller additional viscoelastic contribution to reach the breakup condition.

%%%%%%%%%%%%%%%%%%%%%%%%%%%%%%%%%%%%%%%%%%%%%%%%%%%%%%%%%%%%%%%%%%%%%%%%%%%%%%%%%%%%%%%%
%%%%%%%%%%%%%%%%%%%%%%%%%%%%%%%%%%%% CONCLUSIONS %%%%%%%%%%%%%%%%%%%%%%%%%%%%%%%%%%%%%%%
\section{Conclusion}\label{sec:conclusion}
This study examined the deformation dynamics of an Oldroyd-B drop subjected to an alternating electric field, combining asymptotic analysis in the limit of small deformation ($Ca_E \ll 1$) and weak viscoelasticity ($De \ll 1$) with direct numerical simulations (using the open-source solver Basilisk) to access the strongly deformed, highly viscoelastic regime. Simulations spanned three representative points on the 
($\sigma_r$, $\epsilon_r$) phase plane, $PR_A^+$, $PR_B^+$ and $OB^-$, at three angular frequencies of the applied field (628, 62.8, and 6.28 rad/s). Across every case, the drop deformation oscillates at twice the frequency of the applied field, a direct consequence of the quadratic dependence of the electric stress on field strength.

In the prolate regions ($PR_A^+$ and $PR_B^+$), the character of the response depends strongly on the forcing frequency relative to two internal time scales: the interfacial deformation time and the polymer relaxation time. At high frequency, the forcing period is much shorter than the polymer relaxation time, so polymeric stresses cannot relax within a cycle; both $D_{max}$ and $D_{min}$ remain positive throughout, indicating sustained prolate deformation, the mean deformation decreases monotonically with $De$, and the oscillation amplitude and phase lag both rise from $De = 0$ to $De \approx 1$ before saturating as the relaxation time comes to far exceed the oscillation period. At intermediate frequency, the forcing period becomes comparable to the polymer relaxation time and the drop responds more fully within each cycle; $D_{min}$ can turn slightly negative, indicating a brief prolate-to-oblate excursion during part of the cycle, while the mean deformation varies only weakly with $De$ -- non-monotonically in $PR_A^+$ and monotonically increasing in $PR_B^+$ -- and the oscillation amplitude increases with $De$ once $De \gtrsim 1$. At low frequency, the forcing period far exceeds both the deformation and relaxation time scales, so the drop deforms in a quasi-steady manner: the deformation oscillates between a nearly spherical state and a peak shape that becomes multi-lobed in $PR_A^+$ or pointed in $PR_B^+$ at sufficiently large $Ca_E$, with peak deformation decreasing monotonically with $De$ in $PR_A^+$ but varying non-monotonically in $PR_B^+$, likely reflecting history-dependent elastic effects similar to those observed in the corresponding steady-field problem.

In the oblate region ($OB^-$), the drop deformation remains oblate throughout the oscillation cycle at all three frequencies, with dimpled interfaces appearing at peak deformation for large $Ca_E$ and $De$. Unlike in the prolate regions, increasing $De$ consistently enhances the magnitude of oblate deformation at every frequency, so that elasticity reinforces rather than resists the electrically driven shape change. At low frequency this enhancement becomes severe enough to drive the drop to breakup once $Ca_E$ and $De$ exceed threshold values, with the breakup threshold decreasing as either parameter increases.

Overall, the observed frequency-dependent behavior is governed by the relative time scales of electric forcing, interfacial deformation, and polymer relaxation. At high frequency, polymeric stresses retain strong memory over an oscillation cycle, whereas at intermediate frequency their evolution becomes comparable to the forcing time scale. At low frequency, the drop responds nearly quasi-steadily to the slowly varying electric field.

%%%%%%%%%%%%%%%%%%%%%%%%%%%%%%%%%%%%%%%%%%%%%%%%%%%%%%%%%%%%%%%%%%%%%%%%%%%%%%%%%%%%%%%%
%%%%%%%%%%%%%%%%%%%%%%%%%%%%%%%%%%%%%% APPENDIX %%%%%%%%%%%%%%%%%%%%%%%%%%%%%%%%%%%%%%%%
\appendix
\section{Asymptotic solution}\label{appA}
\subsection{Solution of \texorpdfstring{$\order{1}$}{Lg} equations}
\label{Sec_O1_Sol}
The governing equations at $\order{1}$ are,
\begin{equation}
	\bm{\nabla}^2 V^*_{i_{00}} = 0 \quad \text{where} \quad \bm{E}^*_{i_{00}} = -\bm{\nabla}V^*_{i_{00}} \quad \text{and} \quad \bm{\nabla}^2 V^*_{e_{00}} = 0 \quad \text{where} \quad \bm{E}^*_{e_{00}} = -\bm{\nabla}V^*_{e_{00}}
\end{equation}

subjected to the boundary conditions given by,
\begin{subequations}
\begin{equation}
    \frac{\partial V^*_{i_{00}}}{\partial r} = 0 \quad \text{at} \quad r = 0 \quad ; \quad \bm{E}^*_{e_{00}} \rightarrow 1 \quad \text{at} \quad r \rightarrow \infty
\end{equation}
\begin{equation}
    V^*_{i_{00}} = V^*_{e_{00}} \quad \text{at} \quad r = 1 \quad ; \quad (\sigma_r + i\omega a \epsilon_r) E^*_{i_{r_{00}}} = (1 + i\omega a) E^*_{e_{r_{00}}} \quad \text{at} \quad r = 1.
\end{equation}
\end{subequations}
General solution of the Laplace equation in spherical axisymmetric coordinate system gives,
\begin{equation}
    V^*_{i_{00}} = \sum_n \left( A^{E^*}_{i_{{00}_n}} r^n + \frac{B^{E^*}_{i_{{00}_n}}}{r^{n+1}} \right) P_n(\cos\theta) \quad \text{and} \quad V^*_{e_{00}} = \sum_n \left( A^{E^*}_{e_{{00}_n}} r^n + \frac{B^{E^*}_{e_{{00}_n}}}{r^{n+1}} \right) P_n(\cos\theta) 
\end{equation}
where implementation of boundary conditions gives the following constants with all other constants being zero.
\begin{equation}
    \label{Eqn_Const_AE_e001}
	A^{E^*}_{e_{{00}_1}} = - 1 \quad ; \quad A^{E^*}_{{i_{00}}_1} = -\frac{3(a\omega - i)}{a\omega(\epsilon_r + 2) -i(\sigma_r + 2)} \quad ; \quad B^{E^*}_{{e_{00}}_1} = \frac{(a\omega(\epsilon_r-1) - i (\sigma_r - 1))}{a\omega(\epsilon_r + 2) - i(\sigma_r + 2)}. 
\end{equation}
Thus, complex amplitudes of electric potential for drop and ambient phases are given by,
\begin{equation}
    {V^*_{i_{00}}} = -\frac{3 (a\omega - i)}{a\omega(\epsilon_r + 2) -i(\sigma_r + 2)}  r \cos\theta \quad ; \quad {V^*_{e_{00}}} = - \left(r - \frac{1}{r^{2}}\frac{(a\omega(\epsilon_r-1) - i (\sigma_r - 1))}{a\omega(\epsilon_r + 2) - i(\sigma_r + 2)} \right) \cos\theta
\end{equation}
and by taking gradient of electric potential, the complex amplitudes of components of electric field are,
\begin{subequations}
\begin{equation}
    E^*_{i_{{00}_r}} = \frac{3  (a\omega - i)}{a\omega(\epsilon_r + 2) -i(\sigma_r + 2)} \cos\theta 
\end{equation}
\begin{equation}
    E^*_{i_{{00}_\theta}} = -\frac{3 (a\omega - i)}{a\omega(\epsilon_r + 2) -i(\sigma_r + 2)}  r \sin\theta
\end{equation}
\begin{equation}
    E^*_{e_{{00}_r}} = \left(1 + \frac{2}{r^{3}}\frac{(a\omega(\epsilon_r-1) - i (\sigma_r - 1))}{a\omega(\epsilon_r + 2) - i(\sigma_r + 2)} \right) \cos\theta
\end{equation}
\begin{equation}
    E^*_{e_{{00}_\theta}} = - \left(r - \frac{1}{r^{2}}\frac{(a\omega(\epsilon_r-1) - i (\sigma_r - 1))}{a\omega(\epsilon_r + 2) - i(\sigma_r + 2)} \right) \sin\theta.
\end{equation}
\end{subequations}
Real parts of the time-dependent coefficients are:
\begin{subequations}
\begin{equation}
	\label{Const_AE_e00_1}
	A^E_{{e_{00}}_1} = \real({A^{E^*}_{{e_{00}}_1}e^{it_R t}}) = - \cos(t_Rt)
\end{equation}
\begin{equation}
    \label{Const_AE_i00_1}
	A^E_{{i_{00}}_1} = \real(A^{E^*}_{{i_{00}}_1}e^{it_R t}) = -\frac{3a\omega (\epsilon_r - \sigma_r)\sin(t_Rt) - 3 \left(a^2\omega^2(\epsilon_r + 2) + \sigma_r + 2\right)\cos(t_Rt)}{a^2\omega ^2 (\epsilon_r + 2)^2 + (\sigma_r + 2)^2} 
\end{equation}
\begin{equation}
	\label{Const_BE_e00_1}
	B^E_{{e_{00}}_1} = \real(B^{E^*}_{{e_{00}}_1}e^{it_R t}) = \frac{\left(a^2\omega^2 \left(\epsilon_r^2 + \epsilon_r - 2\right)\cos(t_Rt) +\sigma_r^2 + \sigma_r - 2\right) - 3 a \omega (\epsilon_r - \sigma_r) \sin(t_Rt)}{a^2\omega^2(\epsilon_r + 2)^2  + (\sigma_r + 2)^2}.
\end{equation}
\end{subequations}
Electric stress is given by Maxwell's stress tensor as,
\begin{equation}
	\bm{\tau}_{i_{00}}^E = \epsilon_r \left( \bm{E}_{i_{00}}\bm{E}_{i_{00}} - \frac{1}{2}\bm{E}_{i_{00}}.\bm{E}_{i_{00}} \bm{I}\right) \quad \text{and} \quad \bm{\tau}_{e_{00}}^E = \left( \bm{E}_{e_{00}}\bm{E}_{e_{00}} - \frac{1}{2}\bm{E}_{e_{00}}.\bm{E}_{e_{00}} \bm{I}\right)
\end{equation}
The normal and tangential components of electric stress are,
\begin{multline}
	\tau^E_{{00}_{rr}} |_{r=1} = \bigg(\tau_{e_{{00}_{rr}}}^E - \tau_{i_{{00}_{rr}}}^E\bigg)|_{r=1} 
	= \frac{9\cos^2(\theta)}{2 \left(a^2\omega^2 (\epsilon_r + 2)^2 + (\sigma_r+2)^2\right)^2} \\
	\bigg[\bigg(\left(a^2\epsilon_r (\epsilon_r + 2) \omega ^2 + \sigma_r (\sigma_r + 2)\right)\cos(t_Rt) 
	+ 2a\omega(\sigma_r - \epsilon_r)\sin(t_Rt)\bigg)^2  \\+
	(1 - 2 \epsilon_r) \bigg(\left(a^2\omega^2(\epsilon_r + 2) + \sigma_r + 2\right)\cos(t_Rt) 
	+ a\omega(\epsilon_r - \sigma_r) \sin(t_Rt)\bigg)^2\bigg] \\
	+ \underbrace{\frac{9(\epsilon_r - 1) \left(\left(a^2\omega^2(\epsilon_r + 2) 
	+ \sigma_r + 2\right)\cos(t_Rt) + a\omega(\epsilon_r - \sigma_r)\sin(t_Rt)\right)^2}
	{2 \left(a^2\omega^2 (\epsilon_r + 2)^2 +(\sigma_r + 2)^2\right)^2}}_{\text{Independent of $\theta$}}
\end{multline}
\begin{multline}
	\label{Eqn_ElectricShearStress_O1}
	\tau^E_{{00}_{r\theta}} |_{r=1} = \frac{9(\epsilon_r - \sigma_r)\cos(\theta)\sin(\theta)}{2 \left(a^2\omega^2(\epsilon_r + 2)^2 + (\sigma_r + 2)^2\right)}
	\Big[-\left(a^2\omega^2(\epsilon_r + 2)(\epsilon_r - 2 (\sigma_r + 1)) - (\sigma_r + 2)^2\right) \\ \cos(2t_Rt)
	+ a\omega \left((\epsilon_r + 2) \left(a^2\omega^2(\epsilon_r + 2) + \sigma_r + 2\right) + (\sigma_r + 2)(\epsilon_r - \sigma_r)\right)\sin(2t_Rt)\Big].
\end{multline}
Flow equations at $\order{1}$ are reduced to the governing equations for the Newtonian flow. Continuity equations for drop and ambient phases are given by,
\begin{equation}
	\bm{\nabla}.\bm{u}_{i_{00}} = 0 \quad \text{and} \quad \bm{\nabla}.\bm{u}_{e_{00}} = 0 
\end{equation}
and momentum equations for drop and ambient phases at $\order{1}$ are,
\begin{equation}
	-\bm{\nabla}p_{i_{00}} + \bm{\nabla}.\bm{\tau}_{i_{00}} = 0 \quad \text{and} \quad -\bm{\nabla}p_{e_{00}} + \bm{\nabla}.\bm{\tau}_{e_{00}} = 0
\end{equation}
where $\bm{\tau}_{i_{00}}$ and $\bm{\tau}_{e_{00}}$ are given by,
\begin{equation}
	\bm{\tau}_{i_{00}} = \mu_r(\bm{\nabla}\bm{u}_{e_{00}} + (\bm{\nabla}\bm{u}_{e_{00}})^T) \quad \text{and} \quad \bm{\tau}_{e_{00}} = \bm{\nabla}\bm{u}_{e_{00}} + (\bm{\nabla}\bm{u}_{e_{00}})^T
\end{equation}
Governing equations at $\order{1}$ are subjected to the boundary conditions,
\begin{subequations}
\begin{equation}
    \bm{u}_{e_{00}} = 0 \quad \text{as} \quad r \rightarrow \infty \quad ; \quad \frac{\partial \bm{u}_{i_{00}}}{\partial r} = 0 \quad \text{at} \quad r = 0
\end{equation}
\begin{equation}
    u_{i_{{00}_r}} = u_{e_{{00}_r}} = \pdv{r}{t} \quad \text{at} \quad r = R \quad ; \quad u_{i_{{00}_\theta}} = u_{e_{{00}_\theta}} \quad \text{at} \quad r = 1
\end{equation}
\begin{equation}
    \tau_{i_{{00}_{r\theta}}} + \tau^E_{i_{00_{r\theta}}} = \tau_{e_{{00}_{r\theta}} } + {\tau}^E_{e_{{00}_{r\theta}}}  \quad \text{at} \quad r = 1
\end{equation}
\begin{equation}
    (-p_{e_{00}} + \tau_{e_{{00}_{rr}}} + \tau_{e_{{00}_{rr}}}^E) - (-p_{i_{00}} + \tau_{i_{{00}_{rr}}} + \tau_{i_{{00}_{rr}}}^E)  = -\bigg(2 f_{10} + \cot\theta f_{10}^{\prime} + f_{10}^{\prime\prime}\bigg).
\end{equation}
\end{subequations}
Since the flow field is divergence-free and axisymmetric, we can express the velocity field in terms of the Stokes streamfunction.
If $\Psi$ is Stokes streamfunction, components of velocity are given by,
\begin{equation}
	u_r = \frac{1}{r^2\sin\theta} \pdv{\Psi}{\theta} \quad \text{and} \quad u_\theta = -\frac{1}{r\sin\theta}\pdv{\Psi}{r}
\end{equation}
Thus, governing equations for flow inside and outside the drop in terms of streamfunction are given by,
\begin{equation}
	\bigg[\frac{\partial^2 }{\partial r^2} + \frac{\sin\theta}{r^2}\frac{\partial }{\partial \theta}\bigg(\frac{1}{\sin\theta}\frac{\partial }{\partial \theta}\bigg)\bigg]^2\Psi_{i_{00}} = 0 \quad \text{and} \quad \bigg[\frac{\partial^2 }{\partial r^2} + \frac{\sin\theta}{r^2}\frac{\partial }{\partial \theta}\bigg(\frac{1}{\sin\theta}\frac{\partial }{\partial \theta}\bigg)\bigg]^2\Psi_{e_{00}} = 0.
\end{equation}
We choose the streamfunction form as given by,
\begin{equation}
	\Psi_{i_{00}} = f_i(r)\sin^2\theta\cos\theta \quad \text{and} \quad \Psi_{e_{00}} = f_e(r)\sin^2\theta\cos\theta
\end{equation}
Therefore, general solution of streamfunction equation is given by,
\begin{subequations}
\begin{equation}
	\Psi_{i_{00}} = \bigg(\frac{A_{i_{00}}}{r^2} + B_{i_{00}} + C_{i_{00}} r^3 + D_{i_{00}} r^5\bigg)\sin^2\theta\cos\theta 
\end{equation}
\begin{equation}
    \Psi_{e_{00}} = \bigg(\frac{A_{e_{00}}}{r^2} + B_{e_{00}} + C_{e_{00}} r^3 + D_{e_{00}} r^5\bigg)\sin^2\theta\cos\theta.
\end{equation}
\end{subequations}
Components of velocity field are obtained from the streamfunction as given by,
\begin{subequations}
\begin{equation}
	\label{Eqn_uri_O1}
	u_{i_{{00}_r}} = \frac{1}{2} (3 \cos2\theta+1) \left(\frac{A_{i_{00}}}{r^4} + \frac{B_{i_{00}}}{r^2} + C_{i_{00}} r + D_{e_{00}} r^3\right) 
\end{equation}
\begin{equation}
    u_{i_{{00}_\theta}} = \sin\theta\cos\theta \left(\frac{2 A_{i_{00}}}{r^4} - 3C_{i_{00}} r - 5 D_{e_{00}}r^3\right)
\end{equation}
\begin{equation}
	\label{Eqn_ure_O1}
	u_{e_{{00}_r}} = \frac{1}{2} (3 \cos2\theta + 1) \left(\frac{A_{e_{00}}}{r^4}+\frac{B_{e_{00}}}{r^2} + C_{e_{00}} r + D_{e_{00}} r^3\right) 
\end{equation}
\begin{equation}
    u_{e_{{00}_\theta}} = \sin\theta \cos\theta \left(\frac{2 A_{e_{00}}}{r^4} - 3 C_{e_{00}} r - 5 D_{e_{00}} r^3\right).
\end{equation}
\end{subequations}
We substitute the velocity field in the momentum equation and integrate to obtain pressure as given by,
\begin{equation}
    \label{Eqn_p_i00}
	p_{i_{00}} = \frac{7}{2} D_{i_{00}} (1 + 3\cos(2\theta)) \quad \text{and} \quad p_{e_{00}} = \frac{B_{e_{00}}(1 + 3\cos(2\theta))}{r^3}.
\end{equation}
We obtain the constants by implementation of the boundary conditions. To simplify the calculations, 
we make use of the linearity of the problem and split the velocity field as $\bm{u}_{00} = \bm{u}_{00}^{I} + \bm{u}_{00}^{II}$,
where $\bm{u}_{00}^{I}$ considers the tangential stress generated at the interface due to electric field 
and $\bm{u}_{00}^{II}$ considers the effect of rate of change of drop radius on the normal velocity at the interface.
As we will see later, $\bm{u}_{00}^{I}$ contributes to the deformation of drop at $\order{Ca_E}$, i.e. $f_{10}$ and $\bm{u}_{00}^{II}$
contributes to the deformation at $\order{Ca_E^2}$ i.e. $f_{20}$.
$\bm{u}_{00}^{I}$ is subjected to the boundary conditions given by,
\begin{subequations}
\begin{equation}
	\bm{u}^{I}_{e_{00}} = 0 \quad \text{as} \quad r \rightarrow \infty 
    \quad ; \quad \frac{\partial \bm{u}^{I}_{i_{00}}}{\partial r} = 0 \quad \text{at} \quad r = 0
    \end{equation}
\begin{equation}
	u^{I}_{i_{{00}_r}} = u^{I}_{e_{{00}_r}} = 0 \quad \text{at} \quad r = 1
    \quad ; \quad u^{I}_{i_{{00}_\theta}} = u^{I}_{e_{{00}_\theta}} \quad \text{at} \quad r = 1
\end{equation}
\begin{equation}
    \tau^{I}_{i_{{00}_{r\theta}}} + \tau^E_{i_{{00}_{r\theta}}} = {\tau}^{I}_{e_{{00}_{r\theta}}} + {\tau}^E_{e_{{00}_{r\theta}}}  \quad \text{at} \quad r = 1.
\end{equation}
\end{subequations}
Thus, implementation of boundary conditions gives the constants,
% \begin{subequations}
\begin{equation}
    \label{Const_A_I_e00}
    A^{I}_{e_{00}} = U_S + U_{T_S}\sin(2t_Rt) + U_{T_C}\cos(2t_Rt) \quad ; \quad B^{I}_{e_{00}} = - (U_S + U_{T_S}\sin(2t_Rt) + U_{T_C}\cos(2t_Rt)) 
\end{equation}
\begin{equation}
    \label{Const_C_I_i00}
	C^{I}_{i_{00}} = U_S + U_{T_S}\sin(2t_Rt) + U_{T_C}\cos(2t_Rt) \quad ; \quad D^{I}_{i_{00}} = - (U_S + U_{T_S}\sin(2t_Rt) + U_{T_C}\cos(2t_Rt))
\end{equation}
where
\begin{subequations}
\begin{equation}
    \label{Eqn_U_S}
	U_S = \frac{9(\epsilon_r - \sigma_r)}{20(\mu_r + 1) \left(a^2\omega^2 (\epsilon_r + 2)^2 +(\sigma_r + 2)^2\right)}
\end{equation}
\begin{equation}
    \label{Eqn_U_TC}
	U_{T_C} = \frac{\left((\sigma_r + 2)^2 - a^2\omega^2 (\epsilon_r + 2)(\epsilon_r - 2 (\sigma_r + 1))\right)U_S}{a^2\omega^2(\epsilon_r+2)^2 + (\sigma_r + 2)^2}
\end{equation}
\begin{equation}
    \label{Eqn_U_TS}
	U_{T_S} = \frac{a\omega\left(\epsilon_r\left(a^2\omega^2(\epsilon_r + 4) + 4\right) + 4 a^2 \omega^2 + 2\epsilon_r\sigma_r - \sigma_r^2 + 4\right)U_S}{a^2\omega^2(\epsilon_r + 2)^2 + (\sigma_r + 2)^2}.
\end{equation}
\end{subequations}
Thus , the components of velocity field are given by,
\begin{equation}
    \label{Eqn_uI_e00r}
	u^{I}_{e_{{00}_r}} = -U^{I}\left(\frac{3 \cos2\theta + 1}{2}\right) \left(\frac{1}{r^2} - \frac{1}{r^4}\right) \quad ; \quad u^{I}_{e_{{00}_\theta}} = U^{I}\sin2\theta  \left(\frac{1}{r^4}\right) 
\end{equation}
\begin{equation}
\label{Eqn_uI_i00r}
	u^{I}_{i_{{00}_r}} = -U^{I}\left(\frac{3 \cos2\theta + 1}{2}\right) \left(r^3 - r\right) \quad ; \quad u^{I}_{i_{{00}_\theta}} = U^{I}\sin2\theta  \left(\frac{5 r^3}{2} - \frac{3 r}{2}\right)
\end{equation}
where
\begin{equation}
\label{Eqn_U_I}
	U^{I} = U_S + U_{T_S}\sin(2t_Rt) + U_{T_C}\cos(2t_Rt).
\end{equation}
We calculate the pressure using the expressions given by \autoref{Eqn_p_i00}.
Now, we have got the velocity and pressure field in drop and ambient phases. Now, we implement the normal stress condition, which states that, 
\begin{equation}
	\tau_{{00}_{rr}}|_{r = 1} = \tau^{I}_{{00}_{rr}}|_{r = 1} + \tau^E_{{00}_{rr}}|_{r = 1} = -2f_{10}(\theta, t) - \cot\theta \pdv{f_{10}(\theta, t)}{\theta} - \frac{\partial^2f_{10}(\theta, t)}{\partial\theta^2}.
\end{equation}
where $\tau^{I}_{{00}_{rr}}|_{r = 1}$ denotes the jump in hydrodynamic normal stress at the interface, given by,
\begin{equation}
    \tau^{I}_{{00}_{rr}}|_{r = 1} = \left(-p^{I}_{e_{00}} + \tau^{I}_{e_{{00}_{rr}}}\right) - \left(-p^{I}_{i_{00}} + \tau^{I}_{i_{{00}_{rr}}}\right) = -\frac{(2 + 3\mu_r)(1 + 3\cos(2\theta))U^{I}}{2} 
\end{equation}
Net jump in normal stress at the interface due to electric field is given by,
\begin{equation}
	\tau^E_{{00}_{rr}}|_{r = 1} = A^E_{{00}_{{nn}_{P_2}}}P_2(\cos\theta)
\end{equation}
where
\begin{multline}
\label{Eqn_ElectricNormalStressCoeff_O1}
	A^E_{{00}_{{nn}_{P_2}}} = \frac{3\{\epsilon_r\left(a^2\omega^2 (\epsilon_r - 2) - 2\right) + a^2\omega^2 + \sigma_r^2 + 1\}}{2\left(a^2\omega^2 (\epsilon_r + 2)^2 + (\sigma_r + 2)^2\right)} +
   \frac{3}{2\left(a^2\omega^2 (\epsilon_r + 2)^2 + (\sigma_r + 2)^2\right)^2} \\ \bigg[\cos(2t_Rt) \bigg\{(\sigma_r + 2)^2 \left(\sigma_r^2 + 1\right) - 2(\sigma_r + 2)^2\epsilon_r 
   + a^4 \omega^4 (\epsilon_r^4 + 2 \epsilon_r^3 - 3\epsilon_r^2 - 4 \epsilon_r + 4) \\+ a^2 \omega^2 \bigg(2\epsilon_r^3 + \left(2 (\sigma_r - 2) \sigma_r - 13\right)\epsilon_r^2 
   + 6(\sigma_r(\sigma_r + 2) - 2)\epsilon_r + ((4 - 5\sigma_r) \sigma_r + 8)\bigg)\bigg\} \\- \sin(2t_Rt)\bigg\{2a\omega(\epsilon_r - \sigma_r) \bigg(\epsilon_r \left(a^2\omega^2 (4\epsilon_r + 7) + 4\right) - 2 a^2\omega^2 + 2 \epsilon_r \sigma_r + 2\sigma_r^2 + 3 \sigma_r - 2\bigg)\bigg\}\bigg].
\end{multline}
Thus, by implementation of normal stress boundary condition, we get,
\begin{equation}
    \label{Eqn_AS_10}
    f_{10} = A^S_{10} P_2(\cos\theta) \quad \text{where} A^S_{10} = \frac{1}{4}\left(A^E_{{00}_{{nn}_{P_2}}} - 2(2 + 3\mu_r)U^{I}\right) 
\end{equation}
where $A^E_{{00}_{{nn}_{P_2}}}$ is given by \autoref{Eqn_ElectricNormalStressCoeff_O1}.
Thus, we obtain $f_{10}$ and hence deformation of drop at $\order{Ca_E}$.

Now, we will discuss the solution of flow field $\bm{u}_{00}^{II}$, subjected to the boundary conditions given by,
\begin{subequations}
\begin{equation}
	\bm{u}^{II}_{e_{00}} = 0 \quad \text{as} \quad r \rightarrow \infty \quad ; \quad \frac{\partial \bm{u}^{II}_{i_{00}}}{\partial r} = 0 \quad \text{at} \quad r = 0
\end{equation}
\begin{equation}
	u^{II}_{i_{{00}_r}} = u^{II}_{e_{{00}_r}} = \pdv{r}{t} \quad \text{at} \quad r = 1 \quad ; \quad u^{II}_{i_{{00}_\theta}} = u^{II}_{e_{{00}_\theta}} \quad \text{at} \quad r = 1 \quad ; \quad \tau^{II}_{i_{{00}_{r\theta}}} = {\tau}^{II}_{e_{{00}_{r\theta}}} \quad \text{at} \quad r = 1
\end{equation}
\end{subequations}
We have, $u_{e_{0_r}} \rightarrow 0$ and $u_{e_{0_\theta}} \rightarrow 0$ as $r \rightarrow \infty$. 
Thus, we get, $D^{II}_{e_{00}} = 0$ and $C^{II}_{e_{00}} = 0$.
For velocity to be bounded at $r = 0$, we have, $A^{II}_{i_{00}} = B^{II}_{i_{00}} = 0$. 
Then components of velocity field are given by,
\begin{equation}
	u^{II}_{i_{{00}_r}} = \frac{1}{2} (3 \cos2\theta+1) \left(C^{II}_{i_{00}} r + D^{II}_{i_{00}} r^3\right) \quad ; \quad u^{II}_{i_{{00}_\theta}} = \sin\theta\cos\theta \left(- 3C^{II}_{i_{00}} r - 5D^{II}_{i_{00}} r^3\right)  
\end{equation}
\begin{equation}
	u^{II}_{e_{{00}_r}} = \frac{1}{2} (3 \cos2\theta + 1) \left(\frac{A^{II}_{e_{00}}}{r^4}+\frac{B^{II}_{e_{00}}}{r^2} \right) \quad ; \quad  u^{II}_{e_{{00}_\theta}} = \sin\theta \cos\theta \left(\frac{2A^{II}_{e_{00}}}{r^4}\right).
\end{equation}
Interface of the drop is given by the function,
	$r = 1 + Ca_Ef_{10}$
where from \autoref{Eqn_AS_10}, $f_{10}$ can be written as,
\begin{equation}
	f_{10} = \text{Constant} + \real{(H^*e^{2it_Rt})} = H_r\cos(2t_Rt) - H_i\sin(2t_Rt).
\end{equation}
where
\begin{multline}
    \label{Eqn_Hr}
	H_r = \frac{P_2(\cos\theta)}{4}\bigg[\bigg\{a^4\omega^4(\epsilon_r^4 - 3\epsilon_r^2 + 4) + 2\epsilon_r^3\left(a^4\omega^4 + a^2\omega^2\right)
	+ a^2\omega^2\Big(\epsilon_r^2\left(2\sigma_r(\sigma_r - 2) + 13\right) \\+ \sigma_r(4 - 5\sigma_r) + 8\Big) 
	+ (\sigma_r + 2)^2\left(\sigma_r^2 + 1\right)
	- 2 \epsilon_r\left(2 a^4\omega^4 - 3 a^2 \omega^2 (2\sigma_r^2 + 6\sigma_r + 2)\right) \bigg\} 
	\\- 2(2 + 3\mu_r)U_{T_C}\bigg]
\end{multline}
\begin{multline}
    \label{Eqn_Hi}
	H_i = \frac{P_2(\cos\theta)}{4}\bigg[\bigg\{2a\omega (\epsilon_r - \sigma_r)
	\bigg(a^2\omega^2(\epsilon_r(4\epsilon_r + 7) - 2) + 4\epsilon_r 
	+ 2 \sigma_r(\epsilon_r + 2\sigma_r + 3)  - 2\bigg)\bigg\} \\+ 2(2 + 3\mu_r)U_{T_S}\bigg]
\end{multline}
Thus, we get,
\begin{equation}
	\pdv{r}{t} = - 2H_r\sin(2t_Rt) - 2H_i\cos(2t_Rt)
\end{equation}
Constants obtained by implementation of boundary conditions are,
\begin{equation}
	  A^{II}_{e_{00}} = -3 (2 \mu_r + 3) U^{II} \quad ; \quad B^{II}_{e_{00}} = (16 \mu_r + 19) U^{II} 
\end{equation}
\begin{equation} 
    C^{II}_{i_{00}} = (19 \mu_r + 16) U^{II} \quad ; \quad D^{II}_{i_{00}} = -3 (3 \mu_r + 2) U^{II}
\end{equation}
where
\begin{equation}
    \label{Eqn_U_II}
	U^{II} = -\frac{Ca_E (H_i\cos(2t_Rt) + H_r\sin(2t_Rt))}{10(\mu_r + 1)}.
\end{equation}
Pressure field due to $\bm{u}^{II}$ field is calculated by \autoref{Eqn_p_i00}.
Thus, from the velocity and pressure field, we calculate the jump in hydrodynamic normal stress due to $\bm{u}^{II}$ field as given by,
\begin{multline}
	\label{Eqn_NormalHydrodynamicStress_II}
	\tau^{II}_{{10}_{nn}}|_{r = 1} \\= -\frac{U^{II} (3\cos(2\theta) + 1)}{2 r^5}\bigg((9\mu_r + 6) r^7 + 2(19\mu_r + 16) r^5 + 6\mu_r (16\mu_r + 19) r^2  - 24\mu_r (2\mu_r + 3) \bigg).
\end{multline}
As we can see from the above expression, jump in normal stress due to $\bm{u}^{II}$ is proportional to $U^{II}$,
which is proportional to $Ca_E$. Thus, jump in normal stress is of $\order{Ca_E}$, which will contribute to 
the shape of drop interface at order $\order{Ca_E^2}$, i.e. $f_{20}$. 
Calculation of $f_{20}$ will be discussed in the next section.

\subsection{Solution of \texorpdfstring{$\order{Ca_E}$}{Lg} equations}
\label{Sec_Sol_Ca}
From the asymptotic expansion of governing equations, electric field at $\order{Ca_E}$ is governed by the equations,
\begin{subequations}
\begin{equation}
	\bm{\nabla}^2 V_{i_{10}} = 0 \quad \text{where} \quad \bm{E}_{i_{10}} = -\bm{\nabla}V_{i_{10}} \quad \text{and} \quad \bm{\nabla}^2 V_{e_{10}} = 0 \quad \text{where} \quad \bm{E}_{e_{10}} = -\bm{\nabla}V_{e_{10}}
\end{equation}
\end{subequations}
subjected to the boundary conditions as given by,
\begin{subequations}
\begin{equation}
	\frac{\partial V_{i_{10}}}{\partial r} = 0 \quad \text{at} \quad r = 0 \quad ; \quad \bm{E}_{e_{10}} \rightarrow 0 \quad \text{at} \quad r \rightarrow \infty
\end{equation}
\begin{equation}
    E_{i_{{10}_\theta}} + f_{10}\frac{\partial E_{ i_{{00}_\theta}}}{\partial r} + E_{i_{{00}_r}}{f}_{10}^{\prime} 
	= E_{e_{{10}_\theta}} + f_{10}\frac{\partial E_{e_ {{00}_\theta}}}{\partial r} + E_{e_{{00}_r}}{f}_{10}^{\prime} 
	\quad \text{at} \quad r = 1
\end{equation}
\begin{multline}
	(\sigma_r + i\omega a \epsilon_r)\left(E_{i_{{10}_r}} + f_{10}\frac{\partial E_{i_{{00}_r}}}{\partial r} - E_{i_{{00}_\theta}}{f}_{10}^{\prime}\right)
	= (1 + i\omega a)\left(E_{e_{{10}_r}} + f_{10}\frac{\partial E_{e_{{00}_r}}}{\partial r} - E_{e_{{00}_\theta}}{f}_{10}^{\prime}\right)
	\quad \\\text{at} \quad r = 1.
\end{multline}
\end{subequations}
From the general solution of Laplace equation in spherical axisymmetric coordinate system, we get,
\begin{equation}
	V_{i_{10}} = \sum_n \left( A^E_{i_{{10}_n}} r^n + \frac{B^E_{i_{{10}_n}}}{r^{n+1}} \right) P_n(\cos\theta) \; \text{and} \quad V_{e_{10}} = \sum_n \left( A^E_{e_{{10}_n}} r^n + \frac{B^E_{e_{{10}_n}}}{r^{n+1}} \right) P_n(\cos\theta)
\end{equation}
Boundedness of electric potential at the drop center implies that $B^E_{i_{{10}_n}} = 0$ 
and vanishing of electric field as $r \rightarrow \infty$ leads to $A^E_{e_{{10}_n}} = 0$. 
From the interface boundary conditions, terms containing $\order{1}$ electric field, contain only second and fourth modes of Legendre polynomials. Thus, all the constants in the general solution are zero except  $A^E_{i_{{10}_1}}, A^E_{i_{{10}_3}}, B^E_{e_{{10}_1}}, B^E_{e_{{10}_3}}$, which are given by,
\begin{subequations}
\begin{equation}
    \label{Const_AE_i10_1}
     A^E_{i_{{10}_1}} = \frac{-2 A^S_{10}(B^E_{e_{{00}_1}} + A^E_{e{{00}_1}} + A^E_{i_{{00}_1}}(2 - 3\sigma_r))}{5(2 + \sigma_r)}
\end{equation}
\begin{equation}
    \label{Const_AE_i10_3}
    A^E_{i_{{10}_3}} = \frac{-6 A^S_{10} (-3A^E_{e_{{00}_1}} + A^E_{i_{{00}_1}}(2 + \sigma_r))}{5(4 + 3\sigma_r)}
\end{equation}
\begin{equation}
    \label{Const_BE_e10_1}
         B^E_{e_{{10}_1}} = \frac{2 A^S_{10}(B^E_{e_{{00}_1}}(3 + 2\sigma_r) - (-4A^E_{i_{{00}_1}}\sigma_r + A^E_{e_{{00}_1}}(3 + \sigma_r)))}{5(2 + \sigma_r)}
\end{equation}
\begin{equation}
    \label{Const_BE_e10_3}
    B^E_{e_{{10}_3}} = \frac{3A^S_{10}(B^E_{e_{{00}_1}}(8 + 6\sigma_r) + (A^E_{e_{{00}_1}}(2 - 3\sigma_r) + A^E_{i_{{00}_1}}\sigma_r))}{5(4 + 3\sigma_r)}
\end{equation}
\end{subequations}
where $A^E_{e_{{00}_1}}$,$A^E_{i_{{00}_1}}$ and $B^E_{e_{{00}_1}}$ are given by \Cref{Const_AE_e00_1,Const_AE_i00_1,Const_BE_e00_1} respectively.
The components of the electric field are calculated from the electric potential.
Stress tensors due to electric field at $\order{Ca_E}$ for drop and ambient phases are given by,
\begin{equation}
	\bm{\tau}_{i_{10}}^E = \epsilon_r \left( \bm{E}_{i_{00}}\bm{E}_{i_{10}} + \bm{E}_{i_{10}}\bm{E}_{i_{00}} - \bm{E}_{i_{00}}.\bm{E}_{i_{10}} \bm{I}\right) \quad \text{and} \quad \bm{\tau}_{e_{10}}^E = \bm{E}_{e_{00}}\bm{E}_{e_{10}} + \bm{E}_{e_{10}}\bm{E}_{e_{00}} - \bm{E}_{e_{00}}.\bm{E}_{e_{10}} \bm{I}
\end{equation}
Normal and tangential components of stress due to electric field at the interface are given by
\begin{equation}
\tau_{{{10}_{nn}}}^E = \tau_{{10}_{rr}}^E + f_{10}\frac{\partial \tau_{{00}_{rr}}^E}{\partial r} - 2{f}_{10}^{\prime}\tau_{{{00}_{r\theta}}}^E \quad \text{and} \quad \tau_{{{10}_{nt}}}^E = \tau_{{10}_{r\theta}}^E + f_{10}\frac{\partial \tau_{{00}_{r\theta}}^E }{\partial r} + {f}_{10}^{\prime}(\tau_{{00}_{rr}}^E - \tau_{{00}_{\theta\theta}}^E)
\end{equation}
Thus, from the above expressions, we get normal and tangential stress at the interface. Normal stress at the interface due to electric field is,
\begin{equation}
	\label{Eqn_NormalElectricStress_CaE}
	\tau^E_{{{10}_{nn}}} = A^E_{{{00}_{nn}}} + B^E_{{{10}_{nn}}}\cos2\theta + C^E_{{{10}_{nn}}} \cos4\theta
\end{equation}
where
\begin{subequations}
\begin{multline}
    \label{Const_AE_10_nn}
	A^E_{{{10}_{nn}}} = \frac{-3\epsilon_r A^E_{i_{{10}_3}} A^E_{i_{{00}_1}}}{8} - \frac{3A^S_{10} \left({(A^E_{e_{{00}_1}})}^2 - \epsilon_r{(A^E_{i_{{00}_1}})}^2 \right)}{4}  + \frac{3 A^E_{e_{{00}_1}} (13 A^S_{10} B^E_{e_{{00}_1}} - 8B^E_{e_{{10}_1}})}{16} \\- \frac{15 A^E_{e_{{00}_1}} B^E_{e_{{10}_3}}}{16} + \frac{3 B^E_{e_{{00}_1}} (8 B^E_{e_{{10}_1}} - 13 A^S_{10} B^E_{e_{{00}_1}})}{16}+\frac{21 B^E_{e_{{10}_3}} B^E_{e_{{00}_1}}}{16}
\end{multline}
\begin{multline}
    \label{Const_BE_10_nn}
	B^E_{{{10}_{nn}}} = -A^E_{i_{{10}_1}} A^E_{i_{{00}_1}} \epsilon_r - \frac{3A^E_{i_{{10}_3}} A^E_{i_{{00}_1}} \epsilon_r}{4} + \frac{A^E_{e_{{00}_1}} (15 A^S_{10} B^E_{e_{{00}_1}} - 2 B^E_{e_{{10}_1}})}{4} \\- \frac{11 A^E_{e_{{00}_1}} B^E_{e_{{10}_3}}}{4} + \frac{B^E_{e_{{00}_1}} (10 B^E_{e_{{10}_1}} - 21 A^S_{10}B^E_{e_{{00}_1}})}{4} + \frac{13B^E_{e_{{10}_3}}B^E_{e_{{00}_1}}}{4}
\end{multline}
\begin{multline}
    \label{Const_CE_10_nn}
	C^E_{{{10}_{nn}}} = -\frac{15}{8}A^E_{i_{{10}_3}} A^E_{i_{{00}_1}}\epsilon_r + \frac{3A^S_{10}\left({(A^E_{e_{{00}_1}})}^2 - \epsilon_r{(A^E_{i_{{00}_1}})}^2\right)}{4} - \frac{3 A^E_{e_{{00}_1}}B^E_{e_{{00}_1}} A^S_{10}}{16} \\- \frac{5 A^E_{e_{{00}_1}} B^E_{e_{{10}_3}}}{16} - \frac{69 A^S_{10} {(B^E_{e_{{00}_1}})}^2}{16} + \frac{55 B^E_{e_{{10}_3}} B^E_{e_{{00}_1}}}{16}
\end{multline}
\end{subequations}
and tangential stress jump at the interface due to electric field is,
\begin{equation}
	\label{Eqn_TangElectricStress_CaE}
	\tau^E_{{{10}_{nt}}} = A^E_{{{10}_{nt}}}\sin2\theta + B^E_{{{10}_{nt}}} \sin4\theta
\end{equation}
where
\begin{subequations}
\begin{multline}
    \label{Const_AE_10_nt}
	A^E_{{{10}_{nt}}} = A^E_{i_{{10}_1}} {A^E_{i_{{00}_1}}} \epsilon_r + \frac{3A^E_{i_{{10}_3}} {A^E_{i_{{00}_1}}} \epsilon_r}{4}  + \frac{3 {(A^E_{e_{{00}_1}})}^2 A^S_{10}}{4} + \frac{{A^E_{e_{{00}_1}}}{B^E_{e_{{00}_1}}} A^S_{10}  + 2A^E_{e_{{00}_1}} B^E_{e_{{10}_1}}}{2} \\ - \frac{2A^E_{e_{{00}_1}}B^E_{e_{{10}_3}} + 15 A^S_{10} {(B^E_{e_{{00}_1}})}^2 - 8 B^E_{e_{{10}_1}} B^E_{e_{{00}_1}}}{2}  + \frac{7 B^E_{e_{{10}_3}}B^E_{e_{{00}_1}}}{4}
\end{multline}
\begin{multline}
    \label{Const_BE_10_nt}
	B^E_{{{10}_{nt}}} = \frac{15A^E_{i_{{10}_3}} {A^E_{i_{{00}_1}}} \epsilon_r}{8} + \frac{3{(A^E_{i_{{00}_1}})}^2 A^S_{10} \epsilon_r}{4} - \frac{3 {(A^E_{e_{{00}_1}})}^2 A^S_{10}}{8} \\- \frac{9A^E_{e_{{00}_1}}B^E_{e_{{00}_1}}A^S_{10} - 5 A^E_{e_{{00}_1}} B^E_{e_{{10}_3}}}{4} - \frac{33 A^S_{10} {(B^E_{e_{{00}_1}})}^2  - 25 B^E_{e_{{10}_3}} B^E_{e_{{00}_1}}}{8}.
\end{multline}
\end{subequations}
	where $A^E_{e_{{00}_1}}$, $A^E_{i_{{00}_1}}$ and $B^E_{e_{{00}_1}}$ are given by \Cref{Const_AE_e00_1,Const_AE_i00_1,Const_BE_e00_1} respectively 
	and $A^E_{i_{{10}_1}}$, $A^E_{i_{{10}_3}}$, $B^E_{e_{{10}_1}}$ and $B^E_{e_{{10}_3}}$ are given by 
	\Cref{Const_AE_i10_1,Const_AE_i10_3,Const_BE_e10_1,Const_BE_e10_3} respectively.
Now, we have obtained the stress at the interface caused by the electric field. We solve the flow equations to calculate the flow field generated because of electric stress. 
Continuity equation for drop and ambient phases at $\order{Ca_E}$ is given by,
\begin{equation}
	\bm{\nabla}.\bm{u}_{i_{10}} = 0 \quad \text{and} \quad \bm{\nabla}.\bm{u}_{e_{10}} = 0.
\end{equation}
Conservation of momentum in the limit of Stokes flow is given by,
\begin{equation}
	-\bm{\nabla}p_{i_{10}} + \bm{\nabla}.\bm{\tau}_{i_{10}} = 0 \quad \text{and} \quad -\bm{\nabla}p_{e_{10}} + \bm{\nabla}.\bm{\tau}_{e_{10}} = 0.
\end{equation}
where
\begin{equation}
	\quad \quad \quad \bm{\tau}_{i_{10}} = \mu_r(\bm{\nabla}\bm{u}_{i_{10}} + (\bm{\nabla}\bm{u}_{i_{10}})^T) \quad \text{and} \quad \bm{\tau}_{e_{10}} = \bm{\nabla}\bm{u}_{e_{10}} + (\bm{\nabla}\bm{u}_{e_{10}})^T.
\end{equation}
The governing equations are subjected to the boundary conditions as given by,
\begin{subequations}
\begin{equation}
	\bm{u}_{e_{10}} = 0 \quad \text{as} \quad r \rightarrow \infty \quad ; \quad \frac{\partial \bm{u}_{i_{10}}}{\partial r} = 0 \quad \text{at} \quad r = 0
\end{equation}
\begin{equation}
	u_{i_{{10}_r}} + f_{10}\frac{\partial u_{i_{{00}_r}}}{\partial r} - u_{i_{{00}_\theta}}{f}_{10}^{\prime} 
	= u_{e_{{10}_r}} + f_{10}\frac{\partial u_{e_{{00}_r}}}{\partial r} - u_{e_{{00}_\theta}}{f}_{10}^{\prime} = 0 
	\quad \text{at} \quad r = 1
\end{equation}
\begin{equation}
	u_{i_{{10}_\theta}} + f_{10}\frac{\partial u_{i_{{00}_\theta}}}{\partial r} + u_{i_{{00}_r}}{f}_{10}^{\prime} 
	= u_{e_{{10}_\theta}} + f_{10}\frac{\partial u_{e_{{00}_\theta}}}{\partial r} + u_{e_{{00}_r}}{f}_{10}^{\prime} 
	\quad \text{at} \quad r = 1
\end{equation}
\begin{multline}
	\bigg(\tau_{i_{{10}_{r\theta}}} + f_{10}\frac{\partial \tau_{i_{{00}_{r\theta}}}}{\partial r} 
	+ {f}_{10}^{\prime}(\tau_{i_{{00}_{rr}}} - \tau_{i_{{00}_{\theta\theta}}}) + \tau_{i_{nt}}^E\bigg)
	= \bigg(\tau_{e_{{10}_{r\theta}}} + f_{10}\frac{\partial \tau_{e_{{00}_{r\theta}}}}{\partial r} 
	+ {f}_{10}^{\prime}(\tau_{e_{{00}_{rr}}} - \tau_{e_{{00}_{\theta\theta}}}) + \tau_{e_{nt}}^E\bigg) 
	\quad \\\text{at} \quad r = 1
\end{multline}
\begin{multline}
	\bigg[- \bigg(p_{i_{10}} + f_{10}\frac{\partial p_{i_{00}}}{\partial r}\bigg) + \tau_{i_{{10}_{rr}}} 
	+ f_{10}\frac{\partial \tau_{i_{{00}_{rr}}}}{\partial r} - 2{f}_{10}^{\prime}\tau_{i_{{00}_{r\theta}}} 
	+ \tau_{i_{nn}}^E\bigg] \\
	- \bigg[- \bigg(p_{e_{10}} + f_{10}\frac{\partial p_{e_{00}}}{\partial r}\bigg) + \tau_{e_{{10}_{rr}}} 
	+ f_{10}\frac{\partial \tau_{e_{{00}_{rr}}}}{\partial r} - 2{f}_{10}^{\prime}\tau_{e_{{00}_{r\theta}}} 
	+ \tau_{e_{nn}}^E\bigg] \\
	= -\bigg(2{f_{10}^2} - 2f_{20} + 2\cot\theta f_{10}f_{10}^{\prime} - 2\cot\theta f_{20}^{\prime} 
	+ 2f_{10}f_{10}^{\prime\prime} - f_{20}^{\prime\prime}\bigg) \quad \text{at} \quad r = 1.
\end{multline}
\end{subequations}
As discussed in the previous section, due to axisymmetry and incompressibility, we can express velocity field in 
terms of streamfunction and we get the governing equations for streamfunctions as,
\begin{equation}
 	\bigg[\frac{\partial^2 }{\partial r^2} + \frac{\sin\theta}{r^2}\frac{\partial }{\partial \theta}\bigg(\frac{1}{\sin\theta}\frac{\partial }{\partial \theta}\bigg)\bigg]^2\Psi_{i_{10}} = 0 \quad \text{and} \quad \bigg[\frac{\partial^2 }{\partial r^2} + \frac{\sin\theta}{r^2}\frac{\partial }{\partial \theta}\bigg(\frac{1}{\sin\theta}\frac{\partial }{\partial \theta}\bigg)\bigg]^2\Psi_{e_{10}} = 0.
\end{equation}
We choose the form of streamfunction as,
    $\Psi_{10} = (f^{I}_{{10}}(r) + f^{II}_{{10}}(r) \cos2\theta)\sin^2\theta\cos\theta$.
Substitution of streamfunction form in the governing equation gives the ordinary differential equation as,
\begin{multline}
\frac{40 f^{II}_{10}(r)}{r^4} + \frac{24 f_{10}^{I^\prime}(r) + 8f_{10}^{{II}^{\prime}}(r)}{r^3} - \frac{4(3f_{10}^{I^{\prime\prime}}(r) + f_{{10}}^{{{II}^{\prime\prime}}}(r))}{r^2} + f_{10}^{{I}^{(4)}}(r)	 \\ + \cos2\theta \bigg(\frac{280 f_{10}^{II}(r)}{r^4} + \frac{80f_{10}^{{II}^\prime}(r)}{r^3} - \frac{40 f_{10}^{{II}^{\prime\prime}}(r)}{r^2} + f_{10}^{{II}^{(4)}}(r)\bigg) = 0.
\end{multline}
Thus, solving the ODE and satisfying the boundedness of velocity field at the center of the drop and for velocity field to go to zero far away from the drop,
streamfunctions for drop and ambient phases are written as,
\begin{subequations}
\begin{equation}
\Psi_{i_{10}} = \bigg[\bigg(\frac{D_{i_{10}}^{II} r^7}{7} + C^{I}_{i_{{10}}} r^3 + D^{I}_{i_{{10}}} r^5\bigg) + \cos2\theta \bigg(C_{i_{10}}^{II} r^5 + D_{i_{10}}^{II} r^7\bigg)\bigg]\sin^2\theta\cos\theta
\end{equation}
\begin{equation}
\Psi_{e_{10}} = \bigg[\bigg(\frac{A_{e_{10}}^{II}}{7r^4} + \frac{A^{I}_{e_{{10}}}}{r^2} + B^{I}_{e_{{10}}}\bigg) + \cos2\theta\bigg(\frac{A^{II}_{e_{10}}}{r^2} + \frac{B^{II}_{e_{10}}}{r^2} \bigg)\bigg]\sin^2\theta\cos\theta.
\end{equation}
\end{subequations}
Thus, the components of velocity field are given by,
\begin{subequations}
\begin{multline}
    \label{Eqn_u_i10r}
    u_{i_{{10}_r}} = \frac{C^{I}_{i_{{10}}} r}{2} + \frac{r^3 (C^{II}_{i_{{10}}} + 2 D^{I}_{i_{{10}}})}{4}  + \frac{9 D^{II}_{i_{{10}}} r^5}{28} \\+ \bigg(\frac{3 C^{I}_{i_{{10}}} r}{2} + \frac{r^3 (C^{II}_{i_{{10}}} + 3 D^{I}_{i_{{10}}})}{2} + \frac{5 D^{II}_{i_{{10}}} r^5}{7}\bigg)\cos (2\theta) + \left(\frac{5 C^{II}_{i_{{10}}} r^3}{4} + \frac{5 D^{II}_{i_{{10}}} r^5}{4}\right)\cos (4\theta)
\end{multline}
\begin{equation}
    \label{Eqn_u_i10theta}
    u_{i_{{10}_\theta}} = \sin (2\theta) \left(-\frac{3 C^{I}_{i_{{10}}} r}{2}-\frac{5 D^{I}_{i_{{10}}} r^3}{2}-\frac{D^{II}_{i_{{10}}} r^5}{2}\right) + \sin (4\theta) \left(-\frac{5 C^{II}_{i_{{10}}} r^3}{4} - \frac{7 D^{II}_{i_{{10}}} r^5}{4}\right)
\end{equation}
\begin{multline}
    \label{Eqn_u_e10r}
    u_{e_{{10}_r}} = \frac{A^{I}_{e_{{10}}}}{2 r^4} + \frac{9 A^{II}_{e_{{10}}}}{28 r^6}+\frac{B^{I}_{e_{{10}}}}{2 r^2}+\frac{B^{II}_{e_{{10}}}}{4 r^4} + \bigg(\frac{3 A^{I}_{e_{{10}}}}{2 r^4} + \frac{5 A^{II}_{e_{{10}}}}{7 r^6} + \frac{3 B^{I}_{e_{{10}}}}{2 r^2} + \frac{B^{II}_{e_{{10}}}}{2 r^4}\bigg)\cos (2\theta) \\+ \left(\frac{5 A^{II}_{e_{{10}}}}{4
   r^6} + \frac{5 B^{II}_{e_{{10}}}}{4 r^4}\right) \cos (4\theta) 
\end{multline}
\begin{equation}
    \label{Eqn_u_e10theta}
    u_{i_{{10}_\theta}} = \sin (2 \theta) \left(\frac{A^{I}_{e_{{10}}}}{r^4} + \frac{2 A^{II}_{e_{{10}}}}{7 r^6}\right)+\sin (4\theta) \left(\frac{A^{II}_{e_{{10}}}}{r^6}+\frac{B^{II}_{e_{{10}}}}{2 r^4}\right).
\end{equation}
\end{subequations}
After implementing boundary conditions at the interface, we get the expressions for the constants as follows:
\begin{equation}
    \label{Const_A_e10_I}
	A^{I}_{e_{{10}}} = \frac{9 A^S_{10} (71 \mu_r + 119)U^{I} + 28R_0(9A^E_{{{10}_{nt}}} - 4 B^E_{{{10}_{nt}}})}{1260 (\mu_r + 1)} \quad ; \quad A_{e_{10}}^{II} = \frac{7A^S_{10} U^{I}}{4} + \frac{2 B^E_{{{10}_{nt}}}}{9 (\mu_r + 1)}
\end{equation}
\begin{equation}
    \label{Const_B_e10_I}
	B^{I}_{e_{{10}}} = -\frac{3 A^S_{10} (3 \mu_r + 7) U^{I} + 7 A^E_{{{10}_{nt}}} - 2 A^E_{{{10}_{nt}}}}{35 (\mu_r + 1)} \quad ; \quad B_{e_{10}}^{II} = \frac{1}{36} \left(-9 A^S_{10} U^{I} - \frac{8 A^E_{{{10}_{nt}}}}{\mu_r + 1}\right)
\end{equation}
\begin{equation}
    \label{Const_C_i10_I}
	C^{I}_{i_{{10}}} = \frac{-6 A^S_{10} \mu_r U^{I} + 6 A^S_{10} U^{I} + 7 A^E_{{{10}_{nt}}} - 2 A^E_{{{10}_{nt}}}}{35 \mu_r +35} \quad ; \quad C_{i_{10}}^{II} = A^S_{10} U^{I} + \frac{2 B^E_{{{10}_{nt}}}}{9 (\mu_r + 1)}
\end{equation}
\begin{equation}
    \label{Const_D_i10_I}
	D^{I}_{i_{{10}}} = \frac{27 A^S_{10} (7 \mu_r + 3) U^{I} + 7 (4 B^E_{{{10}_{nt}}} - 9 A^E_{{{10}_{nt}}})}{315 (\mu_r + 1)} \quad ; \quad D_{i_{10}}^{II} = \frac{9 A^S_{10} U^{I}(\mu_r + 1) - 4 B^E_{{{10}_{nt}}}}{18(\mu_r + 1)}
\end{equation}
Substitution of velocity field in the momentum equation and integration gives the expression for pressure in drop and ambient phases at $\order{Ca_E}$ as,
\begin{subequations}
\begin{multline}
    \label{Eqn_p_i10}
	p_{i_{10}} = \frac{\mu_r r^2 \left(-28 C_{i_{10}}^{II} + 196 D^{I}_{i_{{10}}} + 99 D_{i_{10}}^{II} r^2\right)}{56} \\ +
	\frac{\mu_r r^2 \cos(2\theta) \left(-84 C_{i_{10}}^{II} + 588 D^{I}_{i_{{10}}} + 220 D_{i_{10}}^{II} r^2\right) }{56} 
	+ \frac{55D_{i_{10}}^{II} \mu_r r^4 \cos(4\theta)}{8}
\end{multline}
\begin{equation}
    \label{Eqn_p_e10}
	p_{e_{10}} =  \frac{10 B^{I}_{e_{{10}}} r^2 + 9 B_{e_{10}}^{II}}{10 r^5} + \frac{\cos(2\theta) \left(3 B^{I}_{e_{{10}}} r^2 + 2 B_{e_{10}}^{II}\right)}{r^5} + \frac{7 B_{e_{10}}^{II} \cos(4\theta)}{2 r^5}.
\end{equation}
\end{subequations}
Thus, we have obtained the velocity and pressure field at $\order{Ca_E}$.
Net normal stress at the interface at $\order{Ca_E}$ comes from the normal stress due to electric field, hydrodynamic normal stress at $\order{Ca_E}$ and hydrodynamic normal stress due to shape oscillations of the drop at $\order{1}$. Thus, we have,
\begin{equation}
	\tau_{{10}_{nn}} = \tau^E_{{10}_{nn}} + \tau^H_{{10}_{nn}} + \tau^{II}_{{00}_{nn}}
\end{equation}
$\tau^{II}_{{00}_{nn}}$ and $\tau^E_{{10}_{nn}}$ are given by \Cref{Eqn_NormalHydrodynamicStress_II,Eqn_NormalElectricStress_CaE} respectively and
\begin{multline}
    \tau^H_{{10}_{nn}} = \bigg[- \bigg(p_{i_{10}} + f_{10}\frac{\partial p_{i_{00}}}{\partial r}\bigg) + \tau_{i_{{10}_{rr}}} 
	+ f_{10}\frac{\partial \tau_{i_{{00}_{rr}}}}{\partial r} - 2{f}_{10}^{\prime}\tau_{i_{{00}_{r\theta}}} \bigg]_{r=1} \\- \bigg[- \bigg(p_{e_{10}} + f_{10}\frac{\partial p_{e_{00}}}{\partial r}\bigg) + \tau_{e_{{10}_{rr}}} 
	+ f_{10}\frac{\partial \tau_{e_{{00}_{rr}}}}{\partial r} - 2{f}_{10}^{\prime}\tau_{e_{{00}_{r\theta}}}\bigg]_{r=1}
\end{multline}
which is given by,
\begin{multline}
	\tau^H_{{{10}_{nn}}} = -\frac{27 A_{e_{10}}^{II}}{7} - 4A_{e_{{10}_{I}}} + 29 B_{e_{10}}^{II} - 3 B_{e_{{10}_{I}}} - \frac{ A^S_{10} U^{I}(71\mu_r - 61)}{8} - C_{i_{{10}_{I}}} \mu_r 
	\\- \frac{ U^{II} (16 + \mu_r(48 - 64\mu_r))}{8}  - \frac{\mu_r (4 C_{i_{10}}^{II} - D_{i_{{10}_{I}}})}{2}  - \frac{81D_{i_{10}}^{II} \mu_r}{56}
	\\+ \bigg(- \frac{60 A_{e_{10}}^{II}}{7} - 12 A_{e_{{10}_{I}}} + 6 B_{e_{10}}^{II} - 9B_{e_{{10}_{I}}} + \frac{3A^S_{10} U^{I} (11 - \mu_r)}{2}  - 3 C_{i_{{10}_{I}}} \mu_r 
	\\+ \frac{(3 \mu_r (48 \mu_r + 89) + 114) U^{II}}{2} - \frac{3\mu_r (3 C_{i_{10}}^{II}- D_{i_{{10}_{I}}})}{2}   - \frac{45D_{i_{10}}^{II} \mu_r}{14}\bigg)\cos(2\theta)  
	\\+ 3 \bigg( - 5 A_{e_{10}}^{II} - \frac{9 B_{e_{10}}^{II}}{2} + \frac{A^S_{10} (17 \mu_r + 53) U^{I}}{8} - \frac{5C_{i_{10}}^{II} \mu_r}{2} - \frac{15D_{i_{10}}^{II} \mu_r}{8}\bigg)\cos(4\theta).
\end{multline}
Implementation of the normal stress boundary condition at the interface gives the 
correction to the interface shape at $\order{Ca_E^2}$, given by $f_{20}$.
We write $f_{20}$ in terms of Legendre modes as,
\begin{multline}
	f_{20} = A_{20}^S + B_{20}^S\cos2\theta + C_{20}^S\cos4\theta \bigg(A_{20}^{SP} - \frac{B_{20}^{SP}}{3} - \frac{C_{20}^{SP}}{15}\bigg)  + \bigg(\frac{4B_{20}^{S}}{3} - \frac{16C_{20}^{S}}{21}\bigg)P_2(\cos\theta) \\+ \frac{64C_{20}^{S}}{35}P_4(\cos\theta) 
\end{multline}
where $A^S_{{20}_{P_0}}$ is calculated based on the volume conservation condition and is given by, $A^S_{{20}_{P_0}} = - \frac{1}{5}(A^S_{10})^2$
and remaining constants are given by,
\begin{subequations}
\begin{multline}
   \label{Const_AS_20_P2}
	A^S_{{20}_{P_2}} = \frac{4B^{II}_{e_{10}} - 28A^{I}_{e_{10}}}{7} - 3 B^{I}_{e_{10}}  + \frac{1}{14} (10 (A^S_{10})^2 - 24 (A^S_{10}) (\mu_r - 1) U^{I} \\+ 7(3 \mu_r + 2) (16 \mu_r + 19) U^{II}) + \frac{(7 B^E_{{{10}_{nn}}} - 4 C^E_{{{10}_{nn}}} - 21 C^{I}_{i_{10}} \mu_r)}{21}  - \frac{\mu_r (C^{II}_{i_{10}} - 7 D^{I}_{i_{10}})}{14} 
\end{multline}
\begin{multline}
   \label{Const_AS_20_P4}
	A^S_{{20}_{P_4}} = -\frac{32 A^{II}_{e_{10}}}{21} + \frac{2}{315} \left(45 (A^S_{10})^2 + 6 (A^S_{10}) (17 \mu_r + 53) U^{I}\right) - \frac{48 B^{II}_{e_{10}}}{35} - \frac{16}{21} C^{II}_{i_{10}} \mu_r \\- \frac{4}{7} D^{II}_{i_{10}} \mu_r + \frac{32 C^E_{{{10}_{nn}}}}{315}.
\end{multline}
\end{subequations}

\subsection{Solution of \texorpdfstring{$\order{De}$}{Lg} equations}
\label{Sec_Sol_De}
Both the governing equations and boundary conditions for electric field are homogeneous leading to  zero electric field components at $\order{De}$.
Considering the flow equations, at $\order{De}$, continuity equations are,
\begin{equation}
    \bm{\nabla}.\bm{u}_{i_{01}} = 0 \quad \text{and} \quad \bm{\nabla}.\bm{u}_{e_{01}} = 0
\end{equation}
and momentum equations are,
\begin{equation}
    -\bm{\nabla}p_{i_{01}} + \bm{\nabla}.\bm{\tau}_{i_{01}} = 0 \quad \text{and} \quad  -\bm{\nabla}p_{e_{01}} + \bm{\nabla}.\bm{\tau}_{e_{01}} = 0
\end{equation}
where
\begin{subequations}
\begin{equation}
	\bm{\tau}_{i_{01}} = \mu_r(\nabla \bm{u}_{i_{01}} + (\nabla \bm{u}_{i_{01}})^T) - \Lambda_{i} (\bm{u}_{i_{00}}.\bm{\nabla}\bm{\tau}_{p_{i_{00}}} - \bm{\tau}_{p_{i_{00}}}\bm{\nabla}\bm{u}_{i_{00}} - (\bm{\nabla}\bm{u}_{i_{00}})^T \bm{\tau}_{p_{i_{00}}})
\end{equation}
\begin{equation}
	\bm{\tau}_{e_{01}} = (\nabla \bm{u}_{e_{01}} + (\nabla \bm{u}_{e_{01}})^T) - \Lambda_{e}(\bm{u}_{e_{00}}.\bm{\nabla}\bm{\tau}_{p_{e_{00}}} - \bm{\tau}_{p_{e_{00}}}\bm{\nabla}\bm{u}_{e_{00}} - (\bm{\nabla}\bm{u}_{e_{00}})^T \bm{\tau}_{p_{e_{00}}}).
\end{equation}
\end{subequations}
Governing equations are subjected to the boundary conditions, 
\begin{subequations}
\begin{equation}
	\bm{u}_{e_{01}} = 0 \quad \text{as} \quad r \rightarrow \infty \quad ; \quad \frac{\partial \bm{u}_{i_{01}}}{\partial r} = 0 \quad \text{at} \quad r = 0
\end{equation}
\begin{equation}
	u_{i_{{01}_r}} = u_{e_{{01}_r}} = 0 \quad \text{at} \quad r = 1 \quad ; \quad u_{i_{{01}_\theta}} = u_{e_{{01}_\theta}} \quad \text{at} \quad r = 1
\end{equation}
\begin{equation}
	\tau_{i_{{01}_{r\theta}}} = {\tau}_{ e_{{01}_{r\theta}}} \quad \text{at} \quad r = 1 \quad ; \quad  (-p_{e_{01}} + \tau_{e_{{01}_{rr}}}) - (-p_{i_{01}} + \tau_{i_{{01}_{rr}}}) = -(2 f_{11} + \cot\theta f_{11}^{\prime} + f_{11}^{\prime\prime})
\end{equation}
\end{subequations}
Representing velocity field in terms of Stokes streamfunction, the governing equations for streamfunctions are,
\begin{subequations}
\begin{multline}
	\bigg[\frac{\partial^2 }{\partial r^2} + \frac{\sin\theta}{r^2}\frac{\partial }{\partial \theta}\bigg(\frac{1}{\sin\theta}\frac{\partial }{\partial \theta}\bigg)\bigg]^2\Psi_{i_{01}} \\= - \frac{\Lambda_{i}}{\mu_r} r\sin\theta \bm{\nabla}\times\bigg( \bm{\nabla}.\bigg(\pdv{\bm{\tau}_{p_{i_{00}}}}{t} + \bm{u}_{i_{00}}.\bm{\nabla}\bm{\tau}_{p_{i_{00}}} - \bm{\tau}_{p_{i_{00}}}\bm{\nabla}\bm{u}_{i_{00}} - (\bm{\nabla}\bm{u}_{i_{00}})^T \bm{\tau}_{p_{i_{00}}}\bigg)\bigg) \\
	= -\Lambda_{i} r \sin\theta\bm{\nabla}\times \left(\bm{\nabla}.\left(\bm{\tau}_{p_{i_{{00}_{(2)}}}}\right)\right)
\end{multline}
\begin{multline}
	\bigg[\frac{\partial^2 }{\partial r^2} + \frac{\sin\theta}{r^2}\frac{\partial }{\partial \theta}\bigg(\frac{1}{\sin\theta}\frac{\partial }{\partial \theta}\bigg)\bigg]^2\Psi_{e_{01}} \\= -\Lambda_{e} r \sin\theta\bm{\nabla}\times\bigg(\bm{\nabla}.\bigg(\pdv{\bm{\tau}_{p_{e_{00}}}}{t} + \bm{u}_{e_{00}}.\bm{\nabla}\bm{\tau}_{p_{e_{00}}} - \bm{\tau}_{p_{e_{00}}}\bm{\nabla}\bm{u}_{e_{00}} - (\bm{\nabla}\bm{u}_{e_{00}})^T \bm{\tau}_{p_{e_{00}}}\bigg)\bigg) \\
	= -\Lambda_{e} r \sin\theta\bm{\nabla}\times \left(\bm{\nabla}.\left(\bm{\tau}_{p_{e_{{00}_{(2)}}}}\right)\right)
\end{multline}
\end{subequations}
where $\bm{\tau}_{p_{{{00}_{(2)}}}}$ is the convective derivative of polymeric stress at $\order{1}$. 
Polymeric stress at $\order{1}$ is given by,
\begin{equation}
\bm{\tau}_{p_{i_{00}}} = \mu_r(1 - \beta_i)(\bm{\nabla}u_{i_{00}} + (\bm{\nabla}u_{i_{00}})^T) \quad \text{and} \quad \bm{\tau}_{p_{e_{00}}} = (1 - \beta_e)(\bm{\nabla}u_{e_{00}} + (\bm{\nabla}u_{e_{00}})^T)
\end{equation}
Convective derivative of polymeric stress tensor is given by,
\begin{equation}
    \bm{\tau}_{p_{(2)}} = \pdv{\bm{\tau}_p}{t} + \bm{u}.\bm{\tau}_p - \bm{\tau}_p\grad\bm{u} - \bm{\tau}_p(\grad\bm{u})^T
\end{equation}
Taking curl of the divergence of convective derivative of $\bm{\tau}_{p_{{00}}}$ and substituting in streamfunction equations,
\begin{subequations}
\begin{multline}
    \bigg[\frac{\partial^2}{\partial r^2} + \frac{\sin\theta}{r^2}\frac{\partial }{\partial \theta}\bigg(\frac{1}{\sin\theta}\frac{\partial}{\partial}\bigg)\bigg]^2\Psi_{e_{01}} = 24(\beta_e - 1) \Lambda_{e} \bigg[\left(\frac{9}{r^7} - \frac{35}{r^9}\right) \\+ \left(\frac{15}{r^7} - \frac{45}{r^9}\right)\cos(2\theta)\bigg](U^{I})^2\cos\theta\sin^2\theta
\end{multline}
\begin{equation}
    \bigg[\frac{\partial^2}{\partial r^2} + \frac{\sin\theta}{r^2}\frac{\partial }{\partial \theta}\bigg(\frac{1}{\sin\theta}\frac{\partial}{\partial}\bigg)\bigg]^2\Psi_{i_{01}} = 336(\beta_i - 1) \Lambda_{i} r^3 (U^{I})^2 \cos\theta\sin^2\theta.
\end{equation}
\end{subequations}
Thus, streamfunction equations for $\order{De}$ are inhomogeneous with homogeneous boundary conditions. Here, we choose the form of streamfunction based on the right-hand side of the streamfunction equations, as, $\Psi_{01} = (f^{I}_{01}(r) + f^{II}_{01}(r) \cos2\theta)\sin^2\theta\cos\theta$. 
Considering that velocity should go to zero far away from the drop and should be bounded at the center, we have, $A_{i_{01}}^{I} = B_{i_{01}}^{I} = C_{e_{01}}^{I} = D_{e_{01}}^{I} = 0$ and $A_{i_{01}}^{II} = B_{i_{01}}^{II} = C_{e_{01}}^{II} = D_{e_{01}}^{II} = 0$. 
Thus, the equations for streamfunctions become,
\begin{subequations}
\begin{equation}
    \Psi_{i_{01}} = \bigg[\bigg(\frac{2 (\beta_i - 1) \Lambda_{i} r^7 (U^I)^2}{3} + C_{i_{01}}^{I}r^3 + D_{i_{01}}^{I}r^5 + \frac{D_{i_{01}}^{II}r^7}{7}\bigg) + \bigg(C_{i_{01}}^{II} r^5 + D_{i_{01}}^{II} r^7\bigg)\cos 2\theta\bigg]\sin^2\theta\cos\theta
\end{equation}
\begin{multline}
    \Psi_{e_{01}} = \bigg[\bigg(-\Lambda_{e}(\beta_e - 1)(U^I)^2\bigg(\frac{1}{r^5} - \frac{1}{2 r^3}\bigg) + \frac{A_{e_{01}}^{I}}{r^2} + B_{e_{01}}^{I} + \frac{A_{e_{01}}^{II}}{7r^4}\bigg) \\ + \bigg(-3\Lambda_{e}(\beta_e - 1)(U^I)^2\bigg(\frac{3}{2r^3} + \frac{1}{r^5}\bigg) + \frac{A_{e_{01}}^{II}}{r^4} + \frac{B_{e_{01}}^{II}}{r^2}\bigg)\cos 2\theta\bigg]\sin^2\theta\cos\theta.
\end{multline}
\end{subequations}
where $U^I$ is given by \autoref{Eqn_U_I}.
Thus, the components of velocity are given by,
\begin{subequations}
\begin{multline}
    \label{Eqn_u_i01r}
     u_{i_{{01}_r}} = \frac{C_{i_{01}}^I r}{2} + \frac{(C_{i_{01}}^{II} + 2D_{i_{01}}^{I}) r^3}{4} + \frac{9D_{i_{01}}^{II}r^5}{28} + \frac{(-1 + \beta_i)\Lambda_{i}r^5(U^I)^2}{3} + \bigg(\frac{3C_{i_{01}}^{I}r}{2} + \frac{(C_{i_{01}}^{II} + 3D_{i_{01}}^{I})r^3}{2} \\+ \frac{5D_{i_{01}}^{II} r^5}{7} + ((-1 + \beta_i)\Lambda_{i}r^5(U^I)^2)\bigg)\cos(2\theta) + \bigg(\frac{5C_{i_{01}}^{II}r^3}{4} + \frac{5D_{i_{01}}^{II}r^5}{4}\bigg)\cos(4\theta)
\end{multline}
\begin{multline}
   \label{Eqn_u_i01theta}
    u_{i_{{01}_\theta}} = -\bigg(\frac{3C_{i_{01}}^{I}r}{2} + \frac{5D_{i_{01}}^{I}r^3}{2} + \frac{D_{i_{01}}^{II}r^5}{2} + \frac{7(-1 + \beta_i)\Lambda_{i}r^5(U^I)^2)}{3}\bigg)\sin(2\theta) \\- \bigg(\frac{5C_{i_{01}}^{II}r^3}{4} + \frac{7D_{i_{01}}^{II}r^5}{4}\bigg)\sin(4\theta)
\end{multline}
 \begin{multline}
    \label{Eqn_u_e01r}
     u_{e_{{01}_r}} = - \frac{7(-1 + \beta_e)\Lambda_{e}(U^I)^2}{8r^5} - \frac{5(-1 + \beta_e)\Lambda_{e}(U^I)^2}{4r^7}  + {9A_{e_{01}}^{II}}{28r^6} + \frac{A_{e_{01}}^{I}}{2r^4} + \frac{B_{e_{01}}^{II}}{4r^4} + \frac{B_{e_{01}}^{I}}{2r^2}  
      \\+ \bigg(- \frac{3(-1 + \beta_e)\Lambda_{e}(U^I)^2}{2r^5} - \frac{3(-1 + \beta_e)\Lambda_{e}(U^I)^2}{r^7} + \frac{5A_{e_{01}}^{II}}{7r^6} + \frac{3A_{e_{01}}^{I}}{2r^4} + \frac{B_{e_{01}}^{II}}{2r^4} + \frac{3B_{e_{01}}^{I}}{2r^2}\bigg)\cos(2\theta) \\ 
      + \bigg(- \frac{45(-1 + \beta_e)\Lambda_{e}(U^I)^2}{8r^5} - \frac{15(-1 + \beta_e)\Lambda_{e}(U^I)^2}{4r^7} + \frac{5A_{e_{01}}^{II}}{4r^6} + \frac{5B_{e_{01}}^{I}}{4r^4} \bigg)\cos(4\theta)
 \end{multline}
 \begin{multline}
    \label{Eqn_u_e01theta}
    u_{e_{{01}_\theta}} = \bigg(\frac{2A_{e_{01}}^{II}}{7r^6} + \frac{A_{e_{01}}^{I}}{r^4} + \frac{3( \beta_e - 1)\Lambda_{e}(U^I)^2}{4r^5} - \frac{5(\beta_e - 1)\Lambda_{e}(U^I)^2}{2r^7}\bigg)\sin(2\theta) \\+ \bigg(\frac{A_{e_{01}}^{II}}{r^6} + \frac{B_{e_{01}}^{II}}{2r^4} - \frac{(27 (\beta_e - 1)\Lambda_{e}(U^I)^2}{8r^5} - \frac{15(\beta_e - 1)\Lambda_{e}(U^I)^2}{4r^7}\bigg)\sin(4\theta).
 \end{multline}
\end{subequations}
We find eight constants in above expressions by implementing the interface boundary conditions as, 
\begin{subequations}
\begin{equation}
    \label{Const_A_e01_I}
    A_{e_{01}}^{I} = -\frac{(\beta_i - 1)\Lambda_{i}\mu_r}{15(\mu_r + 1)}\left(15 t_R\pdv{U^{I}}{t} + 7(U^{I})^2\right) - \frac{(\beta_e - 1)\Lambda_{e}}{140(\mu_r + 1)}\left(140 t_R\pdv{U^{I}}{t} + (98 + 25\mu_r)(U^{I})^2\right)
\end{equation}
\begin{equation}
   \label{Const_A_e01_II}
    A_{e_{01}}^{II} = -\frac{5(\beta_i - 1) \Lambda_{i} \mu_r (U^{I})^2}{ (\mu_r + 1)} + \frac{(\beta_e - 1) \Lambda_{e} (8 + 27\mu_r) (U^{I})^2}{ 4(\mu_r + 1)}
\end{equation}
\begin{equation}
   \label{Const_B_e01_I}
    B_{e_{01}}^{I} = \frac{(\beta_i - 1)\Lambda_{i}\mu_r}{105(\mu_r + 1)}\left(105 t_R\pdv{U^{I}}{t} + 124(U^{I})^2\right) + \frac{(\beta_e - 1)\Lambda_{e}}{35(\mu_r + 1)}\left(35 t_R\pdv{U^{I}}{t} + 2(16 - 5\mu_r)(U^{I})^2\right)
\end{equation}
\begin{equation}
   \label{Const_B_e01_II}
    B_{e_{01}}^{II} = \frac{5(\beta_i - 1) \Lambda_{i} \mu_r (U^{I})^2}{ (\mu_r + 1)} + \frac{(\beta_e - 1) \Lambda_{e} (22 + 3\mu_r) (U^{I})^2}{ 4(\mu_r + 1)}
\end{equation}
\begin{equation}
   \label{Const_C_i01_I}
    C_{i_{01}}^{I} = - \frac{(\beta_e - 1)\Lambda_{e}}{5(\mu_r + 1)}\left(5 t_R\pdv{U^{I}}{t} + 6(U^{I})^2\right) - \frac{(\beta_i - 1)\Lambda_{i}}{105(\mu_r + 1)}\left(105 t_R\mu_r\pdv{U^{I}}{t} + 2(- 35 + 27\mu_r)(U^{I})^2\right)
\end{equation}
\begin{equation}
   \label{Const_C_i01_II}
    C_{i_{01}}^{II} = - \frac{19(\beta_e - 1) \Lambda_{e} (U^{I})^2}{ 4(\mu_r + 1)} - \frac{5(\beta_i - 1) \Lambda_{i} \mu_r (U^{I})^2}{(\mu_r + 1)} 
\end{equation}
\begin{equation}
   \label{Const_D_i01_I}
    D_{i_{01}}^{I} =\frac{(\beta_e - 1)\Lambda_{e}}{140(\mu_r + 1)}\left(140 t_R\pdv{U^{I}}{t} + 73(U^{I})^2\right) + \frac{(\beta_i - 1)\Lambda_{i}}{15(\mu_r + 1)}\left(15 t_R\mu_r\pdv{U^{I}}{t} - (20 + 13\mu_r)(U^{I})^2\right) 
\end{equation}
\begin{equation}
   \label{Const_D_i01_II}
    D_{i_{01}}^{II} = \frac{19(\beta_e - 1) \Lambda_{e} (U^{I})^2}{ 4(\mu_r + 1)} + \frac{5(\beta_i - 1) \Lambda_{i} \mu_r (U^{I})^2}{(\mu_r + 1)}.
\end{equation}
\end{subequations}
By substitution of velocity field in momentum equation and integrating we get pressure at $\order{De}$ as,
\begin{subequations}
\begin{multline}
    \label{Eqn_p_i01}
	p_{i_{01}} = \frac{55D_{i_{01}}^{II} \mu_r r^4 \cos(4\theta)}{8}  
	+ \cos (2\theta) \left(\mu_r r^2 \left(\frac{21 D_{i_{01}}^{I}}{2} - \frac{3 C_{i_{01}}^{II}}{2}\right) + \frac{55}{14} D_{i_{01}}^{II} \mu_r r^4\right)
	\\+ \mu_r r^2 \left(\frac{7 D_{i_{01}}^{I}}{2} - \frac{C_{i_{01}}^{II}}{2}\right) + \frac{99}{56} D_{i_{01}}^{II} \mu_r r^4
	(\beta_i - 1)\Lambda_{i} \bigg[
	- \frac{111 \mu_r r^4 (U^{I})^2 \cos (4\theta)}{8} 
	+ \cos(2\theta) \\\left((U^{I})^2 \left(6 \mu_r r^2-\frac{17 \mu_r r^4}{2}\right) - \frac{21\mu_r r^2}{2}\pdv{U^{I}}{t}\right)
	+ (U^{I})^2 \left(2 \mu_r r^2 - \frac{45 \mu_r r^4}{8}\right)-\frac{7 \mu_r r^2}{2}\pdv{U^{I}}{t}\bigg]
\end{multline}
\begin{multline}
    \label{Eqn_p_e01}
	p_{e_{01}} = \frac{7 B_{e_{01}}^{II} \cos(4\theta)}{2 r^5} + \left(\frac{3 B_{e_{01}}^{I}}{r^3} + \frac{2 B_{e_{01}}^{II}}{r^5}\right)\cos(2\theta) + \frac{B_{e_{01}}^{I}}{r^3} + \frac{9 B_{e_{01}}^{II}}{10 r^5}
    + (\beta_e - 1)\Lambda_{e} \\ \bigg[ (U^{I})^2 
	\left(-\frac{45}{4 r^{10}} + \frac{75}{4 r^8} - \frac{63}{4 r^6}\right) \cos(4\theta) +  \bigg(- \frac{3}{r^3}\pdv{U^{I}}{t} + (U^{I})^2 \left( - \frac{35}{r^{10}} + \frac{45}{r^8} - \frac{21}{r^6}\right) \bigg)\cos(2\theta) \\- \frac{1}{r^3}\pdv{U^{I}}{t} + (U^{I})^2 \left(-\frac{199}{4 r^{10}} + \frac{225}{4 r^8} - \frac{77}{4 r^6}\right)\bigg].
\end{multline}
\end{subequations}
Thus, we have obtained the velocity and pressure field at $\order{De}$. Now, to calculate the deformation at $\order{Ca_EDe}$, we implement the normal stress boundary condition. 
Net normal stress at the interface, $\tau^H_{{01}_{nn}} = \left( -p_{e_{01}} + \tau_{e_{{01}_{rr}}}\right)|_{r=1} - \left(- p_{i_{01}} + \tau_{i_{{01}_{rr}}}\right)|_{r=1}$ is given by,
\begin{multline}
	\tau^H_{{01}_{nn}} = \cos(4\theta)\bigg[15 A_{e_{01}}^{II} + \frac{27 B_{e_{01}}^{II}}{2} + \frac{15}{2} C_{i_{01}}^{II} \mu_r + \frac{45}{8} D_{i_{01}}^{II} \mu_r - \frac{81 (\beta_i - 1) \Lambda_{i} \mu_r (U^{I})^2}{8} \\- 93 (\beta_e - 1)\Lambda_{e} (U^{I})^2\bigg] 
	+ \cos(2\theta)\Bigg[\frac{168 A_{e_{01}}^{I} + 84 B_{e_{01}}^{II}}{14} + \frac{60 A_{e_{01}}^{II}}{7} + 9 B_{e_{01}}^{I} + 3 C_{i_{01}}^{I} \mu_r \\+ \frac{3}{2} \mu_r (3 C_{i_{01}}^{II} - D_{i_{01}}^{I}) + \frac{45}{14} D_{i_{01}}^{II} \mu_r 
	+ \frac{(\beta_e - 1) \Lambda_{e}}{14}  \left(42 \pdv{U^{I}}{t} - 616  (U^{I})^2\right) \\+ \frac{(\beta_i - 1) \Lambda_{i}}{14} \left(63 \pdv{U^{I}}{t} \mu_r - 161 \mu_r (U^{I})^2\right)\Bigg]
	- \frac{3360 A_{e_{01}}^{I} + 2436 B_{e_{01}}^{II}}{840} \\- \frac{27 A_{e_{01}}^{II}}{7} - 3 B_{e_{01}}^{I} - C_{i_{01}}^{I} \mu_r + \frac{1}{2} \mu_r (D_{i_{01}}^{I} - 4 C_{i_{01}}^{II}) - \frac{81}{56} D_{i_{01}}^{II} \mu_r 
	\\+ \frac{(\beta_i - 1) \Lambda_{i}}{840} \left(7595 \mu_r (U^{I})^2-1260 \pdv{U^{I}}{t} \mu_r \right) + \frac{(\beta_e-1) \Lambda_{e}}{840} \left(19320 (U^{I})^2 - 840 \pdv{U^{I}}{t}\right).
\end{multline}
Implementation of normal stress boundary condition gives $f_{11}$.
Interface correction $f_{11}$ can be written in terms of Legendre modes as follows.
\begin{multline}
f_{11} = A_{11}^S + B_{11}^S\cos2\theta + C_{11}^S\cos4\theta = 
% A^S_{{11}_{P_0}} + A^S_{{11}_{P_2}}P_2(\cos\theta) + A^S_{{11}_{P_4}} P_4(\cos\theta) =\\
\bigg(A_{11}^{S} - \frac{B_{11}^{S}}{3} - \frac{C_{11}^{S}}{15}\bigg) + \bigg(\frac{4B_{11}^{S}}{3} - \frac{16C_{11}^{S}}{21}\bigg)P_2(\cos\theta) \\+ \frac{64C_{11}^{S}}{35}P_4(\cos\theta).
\end{multline}
where from the volume conservation condition, we have, $A^S_{{11}_{P_0}} = 0$ and $A^S_{{11}_{P_2}}$ and $A^S_{{11}_{P_4}}$ are given by,
\begin{subequations}
\begin{multline}
    \label{Const_AS_11_P2}
	A^S_{{11}_{P_2}} = (\beta_i - 1)\Lambda_{i} \left(\frac{\mu_r (3 \mu_r (t_R - 1) + 2 t_R - 3)}{2 (\mu_r + 1)}\pdv{U^{I}}{t} + \frac{2 \mu_r (123 \mu_r + 92) (U^{I})^2}{105 (\mu_r + 1)}\right) \\
	+ (\beta_e - 1) \Lambda_{e}\bigg(\frac{(3 \mu_r t_R - 2 \mu_r + 2 t_R - 2)}{2 (\mu_r + 1)} + \frac{(79 \mu_r + 16) (U^{I})^2}{105 (\mu_r+1)}\bigg)
\end{multline}
\begin{equation}
    \label{Const_AS_11_P4}
	A^S_{{11}_{P_4}} = \frac{4 (\beta_i - 1) \Lambda_{i} \mu_r (52 \mu_r + 47) (U^{I})^2}{105 (\mu_r + 1)} + \frac{(\beta_e - 1) \Lambda_{e} (- 101 \mu_r - 120) (U^{I})^2}{105 (\mu_r + 1)}.
\end{equation}
\end{subequations}
Thus, we have obtained the deformation of drop at $\order{Ca_EDe}$.

Thus, we have obtained the electric field and the flow field at $\order{1}$, $\order{Ca_E}$ and $\order{De}$ and we have also calculated the corrections to the drop interface at $\order{Ca_E}$, $\order{Ca_E^2}$ and $\order{Ca_EDe}$.

\bibliography{references}

@article{basaran2013nonstandard,
  title={Nonstandard inkjets},
  author={Basaran, Osman A and Gao, Haijing and Bhat, Pradeep P},
  journal={Annual Review of Fluid Mechanics},
  volume={45},
  number={1},
  pages={85--113},
  year={2013},
  publisher={Annual Reviews}
}

@article{lau2017ink,
  title={Ink-jet printing of micro-electro-mechanical systems (MEMS)},
  author={Lau, Gih-Keong and Shrestha, Milan},
  journal={Micromachines},
  volume={8},
  number={6},
  pages={194},
  year={2017},
  publisher={MDPI}
}

@article{eow2002electrostatic,
  title={Electrostatic enhancement of coalescence of water droplets in oil: a review of the technology},
  author={Eow, John S and Ghadiri, Mojtaba},
  journal={Chemical Engineering Journal},
  volume={85},
  number={2-3},
  pages={357--368},
  year={2002},
  publisher={Elsevier}
}

@article{alvarado2010enhanced,
  title={Enhanced oil recovery: an update review},
  author={Alvarado, Vladimir and Manrique, Eduardo},
  journal={Energies},
  volume={3},
  number={9},
  pages={1529--1575},
  year={2010},
  publisher={MDPI}
}

@article{zhang2011application,
  title={Application of variable frequency technique on electrical dehydration of water-in-oil emulsion},
  author={Zhang, Yanzhen and Liu, Yonghong and Ji, Renjie and Wang, Fei and Cai, Baoping and Li, Hang},
  journal={Colloids and Surfaces A: Physicochemical and Engineering Aspects},
  volume={386},
  number={1-3},
  pages={185--190},
  year={2011},
  publisher={Elsevier}
}

@article{kelly1984electrostatic,
  title={The electrostatic atomization of hydrocarbons},
  author={Kelly, AJ},
  journal={J. Inst. Energy;(United Kingdom)},
  volume={57},
  number={431},
  year={1984}
}

@incollection{law2018electrostatic,
  title={Electrostatic atomization and spraying},
  author={Law, S Edward},
  booktitle={Handbook of electrostatic processes},
  pages={429--456},
  year={2018},
  publisher={CRC Press}
}

@article{hines1966electrostatic,
  title={Electrostatic atomization and spray painting},
  author={Hines, RL},
  journal={Journal of Applied Physics},
  volume={37},
  number={7},
  pages={2730--2736},
  year={1966},
  publisher={American Institute of Physics}
}

@article{moreau2015electrohydrodynamic,
  title={Electrohydrodynamic force produced by a corona discharge between a wire active electrode and several cylinder electrodes--Application to electric propulsion},
  author={Moreau, Eric and Benard, Nicolas and Alicalapa, Fr{\'e}d{\'e}ric and Douy{\`e}re, Alexandre},
  journal={Journal of Electrostatics},
  volume={76},
  pages={194--200},
  year={2015},
  publisher={Elsevier}
}

@inproceedings{huh2019numerical,
  title={Numerical simulation of electrospray thruster extraction},
  author={Huh, Henry and Wirz, Richard E},
  booktitle={Proceedings of the 36th International Electric Propulsion Conference, Vienna, Austria},
  pages={15--20},
  year={2019}
}

@article{laser2004review,
  title={A review of micropumps},
  author={Laser, Daniel J and Santiago, Juan G},
  journal={Journal of micromechanics and microengineering},
  volume={14},
  number={6},
  pages={R35},
  year={2004},
  publisher={IOP Publishing}
}

@article{stone2004engineering,
  title={Engineering flows in small devices: microfluidics toward a lab-on-a-chip},
  author={Stone, Howard A and Stroock, Abraham D and Ajdari, Armand},
  journal={Annu. Rev. Fluid Mech.},
  volume={36},
  number={1},
  pages={381--411},
  year={2004},
  publisher={Annual Reviews}
}

@article{phan2025demand,
  title={On-demand electrostatic droplet sorting and splitting},
  author={Phan, Hoang Anh and Nguyen, Kien and Pham, Phong Tuan and Do Quang, Loc and Thu, Hang Bui and Lam, Dang Bao and Jen, Chun-Ping and Thanh, Tung Bui and Duc, Trinh Chu},
  journal={Sensors and Actuators A: Physical},
  volume={385},
  pages={116311},
  year={2025},
  publisher={Elsevier}
}

@article{wilson1921iii,
  title={III. Investigations on lighting discharges and on the electric field of thunderstorms},
  author={Wilson, Charles Thomson Rees},
  journal={Philosophical Transactions of the Royal Society of London. Series A, Containing Papers of a Mathematical or Physical Character},
  volume={221},
  number={582-593},
  pages={73--115},
  year={1921},
  publisher={The Royal Society London}
}

@article{blanchard1963electrification,
  title={The electrification of the atmosphere by particles from bubbles in the sea},
  author={Blanchard, Duncan C},
  journal={Progress in oceanography},
  volume={1},
  pages={73--202},
  year={1963},
  publisher={Elsevier}
}

@article{simpson1909electricity,
  title={On the electricity of rain and its origin in thunderstorm},
  author={Simpson, GC},
  journal={Phil. Trans., A},
  volume={209},
  pages={397--413},
  year={1909}
}

@article{o1953distortion,
  title={The distortion of aerosol droplets by an electric field},
  author={O'Konski, Chester T and Thacher Jr, Henry C},
  journal={The Journal of Physical Chemistry},
  volume={57},
  number={9},
  pages={955--958},
  year={1953},
  publisher={ACS Publications}
}

@article{taylor1964disintegration,
  title={Disintegration of water drops in an electric field},
  author={Taylor, Geoffrey Ingram},
  journal={Proceedings of the Royal Society of London. Series A. Mathematical and Physical Sciences},
  volume={280},
  number={1382},
  pages={383--397},
  year={1964},
  publisher={The Royal Society London}
}

@article{allan1962particle,
  title={Particle behaviour in shear and electric fields I. Deformation and burst of fluid drops},
  author={Allan, RS and Mason, SG},
  journal={Proceedings of the Royal Society of London. Series A. Mathematical and Physical Sciences},
  volume={267},
  number={1328},
  pages={45--61},
  year={1962},
  publisher={The Royal Society London}
}

@article{o1957electric,
  title={Electric free energy and the deformation of droplets in electrically conducting systems},
  author={O'Konski, Chester T and Harris, Frank E},
  journal={The Journal of Physical Chemistry},
  volume={61},
  number={9},
  pages={1172--1174},
  year={1957},
  publisher={ACS Publications}
}

@article{taylor1966studies,
  title={Studies in electrohydrodynamics. I. The circulation produced in a drop by an electric field},
  author={Taylor, Geoffrey Ingram},
  journal={Proceedings of the Royal Society of London. Series A. Mathematical and Physical Sciences},
  volume={291},
  number={1425},
  pages={159--166},
  year={1966},
  publisher={The Royal Society London}
}

@article{melcher1969electrohydrodynamics,
  title={Electrohydrodynamics: a review of the role of interfacial shear stresses},
  author={Melcher, JR and Taylor, GI},
  journal={Annual review of fluid mechanics},
  volume={1},
  number={1},
  pages={111--146},
  year={1969},
  publisher={Annual Reviews 4139 El Camino Way, PO Box 10139, Palo Alto, CA 94303-0139, USA}
}

@article{torza1971electrohydrodynamic,
  title={Electrohydrodynamic deformation and bursts of liquid drops},
  author={Torza, S and Cox, RG and Mason, SG},
  journal={Philosophical Transactions of the Royal Society of London. Series A, Mathematical and Physical Sciences},
  volume={269},
  number={1198},
  pages={295--319},
  year={1971},
  publisher={The Royal Society London}
}

@article{ajayi1978note,
  title={A note on Taylor’s electrohydrodynamic theory},
  author={Ajayi, OO},
  journal={Proceedings of the Royal Society of London. A. Mathematical and Physical Sciences},
  volume={364},
  number={1719},
  pages={499--507},
  year={1978},
  publisher={The Royal Society London}
}

@article{saville1997electrohydrodynamics,
  title={Electrohydrodynamics: the Taylor-Melcher leaky dielectric model},
  author={Saville, DA1435033},
  journal={Annual review of fluid mechanics},
  volume={29},
  number={1},
  pages={27--64},
  year={1997},
  publisher={Annual Reviews 4139 El Camino Way, PO Box 10139, Palo Alto, CA 94303-0139, USA}
}

@article{sozou1972electrohydrodynamics,
  title={Electrohydrodynamics of a liquid drop: the time-dependent problem},
  author={Sozou, C},
  journal={Proceedings of the Royal Society of London. A. Mathematical and Physical Sciences},
  volume={331},
  number={1585},
  pages={263--272},
  year={1972},
  publisher={The Royal Society London}
}

@article{vlahovska2009electrohydrodynamic,
  title={Electrohydrodynamic model of vesicle deformation in alternating electric fields},
  author={Vlahovska, Petia M and Gracia, Ruben Serral and Aranda-Espinoza, Said and Dimova, Rumiana},
  journal={Biophysical journal},
  volume={96},
  number={12},
  pages={4789--4803},
  year={2009},
  publisher={Elsevier}
}

@article{thaokar2012dielectrophoresis,
  title={Dielectrophoresis and deformation of a liquid drop in a non-uniform, axisymmetric AC electric field},
  author={Thaokar, RM},
  journal={The European Physical Journal E},
  volume={35},
  pages={1--15},
  year={2012},
  publisher={Springer}
}

@article{mohanty2024analytical,
  title={An analytical investigation of liquid droplet deformation in a superposed electric field consisting of an alternating and a constant electric field},
  author={Mohanty, Bikash and Bandopadhyay, Aditya},
  journal={Physics of Fluids},
  volume={36},
  number={11},
  year={2024},
  publisher={AIP Publishing}
}

@article{esmaeeli2018electrohydrodynamics,
  title={Electrohydrodynamics of a liquid drop in AC electric fields},
  author={Esmaeeli, A and Halim, Md Abdul},
  journal={Acta Mechanica},
  volume={229},
  pages={3943--3962},
  year={2018},
  publisher={Springer}
}

@article{sahu2020simulations,
  title={Simulations of a weakly conducting droplet under the influence of an alternating electric field},
  author={Sahu, Kirti Chandra and Tripathi, Manoj Kumar and Chaudhari, Jay and Chakraborty, Suman},
  journal={Electrophoresis},
  volume={41},
  number={23},
  pages={1953--1960},
  year={2020},
  publisher={Wiley Online Library}
}

@article{soni2018electrohydrodynamics,
  title={Electrohydrodynamics of a concentric compound drop in an AC electric field},
  author={Soni, Purushottam and Thaokar, Rochish M and Juvekar, Vinay A},
  journal={Physics of Fluids},
  volume={30},
  number={3},
  year={2018},
  publisher={AIP Publishing}
}

@article{song2024electrokinetic,
  title={Electrokinetic behavior of an individual liquid metal droplet in a rotating electric field},
  author={Song, Chunlei and Tao, Ye and Liu, Weiyu and Chen, Yicheng and Yang, Ruizhe and Guo, Wenshang and Li, Biao and Ren, Yukun},
  journal={Physics of Fluids},
  volume={36},
  number={1},
  year={2024},
  publisher={AIP Publishing}
}

@article{santra2024modulating,
  title={Modulating droplet electrohydrodynamics via the interplay of extensional flow and an alternating current electric field},
  author={Santra, Somnath and Behera, Nalinikanta and Chakraborty, Suman},
  journal={Physics of Fluids},
  volume={36},
  number={10},
  year={2024},
  publisher={AIP Publishing}
}

@article{kireev2025influence,
  title={Influence of constant and alternating electric fields on the deformation and destruction of a liquid droplet},
  author={Kireev, VN and Shalabayeva, BS and Volkov, KN},
  journal={Acta Astronautica},
  volume={226},
  pages={147--156},
  year={2025},
  publisher={Elsevier}
}

@article{ha1999deformation,
  title={Deformation and breakup of a second-order fluid droplet in an electric field},
  author={Ha, Jong-Wook and Yang, Seung-Man},
  journal={Korean Journal of Chemical Engineering},
  volume={16},
  pages={585--594},
  year={1999},
  publisher={Springer}
}

@article{ha2000deformation,
  title={Deformation and breakup of Newtonian and non-Newtonian conducting drops in an electric field},
  author={Ha, Jong-Wook and Yang, Seung-Man},
  journal={Journal of Fluid Mechanics},
  volume={405},
  pages={131--156},
  year={2000},
  publisher={Cambridge University Press}
}

@article{lima2014numerical,
  title={Numerical simulation of electrohydrodynamic flows of Newtonian and viscoelastic droplets},
  author={Lima, NC and d’Avila, MA},
  journal={Journal of Non-Newtonian Fluid Mechanics},
  volume={213},
  pages={1--14},
  year={2014},
  publisher={Elsevier}
}

@article{zhao2025electrohydrodynamic,
  title={Electrohydrodynamic effects on the viscoelastic droplet deformation in shear flows},
  author={Zhao, Jiachen and Dzanic, Vedad and Wang, Zhongzheng and Sauret, Emilie},
  journal={Physics of Fluids},
  volume={37},
  number={1},
  year={2025},
  publisher={AIP Publishing}
}

@article{alves2021numerical,
  title={Numerical methods for viscoelastic fluid flows},
  author={Alves, MA and Oliveira, PJ and Pinho, FT},
  journal={Annual Review of Fluid Mechanics},
  volume={53},
  number={1},
  pages={509--541},
  year={2021},
  publisher={Annual Reviews}
}

@article{lopez2019adaptive,
  title={An adaptive solver for viscoelastic incompressible two-phase problems applied to the study of the splashing of weakly viscoelastic droplets},
  author={L{\'o}pez-Herrera, Jos{\'e}-Mar{\'\i}a and Popinet, St{\'e}phane and Castrej{\'o}n-Pita, Alfonso-Arturo},
  journal={Journal of Non-Newtonian Fluid Mechanics},
  volume={264},
  pages={144--158},
  year={2019},
  publisher={Elsevier}
}

@article{lopez2011charge,
  title={A charge-conservative approach for simulating electrohydrodynamic two-phase flows using volume-of-fluid},
  author={L{\'o}pez-Herrera, JM and Popinet, St{\'e}phane and Herrada, MA2764018},
  journal={Journal of Computational Physics},
  volume={230},
  number={5},
  pages={1939--1955},
  year={2011},
  publisher={Elsevier}
}

@article{fattal2004constitutive,
  title={Constitutive laws for the matrix-logarithm of the conformation tensor},
  author={Fattal, Raanan and Kupferman, Raz},
  journal={Journal of Non-Newtonian Fluid Mechanics},
  volume={123},
  number={2-3},
  pages={281--285},
  year={2004},
  publisher={Elsevier}
}

@article{das2026effect,
  title={Effect of viscoelasticity on electrohydrodynamic drop deformation},
  author={Das, Santanu Kumar and Bangar, Sarika Shivaji and Dalal, Amaresh and Tomar, Gaurav},
  journal={International Journal of Multiphase Flow},
  volume={197},
  pages={105633},
  year={2026},
  publisher={Elsevier}
}

@article{bangar2026large,
  title={On Large Deformations of Oldroyd-B Drops in a Steady Electric Field},
  author={Bangar, Sarika Shivaji and Tomar, Gaurav},
  journal={arXiv preprint arXiv:2602.01891},
  year={2026}
}

@article{BANGAR2026105645,
title = {On the deformation of a shear thinning viscoelastic drop in a steady electric field},
journal = {Journal of Non-Newtonian Fluid Mechanics},
volume = {350},
pages = {105645},
year = {2026},
issn = {0377-0257},
author = {Sarika Shivaji Bangar and Gaurav Tomar}
}
\end{document}